\documentclass[12pt]{article}

\usepackage[utf8]{inputenc}
\usepackage[T1]{fontenc}
\usepackage{amsfonts,amsmath,amssymb,dsfont}
\usepackage{mathtools}      
\usepackage{amsthm}
\usepackage{thm-restate}
\usepackage{graphicx}
\usepackage{subcaption}
\usepackage{booktabs}
\usepackage{algorithmicx}
\usepackage{algorithm}
\usepackage{algpseudocode}
\usepackage{enumerate}
\usepackage{tikz}
\usepackage{xcolor}
\usepackage{xurl}           
\usepackage{natbib}         
\usepackage{hyperref}
\usepackage{cleveref}

\newcommand{\blind}{0}

\DeclareMathOperator*{\argmin}{arg\,min}
\DeclareMathOperator*{\argmax}{arg\,max}
\newcommand{\tr}{\operatorname{tr}}
\newcommand{\sgn}{\operatorname{sgn}}
\newcommand{\supp}{\operatorname{supp}}
\newcommand{\diag}{\operatorname{diag}}

\newcommand{\pkgname}{\texttt{PartialCorrelation\allowbreak SpikeSlab.jl}}

\newtheorem{theorem}{Theorem}
\newtheorem{lemma}[theorem]{Lemma}

\newtheorem{corollary}[theorem]{Corollary}
\theoremstyle{definition}
\newtheorem{definition}[theorem]{Definition}
\newtheorem*{assumptions}{Assumptions}
\theoremstyle{remark}

\begin{document}

\def\spacingset#1{\renewcommand{\baselinestretch}%
{#1}\small\normalsize} \spacingset{1}


\if0\blind
{
  \title{\bf A Partial Correlation Spike-and-Slab Estimator for Sparse Gaussian Graphical Models}
  \author{Bautista Arenaza\\
    CONICET and Department of Physics and Biology Applied to Health,\\
    Centro At\'omico Bariloche and Instituto Balseiro, Argentina\\
    and \\
    In\'es Samengo \\
    CONICET and Department of Physics and Biology Applied to Health,\\
    Centro At\'omico Bariloche and Instituto Balseiro, Argentina\\
    and \\
    Sebastián Risau-Gusman \\
    CONICET and Department of Physics and Biology Applied to Health,\\
    Centro At\'omico Bariloche, Argentina}
  \maketitle
} \fi

\if1\blind
{
  \bigskip
  \bigskip
  \bigskip
  \begin{center}
    {\LARGE\bf A Partial Correlation Spike-and-Slab Estimator for Sparse Gaussian Graphical Models}
  \end{center}
  \medskip
} \fi

\bigskip
\begin{abstract}
We propose the partial correlation spike-and-slab (PCSS) estimator of a sparse Gaussian graphical model, which places a spike-and-slab prior on the partial correlations.
Because partial correlations are bounded, the slab can be uniform, leaving a single hyperparameter, while the mixture weight of the slab is inferred from the data.
Large partial correlations are then nearly unpenalized, and the estimated graph is invariant to rescaling the variables.
We select the hyperparameter so that, when the variables are independent, the estimator favours the empty graph.
The selected value depends only on the sample size and the number of variables, so a single fit suffices.
Under a weak-coupling condition, and from an initial estimate that separates true edges from other pairs, the fitting algorithm recovers the support with errors bounded by a multiple of the sampling error that does not depend on the hyperparameter.
In simulations on three random graph models, wherever PCSS recovers part of the support, its precision stays above $0.97$, and none of six other estimators has both higher precision and higher recall.
PCSS is implemented in the Julia package \pkgname{}.
\end{abstract}

\noindent%
\vfill

\newpage
\spacingset{1.45} 

\section{Introduction}\label{sec:intro}

The partial correlation between two variables measures the dependence between them that is not mediated by the remaining variables.
For Gaussian variables, it vanishes if and only if the two variables are conditionally independent given the rest.
The nonzero partial correlations define a graph on the variables, the partial correlation network, which for Gaussian variables is the Gaussian graphical model of the distribution~\citep{lauritzen1996}.
The Pearson correlation, in contrast, also includes indirect dependencies, transmitted along the paths of this graph.
\citet{arenaza2025expansion} make this explicit by expanding each Pearson correlation as a sum of contributions from the paths that connect the two variables, so that two variables are in general correlated whenever such a path exists.
Partial correlation networks are used, for instance, for the functional connectivity between brain regions, where they are among the most accurate methods~\citep{smith2011}, and for the relations between psychological variables~\citep{epskamp2018}.

The number of partial correlations grows quadratically with the number of variables, so estimating them from a sample usually requires a structural assumption.
The usual one, which goes back to the covariance selection models of~\citet{dempster1972}, is that the graph is sparse, with few edges compared with the number of pairs of variables, as expected when each variable depends directly on few others.
A sparse graph remains compatible with dense Pearson correlations, since its paths connect variables that are not joined by an edge.

The standard estimator of a sparse Gaussian graphical model is the graphical lasso~\citep{yuan2007,banerjee2008,friedman2008}, which maximizes the Gaussian log-likelihood with an $\ell_1$ penalty on the entries of the precision matrix.
Three of its limitations motivate this work.
First, the $\ell_1$ penalty shrinks every nonzero entry toward zero by the same amount, so the estimates of large entries are biased~\citep{fan2001,fan2009}, and the bound on the estimation error grows with the hyperparameter that sets the size of the penalty~\citep{ravikumar2011}.
Second, the penalty acts on the entries of the precision matrix, which depend on the scales of the variables, so the estimated graph can change when the variables are rescaled~\citep{carter2023pcglasso}.
Standardizing the variables to unit sample variance removes this dependence, but it can degrade the estimate, in particular for graphs whose vertices have very different degrees~\citep{carter2023pcglasso}.
Third, the hyperparameter is in practice chosen by computing the estimator at each value of a grid and selecting one value with an information criterion, such as the extended Bayesian information criterion~\citep{foygel2010,epskamp2018}.

Each of these limitations has been addressed separately.
Folded-concave penalties such as SCAD~\citep{fan2001} and MCP~\citep{zhang2010}, used for graphical models in~\citet{fan2009}, have a derivative that vanishes for large entries, so the estimates of large entries are nearly unbiased.
They still act on the entries of the precision matrix, however, and are not scale invariant~\citep{carter2023pcglasso}.
Spike-and-slab priors on the entries of the precision matrix~\citep{wang2015,gan2018} also shrink large entries less than small ones, but their slab carries a hyperparameter of its own, in addition to that of the spike.
The partial correlation graphical lasso~\citep{carter2023pcglasso} moves the $\ell_1$ penalty from the entries of the precision matrix to the partial correlations, which makes it scale invariant, but it keeps the bias of the $\ell_1$ penalty, as does the joint sparse regression estimator SPACE~\citep{peng2009}.
Nodewise regression~\citep{meinshausen2006} selects the neighbours of each variable with a separate lasso regression, and its hyperparameter can be set without a grid, so as to bound the probability of falsely connecting two connected components of the graph.
In its original form, however, it estimates the graph and not the values of the partial correlations.
None of these estimators combines nearly unpenalized large partial correlations, scale invariance, and a hyperparameter set without computing the estimator over a grid.

We propose the partial correlation spike-and-slab (PCSS) estimator, which places a spike-and-slab prior on the partial correlations, and thus combines the spike-and-slab priors of~\citet{wang2015,gan2018} with the parametrization of the partial correlation graphical lasso.
The spike is a Laplace density, as in the spike-and-slab lasso~\citep{rockova2018}.
Since partial correlations are bounded, the slab can be uniform, so the prior has a single hyperparameter $\gamma$, the inverse width of the spike, while the mixture weight of the slab is inferred together with the partial correlations.
The derivative of the resulting penalty decays exponentially with the magnitude of the partial correlation, so large partial correlations are nearly unpenalized.
The log-posterior is not concave, and it has no global maximum when $n < p$ (\Cref{prop:existence}), so we define the PCSS estimate as the output of a procedure rather than as a maximizer.
The procedure ascends the log-posterior with an expectation--conditional maximization algorithm~\citep{meng1993}, whose E-step turns the probability that each partial correlation belongs to the slab into a penalty weight for that entry, as the local linear approximation~\citep{zou2008} does for folded-concave penalties.
Since the result of this algorithm depends on its starting point, the procedure also specifies the initial estimate, which it obtains from fits at increasing values of $\gamma$.

Our theoretical contributions are the following.
\begin{itemize}
  \item The estimates of the partial correlations at every local maximum of the log-posterior are unchanged when the variables are rescaled, and so are the estimated graph and the PCSS estimate (\Cref{thm:scale-invariance}).
    The estimator therefore has the selection scale invariance of~\citet{carter2023pcglasso}.
  \item We select $\gamma$ so that, under a diagonal ground truth, the estimator favours the diagonal model.
    Above an explicit lower bound on $\gamma$, the diagonal estimate is a strict local maximum of the log-posterior and is preferred to every estimate with a bounded number of nonzero partial correlations, no two of which involve the same variable (\Cref{thm:matching}).
    The selected value $\gamma_q$ depends on the data only through $n$ and $p$, and exceeds this lower bound with probability at least $1 - q$ in the limit $p \to \infty$, uniformly in $n$ (\Cref{thm:calibration}).
    Both results extend to estimates whose nonzero partial correlations form a forest with a bounded number of edges and bounded degrees (\Cref{app:forest}).
    Like the hyperparameter of nodewise regression~\citep{meinshausen2006}, it is set before computing the estimator, which then requires a single fit.
  \item Under a weak-coupling condition on the ground truth, together with conditions on the sample, on $\gamma$ and on the initial estimate, the algorithm recovers the support in its first iteration and keeps it in every later one, and its errors are eventually bounded by multiples of the sampling error (\Cref{thm:convergence}).
    These bounds do not depend on $\gamma$, whereas the bound for the graphical lasso grows with its hyperparameter~\citep{ravikumar2011}.
    The weak-coupling condition takes the place of the irrepresentability condition of~\citet{ravikumar2011}, and neither condition implies the other (\Cref{tab:thresholds}).
    The theorem requires an initial estimate that separates the true edges from the remaining pairs, and the path of increasing values of $\gamma$ is a heuristic designed to produce one.
\end{itemize}

We compare the recovery of the support by PCSS and by six other estimators on simulated data (\Cref{sec:simulations}).
The ground truths have $p = 256$ variables and are drawn from three random graph models with two mean degrees and five magnitudes of the partial correlations, and the sample size ranges from $n = 2^{-3} p$ to $n = 2^4 p$.
Wherever PCSS recovers part of the support, no other estimator has at the same time a higher precision and a higher recall, and the precision of PCSS stays above $0.97$.
With $\gamma = \gamma_q$, PCSS gives almost the same precision and recall as with $\gamma$ selected by EBIC in most of the simulated conditions, and is the more conservative of the two in the others, with a single fit instead of many fits over a grid of values.

The PCSS estimator and the computation of $\gamma_q$ are implemented in the Julia package \pkgname{}~\citep{pcssjl}, and the scripts and data that reproduce our simulations are listed in the code and data availability statement at the end of the paper.
\Cref{sec:setup} defines the model and the PCSS estimate, and shows that the estimate is scale invariant.
\Cref{sec:gamma_selection} presents the two choices of $\gamma$, together with the guarantee behind $\gamma_q$.
\Cref{sec:algorithm} describes the procedure that computes the estimate.
\Cref{sec:theory} presents the convergence theorem, \Cref{sec:simulations} the simulations, and \Cref{sec:conclusion} our conclusions and open problems.
\Cref{app:notation} collects the notation used throughout.

\section{Bayesian inference of the partial correlation matrix}\label{sec:setup}

We consider the problem of estimating the partial correlation matrix of a multivariate Gaussian distribution from a sample of $n$ independent and identically distributed observations $\{\boldsymbol{x}_1,\dots,\boldsymbol{x}_n\}$, with $\boldsymbol{x}_k\in\mathbb{R}^p$.
We assume $\boldsymbol{x}_k\sim N_p(\boldsymbol{\mu},\Omega^{-1})$, where $\boldsymbol{\mu}$ is the mean and $\Omega$ the precision matrix (the inverse of the covariance matrix).
We use this Gaussian model to derive the estimator, but among our theoretical results only the selection of its hyperparameter in \Cref{subsec:gamma_q} requires the observations to be Gaussian.
Throughout, a hat denotes an estimate and a star the corresponding ground-truth quantity, as in $\widehat R$ and $R^*$.

The partial correlation coefficient $r_{ij}$ between variables $i$ and $j$ is the correlation between the residuals of the linear regressions of these two variables on all the remaining ones~\citep{Cramer1946}, and it can be read off directly from the precision matrix as
\begin{equation} \label{eq:partial}
  r_{ij} = -\frac {\Omega_{ij}}{\sqrt{\Omega_{ii}\Omega_{jj}}}.
\end{equation}
We collect these coefficients into the partial correlation matrix $R$, with $(R)_{ij}=r_{ij}$ for $i\neq j$ and $(R)_{ii}=0$, which is the weighted adjacency matrix of the Gaussian graphical model.
In terms of $R$, the precision matrix factorizes as
\begin{equation} \label{eq:omega-factorization}
  \Omega = \Lambda \left( I_p - R  \right) \Lambda,
\end{equation}
where $I_p$ is the $p\times p$ identity matrix and $\Lambda\in\mathbb{R}^{p\times p}$ is the diagonal matrix with $\Lambda_{ii} = \sqrt{\Omega_{ii}}$.
Because $\Lambda_{ii}^{-2}=\Omega_{ii}^{-1}$ is the conditional variance of $i$ given the remaining variables, $\Lambda$ encodes the scale of the data while $R$ carries the scale-free dependence structure.
The factorization \eqref{eq:omega-factorization} keeps these two roles separate, and it is the parametrization we use throughout.
In terms of $\Lambda$ and $R$, the probability density of a single observation reads
\begin{equation} \label{eq:likelihood}
  p(\boldsymbol{x} | \, \boldsymbol{\mu}, \Lambda, R) = \det(\Lambda) \sqrt{\frac{\det(I_p-R)}{(2\pi)^p}} \, \exp\left\{-\frac{1}{2} \left[\Lambda (\boldsymbol{x}-\boldsymbol{\mu})\right]^\top ( I_p - R ) \left[\Lambda (\boldsymbol{x}-\boldsymbol{\mu})\right]\right\}.
\end{equation}
To infer $\boldsymbol{\mu}$, $\Lambda$ and $R$, we now specify their prior distributions.

\subsection{Ignorance of location and scale} \label{subsec:ignorance}

We assume no a priori knowledge of the location $\boldsymbol{\mu}$ or the scale $\Lambda$ of the observations.
Requiring the prior to keep the same functional form under the change of variables $\boldsymbol\mu' = \boldsymbol\mu + \boldsymbol b$, $\Lambda' = D^{-1}\Lambda$, $R' = R$, where $D \in \mathbb{R}^{p \times p}$ is a diagonal matrix with positive entries and $\boldsymbol b \in \mathbb{R}^p$, fixes its dependence on $\boldsymbol\mu$ and $\Lambda$~(\citealp{jaynes2003probability}; see \Cref{app:prior}), and gives
\begin{equation*}
  p(\boldsymbol{\mu}, \Lambda, R) = \frac{1}{\det(\Lambda)} f(R),
\end{equation*}
which shows that $\boldsymbol{\mu}$, $\Lambda$ and $R$ are a priori independent, with
\begin{equation} \label{eq:prior-mu-lambda}
  \begin{aligned}
    p(\boldsymbol{\mu}) &\propto 1, \\
    p(\Lambda) &\propto \frac{1}{\det(\Lambda)},
  \end{aligned}
\end{equation}
both improper priors.
The block $R$, in contrast, receives a proper prior by construction (\Cref{subsec:sslab_prior}).
For $\boldsymbol\mu$ the impropriety has no consequence, since $\boldsymbol\mu$ is profiled out exactly by the sample mean for any sample size (\Cref{subsec:objective}).
For $\Lambda$ it affects whether the posterior has a global maximum (\Cref{prop:existence}).

\subsection{Spike-and-slab prior on the partial correlations}\label{subsec:sslab_prior}

The prior on $R$ expresses the assumption that the graph is sparse, so that many partial correlations are expected to be close to zero, although it is not known which ones.
We model this state of knowledge by assigning each partial correlation a two-component mixture density, in the spirit of the spike-and-slab construction~\citep{mitchell1988bayesian,george1993variable}.
One component, the spike, represents the entries believed to be close to zero, and the other, the slab, represents the entries about which nothing is known.
The mixture weight $\eta \in [0,1]$ weighs the slab against the spike, and thereby controls the sparsity of $R$.
It is usually treated as an external hyperparameter.
We prefer to infer it, letting the data reveal the appropriate level of sparsity, so we treat $\eta$ as an unknown parameter and specify the joint prior of $R$ and $\eta$,
\begin{equation} \label{eq:marginal-prior}
  p(R,\eta) \propto \mathds{1}_{\mathcal R_p}(R)\, \mathds{1}_{[0,1]}(\eta)
  \prod_{i<j} \Big[ \eta\, p_1(r_{ij}) + (1-\eta)\, p_0(r_{ij}) \Big],
\end{equation}
where $p_0$ and $p_1$ are the densities of the spike and of the slab, and $\mathcal R_p := \{ R \in \mathbb{R}^{p\times p} : R = R^\top,\ R_{ii} = 0,\ I_p - R \succ 0 \}$ is the set of partial correlation matrices.
The indicator $\mathds{1}_{\mathcal R_p}(R)$ restricts $R$ to values for which $\Omega = \Lambda(I_p-R)\Lambda \succ 0$.
It also couples the partial correlations, which would otherwise be independent under \eqref{eq:marginal-prior}.
The product in \eqref{eq:marginal-prior} runs only over the entries of $R$ above the diagonal.
Since $R$ is symmetric, these entries are the free parameters of the prior, and they determine the entries below the diagonal.

The spike concentrates near zero.
Rather than a degenerate spike (a Dirac delta) as in the original formulation~\citep{mitchell1988bayesian,george1993variable}, we follow~\citet{rockova2018,bai2021spike} and use a Laplace spike, here truncated to $(-1,1)$ since $|r_{ij}|<1$,
\begin{equation*}
  p_0(r_{ij}) = \frac{\tilde\gamma}{2}\, e^{-\gamma\,|r_{ij}|}\, \mathds{1}_{(-1,1)}(r_{ij}),
  \qquad
  \tilde\gamma := \frac{\gamma}{1-e^{-\gamma}},
\end{equation*}
where $\gamma > 0$ and $\tilde\gamma$ is the normalizing constant of the truncated Laplace.
The hyperparameter $\gamma$ controls the concentration of the spike.
The width of the spike is of order $1/\gamma$, so larger values of $\gamma$ encode a stronger belief that the entries assigned to the spike are close to zero.
Its selection is discussed in \Cref{sec:gamma_selection}.
When the entries of the precision matrix are given a spike-and-slab prior, the slab is a broad distribution governed by an additional hyperparameter~\citep{wang2015,gan2018}.
The partial correlations are bounded, which instead lets us take a uniform slab,
\begin{equation*}
  p_1(r_{ij}) = \frac{1}{2}\, \mathds{1}_{(-1,1)}(r_{ij}).
\end{equation*}
With these choices, the prior \eqref{eq:marginal-prior} becomes
\begin{equation} \label{eq:prior-R-eta}
  p(R,\eta) \propto \mathds{1}_{\mathcal R_p}(R)\, \mathds{1}_{[0,1]}(\eta)
  \prod_{i<j} \Big[ \eta + (1-\eta)\,\tilde\gamma\, e^{-\gamma|r_{ij}|} \Big],
\end{equation}
which depends on a single hyperparameter, $\gamma$.

If the constraint $R \in \mathcal R_p$ were absent, $\eta$ would be the prior probability that a given partial correlation is drawn from the slab, and hence the expected proportion of nonzero partial correlations.
The constraint makes this interpretation approximate a priori, but it holds at the estimate (\Cref{subsec:edges}).

\subsection{Optimization objective}\label{subsec:objective}

Bayes' rule gives the joint posterior of the parameters as
\begin{equation*}
  p(\boldsymbol\mu,\Lambda,R,\eta \mid X) \propto
  p(X \mid \boldsymbol\mu,\Lambda,R)\, p(\boldsymbol\mu)\, p(\Lambda)\, p(R,\eta),
\end{equation*}
with $p(X\mid\boldsymbol\mu,\Lambda,R)=\prod_{k=1}^n p(\boldsymbol x_k\mid\boldsymbol\mu,\Lambda,R)$ given by \eqref{eq:likelihood}, and $p(\boldsymbol\mu)$, $p(\Lambda)$, $p(R,\eta)$ given by \eqref{eq:prior-mu-lambda} and \eqref{eq:prior-R-eta}.
For fixed $\Lambda$ and $R$, the log-posterior is maximized over $\boldsymbol\mu$ at the sample mean $\bar{\boldsymbol{x}} := n^{-1}\sum_{k=1}^n \boldsymbol{x}_k$, whatever $\Lambda$ and $R$, so $\boldsymbol\mu$ can be profiled out exactly, and we fix $\widehat{\boldsymbol\mu}=\bar{\boldsymbol{x}}$ for the remainder of the paper.
Writing the sum over the observations in terms of the sample covariance $S_n := n^{-1}\sum_{k=1}^n (\boldsymbol{x}_k-\bar{\boldsymbol{x}})(\boldsymbol{x}_k-\bar{\boldsymbol{x}})^\top$, and dropping additive constants, the profiled log-posterior is
\begin{equation} \label{eq:map-objective}
  \mathcal L(\Lambda,R,\eta) = - \frac n2 \tr\!\big[S_n\Lambda(I_p-R)\Lambda\big] + (n-1)\log\det\Lambda + \frac n2\log\det(I_p-R)
  - \sum_{i<j} \mathrm{pen}(r_{ij}, \eta),
\end{equation}
subject to $R\in\mathcal R_p$ and $\eta\in[0,1]$, where
\begin{equation*}
  \mathrm{pen}(r, \eta) := -\log\Big[\eta + (1-\eta)\, \tilde\gamma\, e^{-\gamma |r|}\Big].
\end{equation*}
For $\eta = 0$ the penalty reduces to the $\ell_1$ penalty $\gamma|r|$ up to an additive constant.

Equation \eqref{eq:map-objective} can be read as a penalized Gaussian log-likelihood.
In terms of the precision matrix $\Omega=\Lambda(I_p-R)\Lambda$, the first three terms of \eqref{eq:map-objective} coincide with the Gaussian log-likelihood except for the coefficient of $\log\det\Lambda$, which is $n-1$ rather than $n$.
The function $\mathrm{pen}$ plays the role of a penalty on the partial correlations.
The penalty biases the estimates toward zero, a systematic error known as shrinkage bias, and the estimates of large entries are nearly unbiased when the derivative of the penalty vanishes for large arguments~\citep{fan2001,fan2009}.
For our penalty, writing $\partial_r\mathrm{pen}$ for the derivative of $\mathrm{pen}$ with respect to its first argument,
\begin{equation*}
  \partial_r\mathrm{pen}(r, \eta) = \sgn(r)\, \frac{\gamma\,(1-\eta)\,\tilde\gamma\, e^{-\gamma|r|}}{\eta + (1-\eta)\,\tilde\gamma\, e^{-\gamma|r|}},
  \qquad r \ne 0,
\end{equation*}
whose magnitude decays exponentially as $|r|$ grows, so for $\eta > 0$ large partial correlations are nearly unpenalized.
The decay to zero is a consequence of the uniform slab, whose density is constant.
With a Laplace slab, as in the spike-and-slab lasso~\citep{rockova2018} and in the spike-and-slab prior of~\citet{gan2018} on the entries of the precision matrix, the derivative would instead tend to the rate of the slab, which is positive.

The objective \eqref{eq:map-objective} is not concave, and may have several local maxima.
For fixed $R$ and $\eta$, however, it is a strictly concave function of the diagonal entries of $\Lambda$, with a unique maximizer whenever every $(S_n)_{ii}$ is positive, that is, whenever no variable takes the same value in every observation (\Cref{app:ecm-lambda}).
The joint dependence on $\Lambda$ and $R$, through which the improper scale prior enters, decides whether \eqref{eq:map-objective} has a global maximum.

\begin{restatable}[]{proposition}{propexistence}
  \label{prop:existence}
  Let $n \ge 2$.
  \begin{enumerate}
    \item If $n > p$ and the observations are independent, each with a distribution that assigns probability zero to every affine hyperplane of $\mathbb R^p$, then the objective \eqref{eq:map-objective} attains a global maximum over positive diagonal $\Lambda$, $R \in \mathcal R_p$ and $\eta \in [0,1]$ almost surely.
    \item If $n < p$, the objective \eqref{eq:map-objective} is unbounded above.
  \end{enumerate}
\end{restatable}

The proof is given in \Cref{app:existence}.
Neither part applies when $n = p$, and we leave this case open.
\citet{carter2025existence} study the same question for the partial correlation graphical lasso, and show that its estimate exists for $n \le p$ when the coefficient of its $\log\det\Lambda$ term is small enough, a condition that the coefficient $n-1$ of \eqref{eq:map-objective} does not satisfy.

Since \eqref{eq:map-objective} may have several local maxima, and has no global maximum when $n < p$ (\Cref{prop:existence}), the objective alone does not determine an estimate.
We define the partial correlation spike-and-slab (PCSS) estimate $(\widehat\Lambda, \widehat R, \widehat\eta)$ as the output of the procedure described in \Cref{sec:algorithm}.
This procedure is deterministic and ascends the log-posterior with an algorithm of the EM family, so its output is a function of the sample and of $\gamma$ alone.

\subsection{Scale invariance}\label{subsec:scale-invariance}

Partial correlations are invariant under rescaling of the individual variables, and so is the graph they encode.
A sound estimator should inherit this property.
We show that the PCSS estimator does, with the rescaling absorbed entirely by the scale block $\widehat\Lambda$.

\begin{theorem}\label{thm:scale-invariance}
Let $D$ be a positive diagonal matrix, and consider the rescaled sample $\boldsymbol x'_k = D \boldsymbol x_k$, $k = 1, \dots, n$.
If $(\widehat\Lambda, \widehat R, \widehat\eta)$ is a local maximum of the objective \eqref{eq:map-objective} for the original sample, then
\begin{equation*}
  \big(\widehat\Lambda', \widehat R', \widehat\eta'\big)
  = \big(D^{-1}\widehat\Lambda, \ \widehat R, \ \widehat\eta\big)
\end{equation*}
is a local maximum of the objective for the rescaled sample.
\end{theorem}

\begin{proof}
Write $D_s := \diag\big(\sqrt{(S_n)_{11}}, \dots, \sqrt{(S_n)_{pp}}\big)$, so that $S_n = D_s \rho_n D_s$, where $\rho_n$ is the sample correlation matrix, and $\Lambda_s := D_s \Lambda$ for the standardized scale.
Since diagonal matrices commute, $\tr[S_n \Lambda (I_p - R) \Lambda] = \tr[\rho_n \Lambda_s (I_p - R) \Lambda_s]$, and $\log\det\Lambda = \log\det\Lambda_s - \log\det D_s$, so $\mathcal L(\Lambda, R, \eta) = \mathcal L_\rho(\Lambda_s, R, \eta) - (n-1)\log\det D_s$ with
\begin{equation} \label{eq:standardized-objective}
  \mathcal{L}_\rho(\Lambda_s, R, \eta)
  = - \frac n2 \tr\!\big[\rho_n\Lambda_s(I_p-R)\Lambda_s\big]
  + (n-1)\log\det\Lambda_s + \frac n2\log\det(I_p-R)
  - \sum_{i<j} \mathrm{pen}(r_{ij}, \eta),
\end{equation}
which depends on the data only through $\rho_n$.
The rescaled sample has $S_n' = D S_n D$, hence $\rho_n' = \rho_n$ and $D_s' = D D_s$, so both samples have the same standardized objective.
The map $\Lambda \mapsto D_s\Lambda$ is a bijection of the positive diagonal matrices that leaves $R$ and $\eta$ unchanged, so it maps the local maxima of \eqref{eq:map-objective} to those of $\mathcal L_\rho$, and likewise for $D_s'$.
Hence $(D_s\widehat\Lambda, \widehat R, \widehat\eta)$ is a local maximum of $\mathcal L_\rho$, and $\big((D_s')^{-1} D_s \widehat\Lambda, \widehat R, \widehat\eta\big) = \big(D^{-1}\widehat\Lambda, \widehat R, \widehat\eta\big)$ is a local maximum of the objective for the rescaled sample.
\end{proof}

The PCSS estimate inherits this property.
Written in terms of $(\Lambda_s, R, \eta)$, the procedure of \Cref{sec:algorithm} that defines it depends on the data only through $\rho_n$, so it returns the same $\widehat R$ and $\widehat\eta$ for the original and the rescaled samples, and the scale estimates are related as at the end of the proof.

\citet{carter2023pcglasso} define scale invariance for estimators of the precision matrix in the same way, as $\widehat\Omega' = D^{-1}\widehat\Omega D^{-1}$, and show that the graphical lasso, SCAD and MCP do not have it.
They also show that it holds whenever the penalty is a sum of the same symmetric function of each partial correlation, combined with a logarithmic penalty on the diagonal entries of $\Omega$, as for the partial correlation graphical lasso.
The objective \eqref{eq:map-objective} has this form.
Their result concerns the maximizer of the penalized likelihood, whereas \Cref{thm:scale-invariance} applies to every local maximum, as needed for the procedure that defines the PCSS estimate.
Since the estimated edges depend only on $\widehat R$ and $\widehat\eta$ (\Cref{subsec:edges}), they are unchanged as well, a property that \citet{carter2023pcglasso} call selection scale invariance.

\subsection{Determining the edges of the graphical model}\label{subsec:edges}

Given an estimate $(\widehat\Lambda, \widehat R, \widehat\eta)$, it remains to decide which pairs $(i,j)$ constitute edges of the graphical model.
The mixture prior \eqref{eq:prior-R-eta} offers a natural criterion: each partial correlation can be attributed to one of the two components.
To make this attribution precise, we augment the prior with binary indicators $\sigma_{ij}\in\{0,1\}$, where $\sigma_{ij}=1$ assigns $r_{ij}$ to the slab and $\sigma_{ij}=0$ to the spike, by defining the joint distribution
\begin{equation} \label{eq:prior-R-sigma-eta}
  p(R,\boldsymbol\sigma,\eta) \propto \mathds{1}_{\mathcal R_p}(R)\,
  \mathds{1}_{[0,1]}(\eta) \prod_{i<j} \eta^{\sigma_{ij}}
  \Big[(1-\eta)\,\tilde\gamma\, e^{-\gamma|r_{ij}|}\Big]^{1-\sigma_{ij}},
\end{equation}
with the convention $0^0=1$.
Summing each factor over $\sigma_{ij}\in\{0,1\}$ recovers exactly \eqref{eq:prior-R-eta}.
The augmentation \eqref{eq:prior-R-sigma-eta} is therefore consistent with the prior already specified in \Cref{subsec:sslab_prior}.
Were the positive-definiteness constraint absent, \eqref{eq:prior-R-sigma-eta} would be the classical hierarchical spike-and-slab prior with independent $\sigma_{ij}\sim\text{Bernoulli}(\eta)$ indicators~\citep{mitchell1988bayesian,george1993variable,wang2015}.
With the constraint, it is still the joint prior of a hierarchical model, whose priors on $\boldsymbol\sigma$ and $\eta$ cancel the normalizing constant of $p(R \mid \boldsymbol\sigma)$, as in~\citet{wang2015}.

The augmented prior lets us assign each pair to a component.
Dividing \eqref{eq:prior-R-sigma-eta} by its sum over $\boldsymbol\sigma$, the $\boldsymbol\sigma$-free indicators cancel and the ratio factorizes over pairs, so the indicators are conditionally independent given $(R,\eta)$, each depending on $R$ only through its own entry.
The slab-membership probability of pair $(i,j)$ is
\begin{equation} \label{eq:responsibility}
  p(\sigma_{ij}=1 \mid r_{ij}, \eta)
  = \frac{\eta}{\eta + (1-\eta)\,\tilde\gamma\, e^{-\gamma|r_{ij}|}}
  = \frac{1}{1 + e^{-\gamma\left(|r_{ij}| - r_c(\eta)\right)}},
\end{equation}
a logistic function of $|r_{ij}|$ with critical threshold
\begin{equation} \label{eq:rc}
  r_c(\eta) := \frac{1}{\gamma} \left[ \log \tilde{\gamma} + \log\left(\frac{1-\eta}{\eta}\right)\right].
\end{equation}
The slab is more probable than the spike exactly when the partial correlation exceeds the threshold in magnitude,
\begin{equation*}
  p(\sigma_{ij} = 1 \mid r_{ij}, \eta) > p(\sigma_{ij} = 0 \mid r_{ij}, \eta)
  \qquad \iff \qquad
  |r_{ij}| > r_c(\eta).
\end{equation*}
We therefore define the estimated edge set as the pairs $(i,j)$ with $|\widehat r_{ij}| > r_c(\widehat\eta)$.

The slab-membership probabilities \eqref{eq:responsibility} also justify the inference of $\eta$ announced in \Cref{subsec:sslab_prior}.
Setting the derivative of \eqref{eq:map-objective} with respect to $\eta$ to zero yields
\begin{equation} \label{eq:eta-stationarity}
  \widehat\eta = \frac{1}{\ell} \sum_{i<j} p(\sigma_{ij}=1 \mid \widehat r_{ij}, \widehat\eta),
  \qquad \ell := \frac{p(p-1)}{2},
\end{equation}
so at any stationary point $\widehat\eta$ is the average slab-membership probability over the $\ell$ pairs, that is, the expected proportion of nonzero partial correlations.
The slab-membership probabilities reappear in \Cref{sec:algorithm} as the E-step of the ECM algorithm, where they determine the penalty weights of its $R$-update.

\section{Hyperparameter selection}\label{sec:gamma_selection}

The objective \eqref{eq:map-objective} depends on a single hyperparameter, the inverse width $\gamma$ of the spike, and larger values of $\gamma$ favour sparser estimates.
This section presents two ways of choosing $\gamma$.
The first, in \Cref{subsec:ebic}, is an information criterion, which selects $\gamma$ from a grid of values after computing the estimator at each of them.
The second, in \Cref{subsec:gamma_q}, selects $\gamma$ so that the estimator favours the diagonal model under a diagonal ground truth, in which all partial correlations vanish, and gives its value before any estimate is computed, from $n$ and $p$ alone.

\subsection{Extended Bayesian information criterion}\label{subsec:ebic}

The first criterion is the extended Bayesian information criterion \citep[EBIC;][]{foygel2010}, a version of the Bayesian information criterion~\citep{schwarz1978} designed for the high-dimensional regime, in which the number of candidate edges $\ell = p(p-1)/2$ is large relative to $n$.
The estimator is computed over a grid of $\gamma$ values, and the selected $\gamma$ minimizes
\begin{equation} \label{eq:ebic}
  \mathrm{EBIC}(\gamma) = -n\Big[\log\det\widehat\Omega_\gamma - \tr\big(S_n \widehat\Omega_\gamma\big)\Big]
  + k_\gamma \log n
  + 4\, \gamma_{\mathrm{ebic}}\, k_\gamma \log p,
\end{equation}
where $\widehat\Omega_\gamma = \widehat\Lambda(I_p-\widehat R)\widehat\Lambda$ is the estimate obtained with hyperparameter $\gamma$, and $k_\gamma$ is the number of nonzero entries of $\widehat\Omega_\gamma$ above the diagonal.
The last term is the extended penalty of~\citet{foygel2010}, with a parameter $\gamma_{\mathrm{ebic}} \in [0,1]$, larger values favoring sparser models, and $\gamma_{\mathrm{ebic}} = 0$ recovers the ordinary BIC.
In our simulation studies (\Cref{sec:simulations}) we fix $\gamma_{\mathrm{ebic}} = 1/2$.

\subsection{Selection under a diagonal ground truth}\label{subsec:gamma_q}

If the ground truth is diagonal, we want the estimator to return the diagonal model with probability at least $1-q$, and we call $q$ the level of the selection.
\citet{meinshausen2006} follow an analogous requirement, and set the hyperparameter of their $\ell_1$ penalty so that, under a diagonal ground truth, the probability of falsely estimating an edge is at most a prescribed level.
In their setting the requirement can be translated into an explicit value of the hyperparameter.
The quantile universal threshold of \citet{giacobino2017} applies the same requirement to a broad class of thresholding estimators, including the lasso.
For these estimators the estimate vanishes exactly when the hyperparameter is at least a threshold that depends on the sample, and the quantile universal threshold is the $(1-q)$-quantile of this threshold when all the parameters they penalize vanish, the analogue of a diagonal ground truth.

For our objective \eqref{eq:map-objective}, translating the requirement exactly into a value of $\gamma$ is intractable.
Instead, we compute the values of $\gamma$ for which the diagonal model is preferred, in terms of \eqref{eq:map-objective}, to two classes of alternatives, namely the estimates close to it and the estimates supported on a matching with a bounded number of edges.
We state these requirements below as two conditions, and show that both hold whenever $\gamma$ is at least an explicit lower bound, which depends on the sample only through its largest absolute sample correlations.
The selected hyperparameter exceeds this lower bound with probability at least $1-q$ under a diagonal ground truth, in the limit $p \to \infty$.
\Cref{app:forest} extends the second condition from matchings to forests of bounded degree.

To state the two conditions we use the following terminology.
An edge is an unordered pair $(i,j)$ of distinct indices in $\{1, \dots, p\}$, so that $(i,j)$ and $(j,i)$ denote the same edge.
A matching is a set of edges no two of which share an index, so that a matching has at most $\lfloor p/2 \rfloor$ edges.
A matrix $R$ is supported on a set of edges $\mathcal E$ if $r_{ij} = 0$ for every $i \ne j$ with $(i,j) \notin \mathcal E$.
The diagonal model is represented by the diagonal estimate
\begin{equation} \label{eq:diagonal-estimate}
  \widehat\Lambda_{ii} = \sqrt{\frac{1-1/n}{(S_n)_{ii}}},
  \qquad
  \widehat R = 0,
  \qquad
  \widehat\eta = 0 ,
\end{equation}
which maximizes \eqref{eq:map-objective} over the estimates with $R = 0$.
For a given $\gamma$, and for $\bar m$ with $1 \le \bar m \le \lfloor p/2 \rfloor$, the two conditions are the following.
\begin{enumerate}
  \renewcommand{\theenumi}{C\arabic{enumi}}
  \renewcommand{\labelenumi}{(\theenumi)}
  \item\label{cond:diag-local}
  \emph{Local optimality.}
  The diagonal estimate \eqref{eq:diagonal-estimate} is a strict local maximum of \eqref{eq:map-objective}.
  \item\label{cond:diag-matching}
  \emph{Preference over matchings.}
  $\mathcal L(\Lambda, R, \eta) < \mathcal L(\widehat\Lambda, 0, 0)$ for every positive diagonal $\Lambda$, every $\eta \in [0,1]$ and every nonzero $R \in \mathcal R_p$ supported on a matching with at most $\bar m$ edges.
\end{enumerate}

The lower bound is computed from the ordered absolute sample correlations $|\rho_n|_{(1)} \ge |\rho_n|_{(2)} \ge \dots \ge |\rho_n|_{(\ell)}$ of the $\ell$ pairs, as
\begin{equation}\label{eq:gamma-lower}
  \underline\gamma(\bar m) := \max\Bigg\{(n-1)\,|\rho_n|_{(1)}\, \frac{-\log\big(1-|\rho_n|_{(1)}^2\big)}{2\Big(1-\sqrt{1-|\rho_n|_{(1)}^2}\Big)},\ \ 1 + \frac{1}{\ell} \sum_{k=1}^{\bar m} \big(1-|\rho_n|_{(k)}^2\big)^{-(n-1)/2}\Bigg\} ,
\end{equation}
where the fraction is taken to be $1$ when $|\rho_n|_{(1)} = 0$.
The first term depends only on the largest absolute sample correlation, and the second on the $\bar m$ largest.

The following theorem shows that Conditions~(\ref{cond:diag-local}) and~(\ref{cond:diag-matching}) hold once $\gamma \ge \underline\gamma(\bar m)$.

\begin{theorem}\label{thm:matching}
  Fix $\bar m$ with $1 \le \bar m \le \lfloor p/2 \rfloor$.
  Suppose $\gamma \ge \underline\gamma(\bar m)$, where $\underline\gamma(\bar m)$ is given by \eqref{eq:gamma-lower}.
  Then Conditions~(\ref{cond:diag-local}) and~(\ref{cond:diag-matching}) hold.
\end{theorem}

\Cref{thm:matching} is proved in \Cref{app:forest}, as the case $\bar d = 1$ of \Cref{thm:forest}, which compares the diagonal model with the estimates supported on forests with at most $\bar m$ edges and maximum degree at most $\bar d$.

Choosing $\gamma$ so that $\gamma \ge \underline\gamma(\bar m)$ with probability at least $1-q$ requires the joint law of $|\rho_n|_{(1)}, \dots, |\rho_n|_{(\bar m)}$ under a diagonal ground truth.
Simulating this law directly is costly for large $p$, since each replicate forms and sorts the $\ell$ entries of a $p \times p$ correlation matrix, and the simulation must be repeated whenever $n$ or $p$ changes.
We replace it by a limit whose simulation cost does not depend on $p$.

Under a diagonal ground truth, the tail of an absolute sample correlation is, by Theorem 5.1.3 of \citet{muirhead1982},
\begin{equation*}
  \bar F_n(r) := \mathrm P\big(|(\rho_n)_{12}| > r\big) = \mathrm P\!\left(|t_{n-2}|>\frac{r\sqrt{n-2}}{\sqrt{1-r^2}}\right), \qquad r \in [0,1),
\end{equation*}
where $t_{n-2}$ has the Student distribution with $n-2$ degrees of freedom.
\Cref{lem:joint-order-stats} in \Cref{app:calibration} shows that, as $p \to \infty$, the joint law of the variables $\ell\,\bar F_n(|\rho_n|_{(k)})$, $k \le \bar m$, approaches that of the partial sums $\Gamma_k := E_1 + \dots + E_k$, where $E_1, E_2, \dots$ are independent standard exponential variables.
Since $\bar F_n$ is a bijection from $[0,1)$ onto $(0,1]$, we substitute $|\rho_n|_{(k)} = \bar F_n^{-1}(\Gamma_k/\ell)$ in $\underline\gamma(\bar m)$, and take the $(1-q)$-quantile of the result as the selected hyperparameter, that is,
\begin{equation} \label{eq:gamma-q}
  \gamma_q(\bar m) := \inf\big\{\gamma : \mathrm P\big(\underline\gamma^\star(\bar m) \le \gamma\big) \ge 1-q\big\},
\end{equation}
where $\underline\gamma^\star(\bar m)$ is the right-hand side of \eqref{eq:gamma-lower} with $|\rho_n|_{(k)}$ replaced by
\begin{equation} \label{eq:varrho}
  \varrho_k := \bar F_n^{-1}\big(\min\{\Gamma_k/\ell, 1\}\big), \qquad k \le \bar m .
\end{equation}
The minimum keeps $\Gamma_k/\ell$ in the range of $\bar F_n$.
In particular, $\gamma_q(\bar m)$ depends on the data only through $n$ and $p$.

The next theorem shows that, under a diagonal ground truth, $\gamma_q(\bar m)$ is at least $\underline\gamma(\bar m)$ with probability at least $1-q$ in the limit $p \to \infty$, whatever the sample size $n \ge 3$.

\begin{theorem}\label{thm:calibration}
  Fix $\bar m \ge 1$ and $q \in (0,1)$.
  For $n \ge 3$, let $\boldsymbol x_1,\dots,\boldsymbol x_n$ be independent $N_p(\boldsymbol\mu,\Sigma)$ vectors with $\Sigma$ diagonal and positive definite.
  Then
  \begin{equation}\label{eq:calibration}
    \liminf_{p\to\infty}\ \inf_{n \ge 3}\ \mathrm P\big(\underline\gamma(\bar m) \le \gamma_q(\bar m)\big) \ \ge\ 1-q ,
  \end{equation}
  and the same bound holds for the probability that Conditions~(\ref{cond:diag-local}) and~(\ref{cond:diag-matching}) hold with $\gamma = \gamma_q(\bar m)$.
\end{theorem}

\Cref{thm:calibration}, proved in \Cref{app:forest} as the case $\bar d = 1$ of \Cref{thm:calibration-forest}, does not guarantee that the estimator returns the diagonal model with probability $1-q$, since it covers only the estimates close to the diagonal one and those supported on matchings with at most $\bar m$ edges.
\Cref{thm:calibration-forest} gives the same guarantee for the estimates supported on forests.

We can get an idea of how $\gamma_q(\bar m)$ scales with $n$ and $p$ from a lower bound.
The fraction in \eqref{eq:gamma-lower} is at least $1$, so $\underline\gamma^\star(\bar m) \ge (n-1)\varrho_1$.
Since $\Gamma_1$ is standard exponential and $\bar F_n$ is decreasing, $\mathrm P(\varrho_1 \le r) = \exp\big(-\ell\,\bar F_n(r)\big)$ for $r \in [0,1)$, and taking the $(1-q)$-quantile gives
\begin{equation}\label{eq:gamma-q-lower}
  \gamma_q(\bar m) \ \ge\ (n-1)\,\bar F_n^{-1}\Big(\frac{-\log(1-q)}{\ell}\Big).
\end{equation}
For fixed $q$, the right-hand side is asymptotic to $2\sqrt{n\log p}$ as $n, p \to \infty$ with $(\log p)/n \to 0$.
The selected hyperparameter is therefore at least of order $\sqrt{n\log p}$.

In practice one fixes $\bar m$ and $q$ and computes \eqref{eq:gamma-q} by Monte Carlo.
For each replicate, draw $E_1, \dots, E_{\bar m}$ iid standard exponential, form the partial sums $\Gamma_k$, and evaluate $\underline\gamma^\star(\bar m)$ at
\begin{equation*}
  \varrho_k = \frac{z_k}{\sqrt{n-2+z_k^2}}, \qquad k \le \bar m,
\end{equation*}
where $z_k$ is the $(1-v_k/2)$-quantile of $t_{n-2}$, with $v_k := \min\{\Gamma_k/\ell, 1\}$.
This expression for $\varrho_k$ follows from \eqref{eq:varrho} and the definition of $\bar F_n$.
The empirical $(1-q)$-quantile of $\underline\gamma^\star(\bar m)$ over the replicates approximates $\gamma_q(\bar m)$.

In all the experiments of \Cref{sec:simulations} we take $\bar m = 5$ and $q = 0.05$.
These values are also the defaults of \pkgname{}~\citep{pcssjl}, which computes $\gamma_q(\bar m)$ by the Monte Carlo procedure above.
The edge bound $\bar m$ enters $\gamma_q(\bar m)$ only through the second term of \eqref{eq:gamma-lower}.
For $n \ge 16$ and the values of $p$ we considered, $\gamma_q(\bar m)$ takes the same value for every $\bar m$ from $1$ to $20$, to within Monte Carlo error.
The choice of $\bar m$ thus sets the number of edges of the matchings covered by \Cref{thm:calibration}, but does not change the estimate.
Comparing the diagonal model with forests of larger degree raises the selected hyperparameter, but leaves the precision and recall of the estimated graphs in our simulations almost unchanged (\Cref{app:forest}).

\section{An ECM procedure for the PCSS estimate}\label{sec:algorithm}

This section describes the procedure whose output defines the PCSS estimate (\Cref{subsec:objective}).
Its main component is an algorithm of the expectation--maximization (EM) family~\citep{dempster1977}, which increases the log-posterior \eqref{eq:map-objective} monotonically from a given initial estimate.
The algorithm treats the binary indicators of \Cref{subsec:edges} as latent variables and, rather than maximizing \eqref{eq:map-objective} directly, alternates between computing the conditional distribution of $\boldsymbol\sigma$ given the current parameters (the E-step) and increasing the resulting expected log-posterior over the parameters (the M-step).
Our M-step does not maximize the expected log-posterior.
Instead, it maximizes it over each block of parameters in turn, holding the remaining blocks fixed, which makes the scheme an instance of the expectation--conditional maximization (ECM) algorithm~\citep{meng1993}.
Each such iteration nevertheless increases the marginalized objective \eqref{eq:map-objective} (\Cref{app:ecm-monotonicity}).

\subsection{The E-step}\label{subsec:estep}

Write $\vartheta = (\Lambda, R, \eta)$ for the continuous parameters.
Given an estimate $\vartheta^{(t)}$, the E-step evaluates the slab-membership probabilities of \Cref{subsec:edges} for every pair,
\begin{equation} \label{eq:sigma_t}
  \sigma^{(t)}_{ij} := \mathbb{E}[\sigma_{ij} \mid r_{ij}^{(t)}, \eta^{(t)}]
  = \frac{1}{1 + e^{-\gamma(|r^{(t)}_{ij}| - r_c(\eta^{(t)}))}},
\end{equation}
where $r_c(\eta)$ is given by \eqref{eq:rc}.
These probabilities define the function that the M-step increases, known as the $Q$-function of the EM algorithm~\citep{dempster1977},
\begin{align} \label{eq:Q-function}
  Q(\vartheta \mid \vartheta^{(t)})
  &= -\frac n2 \tr\!\big[S_n\Lambda(I_p-R)\Lambda\big]
  + (n-1)\log\det\Lambda + \frac n2\log\det(I_p-R) \nonumber \\
  &\qquad - \gamma \sum_{i<j} \big(1-\sigma^{(t)}_{ij}\big)\,|r_{ij}|
  + \sum_{i<j} \Big[ \sigma^{(t)}_{ij}\log\eta + \big(1-\sigma^{(t)}_{ij}\big)\log(1-\eta) \Big],
\end{align}
subject to $R\in\mathcal R_p$ and $\eta\in[0,1]$.
Up to an additive constant independent of $\vartheta$, \eqref{eq:Q-function} is the expectation of the log-posterior of the full hierarchy, including the indicators, under their conditional distribution given $\vartheta^{(t)}$ (\Cref{app:ecm-estep}).

\subsection{The M-step}\label{subsec:mstep}

The $Q$-function \eqref{eq:Q-function} is not jointly concave in $\vartheta$, but it is concave in each of the three blocks $\eta$, $R$ and $\Lambda$ when the other two are held fixed (\Cref{app:ecm-mstep}).
The M-step maximizes $Q(\cdot \mid \vartheta^{(t)})$ over each block in turn, holding the remaining blocks at their most recent values, so it does not decrease $Q$.
Since $\eta$ enters $Q$ separately from $(\Lambda, R)$, its update does not depend on $(\Lambda, R)$, and we perform it first.
We then maximize over $R$ with $\Lambda = \Lambda^{(t)}$, which yields $R^{(t+1)}$, and over $\Lambda$ with $R = R^{(t+1)}$, which yields $\Lambda^{(t+1)}$.

\paragraph{$\eta$-update.}

The maximizer of $Q(\cdot \mid \vartheta^{(t)})$ over $\eta$ is the average slab-membership probability over the $\ell$ pairs,
\begin{equation}\label{eq:eta_update}
  \eta^{(t+1)}
  = \frac{1}{\ell} \sum_{i<j} \sigma_{ij}^{(t)}.
\end{equation}
At a fixed point of the algorithm, this update reproduces the stationarity identity \eqref{eq:eta-stationarity}.

\paragraph{$R$-update.}

With $\Lambda = \Lambda^{(t)}$ fixed, maximizing $Q(\cdot \mid \vartheta^{(t)})$ over $R$ is equivalent to a convex problem (\Cref{app:ecm-mstep}).
Writing $\Theta = I_p-R$, $\widehat\Sigma^{(t)} = \Lambda^{(t)} S_n \Lambda^{(t)}$ and $\gamma^{(t)}_{ij} := \gamma\big(1-\sigma^{(t)}_{ij}\big)$ for the penalty weight of the pair $(i,j)$, it reads
\begin{equation} \label{eq:R_subproblem}
  \Theta^{(t+1)} = \argmin_{\Theta : \, \Theta_{ii} = 1, \, \Theta \succ 0} \
  n \left[\tr\big(\widehat\Sigma^{(t)} \Theta\big) - \log\det\Theta\right]
  + \sum_{i \neq j} \gamma^{(t)}_{ij} |\Theta_{ij}|,
\end{equation}
and the update of the partial correlations is recovered as $R^{(t+1)} = I_p - \Theta^{(t+1)}$.
Problem \eqref{eq:R_subproblem} is a graphical lasso with entry-specific penalty weights~\citep{fan2009} and with the diagonal of $\Theta$ constrained to be one.
By \eqref{eq:sigma_t}, the penalty weights are the derivatives of the penalty at the current estimate, $\gamma^{(t)}_{ij} = |\partial_r\mathrm{pen}(r^{(t)}_{ij}, \eta^{(t)})|$, so the $R$-update replaces each $\mathrm{pen}(r_{ij}, \eta^{(t)})$ by its linearization in $|r_{ij}|$ at $|r^{(t)}_{ij}|$, as the local linear approximation of~\citet{zou2008} does for folded-concave penalties.
We solve \eqref{eq:R_subproblem} by the alternating direction method of multipliers \citep[ADMM;][]{boyd2011}, as described in \Cref{app:ecm-admm}.

\paragraph{$\Lambda$-update.}

With $R = R^{(t+1)}$ fixed, collecting the diagonal elements of $\Lambda$ in a vector $\boldsymbol\lambda = (\Lambda_{11}, \dots, \Lambda_{pp})^\top$, maximizing $Q(\cdot \mid \vartheta^{(t)})$ over $\Lambda$ is equivalent to
\begin{equation} \label{eq:lambda_subprob}
  \boldsymbol\lambda^{(t+1)} = \argmin_{\boldsymbol\lambda : \, \lambda_i > 0} \
  \frac{1}{2} \boldsymbol{\lambda}^\top \big[S_n \circ (I_p - R^{(t+1)})\big] \boldsymbol{\lambda}
  - \left(1-\frac{1}{n}\right) \sum_i \log \lambda_i,
\end{equation}
where $\circ$ denotes the Hadamard (entrywise) product, and the update of the scales is recovered as $\Lambda^{(t+1)} = \diag(\boldsymbol\lambda^{(t+1)})$.
Whenever every $(S_n)_{ii}$ is positive, problem \eqref{eq:lambda_subprob} is strictly convex and has a unique minimizer (\Cref{app:ecm-lambda}), which we compute by coordinate descent.

\subsection{The full algorithm}\label{subsec:full-algorithm}

Algorithm~\ref{alg:pcss} assembles the preceding steps.
Each iteration consists of the E-step \eqref{eq:sigma_t} followed by the three conditional maximizations of \Cref{subsec:mstep}.
The log-posterior \eqref{eq:map-objective} increases monotonically along the iterates (\Cref{app:ecm-monotonicity}), and the algorithm stops when both the increase of the log-posterior and the change of the estimate fall below tolerance (\Cref{app:ecm-stopping}).
The initial estimate must satisfy the constraints of the model, namely $R^{(0)} \in \mathcal R_p$, diagonal $\Lambda^{(0)}$ with $\Lambda^{(0)}_{ii} > 0$, and $\eta^{(0)} \in [0,1]$.
The $R$- and $\Lambda$-updates are stated separately as Algorithms~\ref{alg:R-update} and~\ref{alg:lambda-update} in \Cref{app:ecm-details}.

We compare the estimate returned by Algorithm~\ref{alg:pcss} with the diagonal estimate \eqref{eq:diagonal-estimate}, and take as the PCSS estimate whichever of the two has the larger value of \eqref{eq:map-objective}.
The package \pkgname{}~\citep{pcssjl} performs this comparison by default.

\begin{algorithm}[t]
\caption{ECM algorithm for the PCSS estimator, at fixed $\gamma$}
\label{alg:pcss}
\begin{algorithmic}[1]
\Require sample covariance $S_n$, hyperparameter $\gamma$, ADMM parameter $\beta > 0$, initial estimate $(\Lambda^{(0)}, R^{(0)}, \eta^{(0)})$ with $R^{(0)} \in \mathcal R_p$, positive diagonal $\Lambda^{(0)}$, $\eta^{(0)} \in [0,1]$
\State $\tilde\gamma \gets \gamma / (1 - e^{-\gamma})$
\State $\ell \gets p(p-1)/2$
\State $(\Lambda, R, \eta) \gets (\Lambda^{(0)}, R^{(0)}, \eta^{(0)})$
\Repeat
  \State $\sigma_{ij} \gets \big[1 + e^{-\gamma(|r_{ij}| - r_c(\eta))}\big]^{-1}$ for every pair $i \neq j$ \Comment{E-step, \eqref{eq:sigma_t}}
  \State $\eta \gets \dfrac{1}{\ell} \displaystyle\sum_{i<j} \sigma_{ij}$ \Comment{$\eta$-update, \eqref{eq:eta_update}}
  \State $R \gets \Call{RUpdate}{S_n, \Lambda, R, \boldsymbol\sigma, \beta}$ \Comment{Algorithm~\ref{alg:R-update}}
  \State $\Lambda \gets \Call{LambdaUpdate}{S_n, R, \Lambda}$ \Comment{Algorithm~\ref{alg:lambda-update}}
\Until{the increase of $\mathcal L(\Lambda, R, \eta)$ and the change of $(\Lambda, R, \eta)$ fall below tolerance}
\State \Return $(\Lambda, R, \eta)$
\end{algorithmic}
\end{algorithm}

\subsection{Initial condition}\label{subsec:initial_condition}

Since the objective \eqref{eq:map-objective} is not concave and Algorithm~\ref{alg:pcss} only increases it, the estimate it returns can depend on the initial estimate $(\Lambda^{(0)}, R^{(0)}, \eta^{(0)})$.
A simple initial estimate is the diagonal estimate \eqref{eq:diagonal-estimate}, but starting Algorithm~\ref{alg:pcss} from it at the target value of $\gamma$ has two drawbacks (\Cref{app:ecm-path}).
First, the diagonal estimate has $\eta^{(0)} = 0$, which implies $\eta^{(t)} = 0$ for every $t \ge 1$, so the slab plays no role.
Second, for large values of $\gamma$ the algorithm does not leave the diagonal estimate.

We therefore reach the target value of $\gamma$ through fits at increasing values of $\gamma$, as done for the spike-and-slab lasso~\citep{rockova2018,bai2021spike}.
Given a number of stages $n_\mathrm{path} \ge 2$, we set
\begin{equation*}
  \gamma_k := \gamma_0^{\,1-k/(n_\mathrm{path}-1)}\, \gamma^{\,k/(n_\mathrm{path}-1)},
  \qquad k = 0, \dots, n_\mathrm{path}-1,
  \qquad \gamma_0 := \frac{n-1}{\sqrt n} .
\end{equation*}
The fit at $\gamma_0$ starts from the diagonal estimate \eqref{eq:diagonal-estimate}.
The fit at $\gamma_k$, $k \ge 1$, starts from the estimate obtained at $\gamma_{k-1}$, with $\eta$ replaced by the fraction of nonzero off-diagonal entries of its $R$.
At small values of $\gamma$ the spike is broad, so many nonzero entries are assigned to it and $\eta$ is underestimated, which without the reset would push the next fit toward an overly sparse estimate.
Only the estimate at $\gamma_{n_\mathrm{path}-1} = \gamma$ is retained.
In all our experiments we take $n_\mathrm{path} = 4$.

The fit at $\gamma_0$ acts as a screening step, which leaves the diagonal estimate through the pairs with $|(\rho_n)_{ij}| > 1/\sqrt n$, about one standard deviation of the sample correlation of two independent variables, and the later fits prune its support (\Cref{app:ecm-path}).
We expect the path to assign most entries of the true edges to the slab, and most of the remaining entries to the spike, before $\gamma$ reaches its target.
The package \pkgname{}~\citep{pcssjl} implements the whole procedure, with $n_\mathrm{path} = 4$ and $\gamma_0 = (n-1)/\sqrt n$ by default.

\section{Theoretical results}
\label{sec:theory}

The results of this section compare the iterates of Algorithm~\ref{alg:pcss} with the distribution that generated the sample.
For a matrix $M$ we write $\|M\|_{\max} := \max_{i,j} |M_{ij}|$ and $\|M\|_{\infty} := \max_{i} \sum_{j} |M_{ij}|$, and $\otimes$ denotes the Kronecker product.
For a diagonal matrix $D$ with positive diagonal, $\log D$ is the diagonal matrix whose entries are the logarithms of those of $D$.
We write $R^* \in \mathcal{R}_p$ for the ground-truth partial correlation matrix, with entries $r^*_{ij}$, and
\begin{equation*}
  \mathcal S := \supp(R^*) = \{(i,j) : r^*_{ij} \neq 0\},
  \qquad
  \eta^* := \frac{|\mathcal S|}{2\ell},
  \qquad
  d := \max_i \big|\{ j : (i,j) \in \mathcal S \}\big|
\end{equation*}
for its support, the fraction of nonzero partial correlations, and the largest degree of the associated graph.
The smallest nonzero partial correlation in magnitude is $R^*_{\min} := \min_{(i,j) \in \mathcal S} |r^*_{ij}|$.
We write $\Theta^* := I_p - R^*$ and $\Lambda^*$ for the ground-truth scales, so that $\Omega^* = \Lambda^*\Theta^*\Lambda^*$ is the ground-truth precision matrix and $\Sigma^* := (\Omega^*)^{-1}$.
The sampling error is $W := S_n - \Sigma^*$.
Pre- and post-multiplying by $\Lambda^*$ removes the scales from the covariance and from the sampling error alike, and we write $\widetilde\Sigma^* := \Lambda^* \Sigma^* \Lambda^* = (\Theta^*)^{-1}$ and $\widetilde W := \Lambda^* W \Lambda^*$ for the two resulting matrices.
We measure the sampling error with the scales removed.

\begin{definition}\label{def:delta}
  We set
  \begin{equation*}
    \bar\delta := \max\left(\|\widetilde W\|_{\max}, \frac{1}{2(n-1)}\right).
  \end{equation*}
\end{definition}

Since the scales are removed, $\Lambda^*$ does not enter the constants introduced below, which depend on the ground truth through $\Theta^*$ alone, and some of which also depend on $\bar\delta$.

The restriction of $\|\widetilde\Sigma^*\|_\infty$ to $\mathcal S$ is
\begin{equation*}
  \widetilde\Sigma^*_{\infty,\mathcal S} := \max_i \sum_{j \,:\, (i,j) \in \mathcal S} \big|\widetilde\Sigma^*_{ij}\big| \ \le \ \|\widetilde\Sigma^*\|_\infty.
\end{equation*}
Writing $\mathcal S^c := \{(i,j) : i \neq j,\ (i,j) \notin \mathcal S\}$ for the remaining off-diagonal pairs, we also set
\begin{equation*}
  \widetilde\Sigma^*_{\max,\mathcal S} := \max_{(i,j) \in \mathcal S} \big|\widetilde\Sigma^*_{ij}\big|,
  \qquad
  \widetilde\Sigma^*_{\max,\mathcal S^c} := \max_{(i,j) \in \mathcal S^c} \big|\widetilde\Sigma^*_{ij}\big|.
\end{equation*}
We then set $J^* := I_p + \widetilde\Sigma^* \circ \Theta^*$, which is invertible since $\widetilde\Sigma^* \circ \Theta^* \succeq 0$ by the Schur product theorem, and write $C_J := \|(J^*)^{-1}\|_{\infty}$.
We also set $\Gamma^* := \widetilde\Sigma^* \otimes \widetilde\Sigma^*$, write $\Gamma^*_{\mathcal S\mathcal S}$ for its submatrix with rows and columns indexed by $\mathcal S$, and put $C_\Gamma := \|(\Gamma^*_{\mathcal S\mathcal S})^{-1}\|_{\infty}$.
Writing $\Gamma^*_{\mathcal S^c\mathcal S}$ for the submatrix of $\Gamma^*$ with rows indexed by $\mathcal S^c$ and columns indexed by $\mathcal S$, we finally put $C_{\mathrm{irr}} := \|\Gamma^*_{\mathcal S^c\mathcal S}(\Gamma^*_{\mathcal S\mathcal S})^{-1}\|_{\infty}$, the analogue of the quantity that the irrepresentability condition of \citet{ravikumar2011} requires to be less than one.
It differs from that quantity in that it is built from $\widetilde\Sigma^*$ rather than $\Sigma^*$, and in that its index sets contain no diagonal pairs $(i,i)$, which belong to the support in \citet{ravikumar2011}.

We measure the distance between the iterates of Algorithm~\ref{alg:pcss} and the ground truth in units of $\bar\delta$.

\begin{definition}\label{def:errors}
  The normalized errors of the iterates $(\Lambda^{(t)}, R^{(t)}, \eta^{(t)})$ of Algorithm~\ref{alg:pcss}, $t \ge 0$, are
  \begin{equation*}
    a_t := \frac{\|R^{(t)} - R^*\|_{\max}}{\bar\delta},
    \qquad
    b_t := \frac{\|\log\Lambda^{(t)} - \log\Lambda^*\|_{\max}}{\bar\delta}.
  \end{equation*}
\end{definition}

$R^{(t)}$ and $\eta^{(t)}$ enter iteration $t+1$ of Algorithm~\ref{alg:pcss} only through $\sigma^{(t)}_{ij}$, given by \eqref{eq:sigma_t}.
The next definition describes pairs $(R^{(t)}, \eta^{(t)})$ for which $\sigma^{(t)}_{ij}$ is close to one for $(i,j) \in \mathcal S$ and small for $(i,j) \notin \mathcal S$.

\begin{definition}\label{def:separated}
  Let $\gamma > 1$, $\xi > 0$ and $c_0 \in [0,1)$, and set
  \begin{equation}\label{eq:boundseta2}
    \eta_{\min} := \eta^* \big(1-e^{-\gamma\xi}\big),
    \qquad
    \eta_{\max} := \eta^* \, \frac{\gamma^{1-c_0}}{\gamma^{1-c_0}-1}.
  \end{equation}
  A pair $(R, \eta)$ with $R \in \mathcal{R}_p$ and $\eta \in [0,1]$ is $(\xi, c_0)$-separated if $\eta \in [\eta_{\min}, \eta_{\max}]$ and
  \begin{equation}\label{eq:sep_estimate}
    |r_{ij}| \geq r_c(\eta_{\min}) + \xi \quad \text{for } (i,j) \in \mathcal S,
    \qquad
    |r_{ij}| \leq \frac{c_0 \log \gamma}{\gamma} \quad \text{for } (i,j) \notin \mathcal S.
  \end{equation}
\end{definition}

The $R$- and $\Lambda$-updates solve nonlinear stationarity equations, so the bounds we derive on their errors hold only while the iterates stay close to the ground truth.
Here, close means that $\|R^{(t)} - R^*\|_{\max} \leq \bar\varepsilon_R$ and $\|\log\Lambda^{(t)} - \log\Lambda^*\|_{\max} \leq \bar\varepsilon_\Lambda$ for some $\bar\varepsilon_R, \bar\varepsilon_\Lambda > 0$ that satisfy the conditions of the next definition.
This definition also introduces two factors $\kappa_R$ and $\kappa_\Lambda$ that enter our bounds on the errors of the updates.

\begin{definition}\label{def:radii}
  Let $\bar\varepsilon_R, \bar\varepsilon_\Lambda > 0$ satisfy
  \begin{equation}\label{eq:thm_eps}
    2d\,\|\widetilde\Sigma^*\|_\infty\,\bar\varepsilon_R \leq 1,
    \qquad
    2dC_\Gamma \|\widetilde\Sigma^*\|_\infty^3\,\bar\varepsilon_R < 1,
    \qquad
    2 C_J \|\widetilde\Sigma^* \circ \Theta^*\|_\infty\, \bar\varepsilon_\Lambda\, e^{2\bar\varepsilon_\Lambda} < 1.
  \end{equation}
  We call such $\bar\varepsilon_R$ and $\bar\varepsilon_\Lambda$ admissible errors, and set
  \begin{equation}\label{eq:thm_kappas}
    \kappa_R := \frac{1}{1 - 2dC_\Gamma \|\widetilde\Sigma^*\|_\infty^3\,\bar\varepsilon_R},
    \qquad
    \kappa_\Lambda := \frac{e^{2\bar\varepsilon_\Lambda}}{1 - 2 C_J \|\widetilde\Sigma^* \circ \Theta^*\|_\infty\, \bar\varepsilon_\Lambda\, e^{2\bar\varepsilon_\Lambda}}.
  \end{equation}
\end{definition}

The normalized errors $a_t$ and $b_t$ after one sweep of Algorithm~\ref{alg:pcss} can be bounded by affine functions of the errors before it.
The next definition fixes the coefficients of these functions, the fixed points $a_\infty$ and $b_\infty$ of the recursion they define, and the resulting bounds $\varepsilon_R$ and $\varepsilon_\Lambda$ on the errors of the iterates.

\begin{definition}\label{def:recursion}
  Let $\bar\varepsilon_R, \bar\varepsilon_\Lambda$ be admissible errors, and set
  \begin{equation}\label{eq:thm_constants}
    \begin{aligned}
      \alpha_R &:= \kappa_R C_\Gamma\big(e^{2\bar\varepsilon_\Lambda} + 1\big), & \beta_R &:= 2\kappa_R C_\Gamma\, e^{2\bar\varepsilon_\Lambda}\,\widetilde\Sigma^*_{\max,\mathcal S}, \\
      \alpha_\Lambda &:= \kappa_\Lambda C_J\|\Theta^*\|_\infty + 1, & \beta_\Lambda &:= \kappa_\Lambda C_J\big(\widetilde\Sigma^*_{\infty,\mathcal S} + d\bar\delta\big).
    \end{aligned}
  \end{equation}
  If $\beta_R\beta_\Lambda < 1$, set further
  \begin{equation}\label{eq:fixed_points}
    a_\infty := \frac{\alpha_R + \beta_R\alpha_\Lambda}{1 - \beta_R\beta_\Lambda},
    \qquad
    b_\infty := \frac{\alpha_\Lambda + \beta_\Lambda\alpha_R}{1 - \beta_R\beta_\Lambda},
  \end{equation}
  with $B_0 := \max(b_0, b_\infty)$ and $A_0 := \max(a_0, \alpha_R + \beta_R B_0)$, and finally $\varepsilon_R := (\alpha_R + \beta_R B_0)\bar\delta$ and $\varepsilon_\Lambda := \kappa_\Lambda C_J\big[\|\Theta^*\|_\infty + A_0\big(\widetilde\Sigma^*_{\infty,\mathcal S} + d\bar\delta\big)\big]\bar\delta$.
\end{definition}

The following assumptions concern the sample, $\gamma$, the ground truth and the initial estimate.

\begin{assumptions}
  Let $\xi > 0$ and $c_0 \in [0,1)$, and let $\bar\varepsilon_R, \bar\varepsilon_\Lambda$ be admissible errors.
  \begin{enumerate}
    \renewcommand{\theenumi}{A\arabic{enumi}}
    \renewcommand{\labelenumi}{(\theenumi)}
    \item\label{cond:coupling}
    \emph{Weak coupling.}
    $\beta_R\beta_\Lambda < 1$.
    \item\label{cond:weights}
    \emph{Large penalty parameter.}
    \begin{gather*}
      \gamma^{1-c_0} > \frac{1}{1-\eta^*},
      \qquad
      \frac{\tilde\gamma}{1+\tilde\gamma} > \eta_{\min},
      \qquad
      \gamma e^{-\gamma\xi} \leq n\bar\delta,
      \\
      \begin{aligned}
        \gamma\,(1-\eta_{\max}) \geq{}& n\bar\delta\Big[e^{2\bar\varepsilon_\Lambda}\big(1 + 2B_0\widetilde\Sigma^*_{\max,\mathcal S^c}\big) + C_{\mathrm{irr}}\Big(e^{2\bar\varepsilon_\Lambda}\big(1 + 2B_0\widetilde\Sigma^*_{\max,\mathcal S}\big) + 1\Big)\Big] \\
        &+ 2\big(1+C_{\mathrm{irr}}\big)\,n\,d\,\|\widetilde\Sigma^*\|_\infty^3\varepsilon_R^2.
      \end{aligned}
    \end{gather*}
    \item\label{cond:localization}
    \emph{Small errors.}
    \begin{equation*}
      B_0\bar\delta \leq \bar\varepsilon_\Lambda,
      \qquad
      \varepsilon_R \leq \bar\varepsilon_R,
      \qquad
      \varepsilon_\Lambda \leq \bar\varepsilon_\Lambda.
    \end{equation*}
    \item\label{cond:signal}
    \emph{Large nonzero partial correlations.}
    \begin{equation*}
      r_c(\eta_{\min}) + \xi + A_0\bar\delta \leq R^*_{\min}.
    \end{equation*}
    \item\label{cond:init}
    \emph{Separated initial estimate.}
    The initial estimate $(R^{(0)}, \eta^{(0)})$ is $(\xi, c_0)$-separated.
  \end{enumerate}
\end{assumptions}

Assumption~(\ref{cond:coupling}) requires the estimation errors of the partial correlations and of the scales to have a weak effect on each other.
Assumption~(\ref{cond:weights}) requires $\gamma$ to be large, so that the penalty is weak on the pairs in $\mathcal S$ and strong on the remaining ones.
For the values $\gamma = \gamma_q(\bar m) \gtrsim 2\sqrt{n\log p}$ of \Cref{subsec:gamma_q}, its first two inequalities exclude only ground truths in which almost every pair has a nonzero partial correlation.
Assumption~(\ref{cond:localization}) requires the sampling error $\bar\delta$ to be small and the initial estimate to be close to the ground truth, so that the errors of the iterates stay below $\bar\varepsilon_R$ and $\bar\varepsilon_\Lambda$.
Assumption~(\ref{cond:signal}) requires the nonzero partial correlations to be large enough to be distinguished from the spike despite the estimation errors.
Assumption~(\ref{cond:init}) requires the initial estimate to distinguish the pairs in $\mathcal S$ from the remaining ones.
The continuation path of \Cref{subsec:initial_condition} is a heuristic designed to produce such an initial estimate for the fit at the target value of $\gamma$.

Assumptions~(\ref{cond:coupling})--(\ref{cond:init}) are invariant under a rescaling of the variables, since $\bar\delta$ and every constant above depend on the sample and the ground truth only through $\widetilde W$ and $\Theta^*$, and $a_0$ and $b_0$ are unchanged when $\Lambda^{(0)}$ is rescaled together with the data.

The next theorem shows that, under Assumptions~(\ref{cond:coupling})--(\ref{cond:init}), Algorithm~\ref{alg:pcss} recovers the support of $R^*$ in its first iteration and keeps it in every later one.
It also shows that any excess of the normalized errors over the fixed points $a_\infty$ and $b_\infty$ of \Cref{def:recursion} shrinks to zero.
The proof is given in \Cref{app:convergence}.

\begin{restatable}[]{theorem}{thmconvergence}
  \label{thm:convergence}
  Let $\xi > 0$, $c_0 \in [0,1)$ and admissible errors $\bar\varepsilon_R, \bar\varepsilon_\Lambda$ satisfy Assumptions~(\ref{cond:coupling})--(\ref{cond:init}).
  Then the iterates of Algorithm~\ref{alg:pcss} satisfy $\supp(R^{(t)}) = \mathcal S$ and $\eta^{(t)} \in [\eta_{\min}, \eta_{\max}]$ for every $t \geq 1$, and their normalized errors satisfy
  \begin{equation}\label{eq:resultsTheorem}
    a_{t+1} - a_\infty \leq \beta_R\beta_\Lambda\,(a_t - a_\infty) \quad (t \geq 1), \qquad b_{t+1} - b_\infty \leq \beta_R\beta_\Lambda\,(b_t - b_\infty) \quad (t \geq 0).
  \end{equation}
  In particular $\limsup_t a_t \leq a_\infty$ and $\limsup_t b_t \leq b_\infty$.
\end{restatable}

Since the normalized errors are measured in units of $\bar\delta$, \Cref{thm:convergence} gives
\begin{equation*}
  \limsup_{t \to \infty} \|R^{(t)} - R^*\|_{\max} \leq a_\infty\bar\delta,
  \qquad
  \limsup_{t \to \infty} \|\log\Lambda^{(t)} - \log\Lambda^*\|_{\max} \leq b_\infty\bar\delta.
\end{equation*}
These bounds do not depend on $\gamma$, which enters \Cref{thm:convergence} only through Assumptions~(\ref{cond:coupling})--(\ref{cond:init}), all of which hold for $\gamma$ large enough.
For the graphical lasso, in contrast, the bound on the max-norm error of the estimate given by~\citet{ravikumar2011} grows linearly with the hyperparameter, which must also be large enough for the support to be recovered.
When the constants of \eqref{eq:thm_constants} remain bounded as $n$ and $p$ grow and $\beta_R\beta_\Lambda$ stays bounded away from one, $a_\infty$ and $b_\infty$ remain bounded as well.
Algorithm~\ref{alg:pcss} then reduces both errors to the order of the sampling error $\bar\delta$.

Assumptions~(\ref{cond:coupling})--(\ref{cond:init}) involve the ground truth together with $\bar\delta$, $\gamma$ and the initial estimate, and we now determine which ground truths they exclude.
For sufficiently small $\bar\delta$, for sufficiently large $\gamma$, and for a sufficiently accurate $(\xi, c_0)$-separated initial estimate, Assumptions~(\ref{cond:weights})--(\ref{cond:init}) hold for any fixed $\xi \in (0, R^*_{\min})$ and for admissible errors that tend to zero more slowly than $\bar\delta$.
In these conditions only Assumption~(\ref{cond:coupling}) restricts the ground truth, and it holds for every sufficiently small $\bar\delta$ if and only if
\begin{equation}\label{eq:coupling-limit}
  2\,C_\Gamma\,C_J\,\widetilde\Sigma^*_{\max,\mathcal S}\,\widetilde\Sigma^*_{\infty,\mathcal S} < 1.
\end{equation}
In terms of the ground truth, it requires each variable to have small covariances $\widetilde\Sigma^*_{ij}$ with its neighbors in the graph of $R^*$, both individually and summed over the neighbors.

\begin{figure}[t]
  \centering
  \definecolor{gtChain}{HTML}{009AFA}
  \definecolor{gtLattice}{HTML}{E36F47}
  \definecolor{gtStar}{HTML}{3EA44E}
  \definecolor{gtBip}{HTML}{C371D2}
  \begin{tikzpicture}[
      gtnode/.style={circle, fill=black, inner sep=0pt, minimum size=2.6pt},
      gtlabel/.style={font=\footnotesize, anchor=north},
    ]
    \begin{scope}[shift={(0,0)}]
      \foreach \k in {0,...,5} {\node[gtnode] (c\k) at (0.4*\k-1.0,0) {};}
      \foreach \k [evaluate=\k as \l using int(\k+1)] in {0,...,4} {\draw[gtChain, thick] (c\k) -- (c\l);}
      \draw[gtChain, thick, dashed] (c0) -- ++(-0.35,0);
      \draw[gtChain, thick, dashed] (c5) -- ++(0.35,0);
      \node[gtlabel] at (0,-1.25) {Chain};
    \end{scope}
    \begin{scope}[shift={(3.6,0)}]
      \foreach \a in {0,...,4} {
        \foreach \b in {0,...,4} {\node[gtnode] (l\a\b) at (0.45*\a-0.9,0.45*\b-0.9) {};}
      }
      \foreach \a in {0,...,4} {
        \foreach \b [evaluate=\b as \c using int(\b+1)] in {0,...,3} {
          \draw[gtLattice, thick] (l\a\b) -- (l\a\c);
          \draw[gtLattice, thick] (l\b\a) -- (l\c\a);
        }
      }
      \node[gtlabel] at (0,-1.25) {2D lattice};
    \end{scope}
    \begin{scope}[shift={(7.2,0)}]
      \node[gtnode] (s0) at (0,0) {};
      \foreach \k in {1,...,15} {
        \node[gtnode] (s\k) at ({90+24*(\k-1)}:1.0) {};
        \draw[gtStar, thick] (s0) -- (s\k);
      }
      \node[gtlabel] at (0,-1.25) {Star, $d = 15$};
    \end{scope}
    \begin{scope}[shift={(10.8,0)}]
      \node[gtnode] (h1) at (0,0.95) {};
      \node[gtnode] (h2) at (0,-0.95) {};
      \foreach \k in {1,...,14} {
        \node[gtnode] (b\k) at ({0.18*(\k-7.5)},0) {};
        \draw[gtBip, thick] (h1) -- (b\k);
        \draw[gtBip, thick] (h2) -- (b\k);
      }
      \node[gtlabel] at (0,-1.25) {$K_{2,14}$};
    \end{scope}
  \end{tikzpicture}
  \caption{
    The graphs of the four ground truths of \Cref{tab:thresholds}, namely a chain on $256$ vertices, an open $16 \times 16$ lattice, $16$ disjoint stars with $d = 15$ leaves, and $16$ disjoint copies of the complete bipartite graph $K_{2,14}$.
    Only portions of the chain and of the lattice, and a single star and a single copy of $K_{2,14}$, are drawn.
  }
  \label{fig:coupling-graphs}
\end{figure}
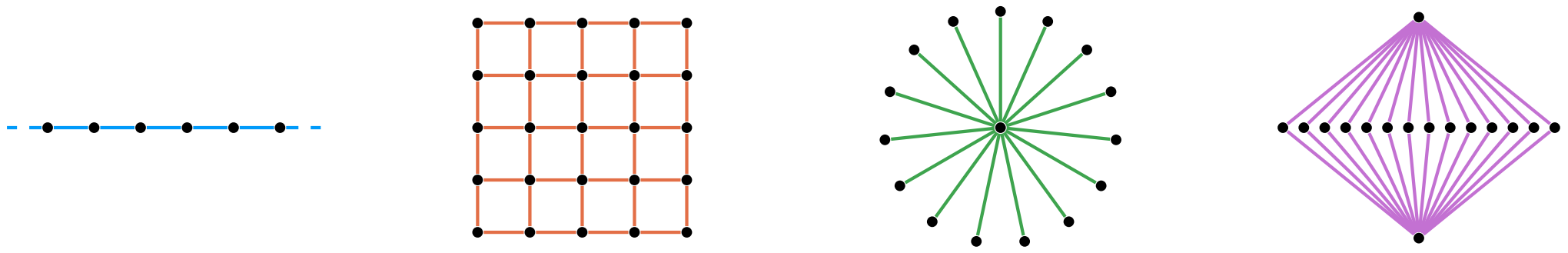

\begin{table}[t]
  \centering
  \caption{
    Value of $r/r_{\mathrm{PD}}$ up to which each condition holds, for the ground truths of \Cref{fig:coupling-graphs}.
    The irrepresentability condition of \citet{ravikumar2011} is evaluated at $\Lambda^* = I_p$.
  }
  \label{tab:thresholds}
  \begin{tabular}{lcc}
    \toprule
    Ground truth & Condition \eqref{eq:coupling-limit} & Irrepresentability \\
    \midrule
    Chain & $0.77$ & $0.71$ \\
    2D lattice & $0.85$ & $0.75$ \\
    Star, $d = 15$ & $0.63$ & $1.00$ \\
    $K_{2,14}$ & $0.65$ & $0.19$ \\
    \bottomrule
  \end{tabular}
\end{table}

To see how restrictive \eqref{eq:coupling-limit} is, we compare it with the irrepresentability condition of \citet{ravikumar2011}, the corresponding condition on the ground truth in the analysis of the graphical lasso.
We consider four ground truths with $p = 256$ in which every nonzero partial correlation equals $r$ (\Cref{fig:coupling-graphs}), and write $r_{\mathrm{PD}}$ for the largest $r$ for which $\Theta^*$ is positive definite.
\Cref{tab:thresholds} reports, for each of the two conditions, the value of $r/r_{\mathrm{PD}}$ up to which it holds.
Condition \eqref{eq:coupling-limit} holds over a larger range than irrepresentability for the chain, the lattice and $K_{2,14}$, and over a smaller one for the star, so neither condition implies the other.
The two conditions differ in two respects.
First, irrepresentability involves the rows of $\Gamma^*$ indexed by $\mathcal S^c$, whereas \eqref{eq:coupling-limit} involves only $\Gamma^*_{\mathcal S\mathcal S}$.
Second, irrepresentability depends on the scales $\Lambda^*$, which is why we evaluate it at $\Lambda^* = I_p$, whereas \eqref{eq:coupling-limit} does not.

\section{Simulation experiments}\label{sec:simulations}

This section compares PCSS with six other estimators of a sparse Gaussian graphical model on simulated data.
\Cref{subsec:sim-methods} presents the estimators and the choice of their hyperparameters, \Cref{subsec:sim-truths} the ground truths and the samples drawn from them, and \Cref{subsec:sim-overview} compares the estimators by how well they recover the support of the ground truth.
\Cref{app:sim-figures} collects the figures for the models and values of $\langle d\rangle$ not shown in \Cref{fig:overview}.

\subsection{Estimators and hyperparameter selection}\label{subsec:sim-methods}

Each estimator depends on a hyperparameter that sets the size of its penalty.
To compare the estimators rather than the rules that choose their hyperparameters, we select the hyperparameter of every estimator with the extended Bayesian information criterion (EBIC), with $\gamma_{\mathrm{ebic}} = 1/2$.
For PCSS and nodewise regression we also use a second choice of hyperparameter, set so as to control the probability of estimating false edges under a diagonal ground truth.

\paragraph{Partial correlation spike-and-slab (PCSS).}
The PCSS estimate $(\widehat\Lambda, \widehat R, \widehat\eta)$ is the output of the procedure of \Cref{sec:algorithm}.
We compute it with the package \pkgname{}~\citep{pcssjl}.
The first choice of hyperparameter minimizes the EBIC \eqref{eq:ebic} over a grid centred at the value $\gamma_q(\bar m)$ of \Cref{subsec:gamma_q},
\begin{equation*}
  \hat\gamma = \argmin_{\gamma \in \mathcal G} \mathrm{EBIC}(\gamma),
  \qquad
  \mathcal G := \big\{ 8^{u/12}\, \gamma_q(\bar m) : u = -12, -11, \dots, 12 \big\}.
\end{equation*}
The estimate at each $\gamma \in \mathcal G$ has its own continuation path, started from the diagonal estimate \eqref{eq:diagonal-estimate}, so it does not depend on the estimates at the other values of the grid.
We refer to PCSS with this choice of hyperparameter as PCSS (EBIC).
The second choice is $\gamma = \gamma_q(\bar m)$, with the values $q = 0.05$ and $\bar m = 5$ given in \Cref{subsec:gamma_q}, and requires a single fit.
In both choices, $\gamma_q(\bar m)$ is computed from $5 \cdot 10^4$ Monte Carlo replicates.
We refer to PCSS with this choice of hyperparameter as PCSS ($\gamma_q$).

\paragraph{Other estimators.}
We compare PCSS with six other estimators, whose definitions, implementations and grids of hyperparameters are given in \Cref{app:sim-estimators}.
The graphical lasso \citep[GLASSO;][]{friedman2008} maximizes the Gaussian log-likelihood with an $\ell_1$ penalty on the off-diagonal entries of the precision matrix, and SCAD~\citep{fan2001,fan2009} and MCP~\citep{zhang2010} replace this penalty by folded-concave ones.
The partial correlation graphical lasso \citep[PCGLASSO;][]{carter2023pcglasso} and the joint sparse regression estimator SPACE~\citep{peng2009} place an $\ell_1$ penalty on the partial correlations instead.
Nodewise regression \citep[NR;][]{meinshausen2006} selects the neighbours of each variable with a separate lasso regression, and joins two variables when either is selected as a neighbour of the other, which is known as the OR rule.
All six estimators are applied to the standardized data, and the hyperparameter of each is selected by EBIC over a grid of values.
For NR the criterion is the EBIC of~\citet{chen2008} for regression, applied separately to each variable, and we refer to this choice as NR (EBIC).
The second choice for NR is the hyperparameter of~\citet{meinshausen2006}, which bounds by $0.05$ the probability that some estimated neighbourhood connects two connected components of the true graph, and we refer to it as NR (MB).
We compute PCGLASSO only for $n \geq p$, since for $n < p$ its computation did not finish within a reasonable time.

\subsection{Ground truths and data}\label{subsec:sim-truths}

All ground truths have $p = 256$ variables and unit scales, $\Lambda^* = I_p$, so that $\Omega^* = I_p - R^*$.
Given a support $\mathcal S$, the ground-truth partial correlations are
\begin{equation*}
  r^*_{ij} \in
  \begin{cases}
    \{0\} & \text{if } (i,j) \notin \mathcal S, \\
    \{-r^*, r^*\} & \text{if } (i,j) \in \mathcal S,
  \end{cases}
\end{equation*}
where $r^* > 0$ is the magnitude of the nonzero entries.
The signs of the nonzero entries above the diagonal are independent, with $\mathrm P(r^*_{ij} = r^*) = \mathrm P(r^*_{ij} = -r^*) = 1/2$.
We identify $\mathcal S$ with the set of edges of a graph on the vertex set $\{1, \dots, p\}$, in which $i$ and $j$ are joined by an edge if $(i,j) \in \mathcal S$.
Since $\mathcal S$ contains both $(i,j)$ and $(j,i)$, the graph has $m = |\mathcal S|/2$ edges and mean degree $\langle d\rangle = |\mathcal S|/p$.
The ground truths differ in their support $\mathcal S$, which we draw from three random graph models with $\langle d\rangle \in \{2, 17.5\}$.
\begin{itemize}
  \item \emph{Erd\H{o}s--R\'enyi.}
    The $m$ edges are drawn uniformly at random, without replacement, from the $\ell$ pairs of vertices~\citep{erdos1959}.
  \item \emph{Torus.}
    Each vertex $i$ is associated with a point $\boldsymbol z_i \in [0,1]^2$, drawn independently and uniformly, and the distance between two vertices is that between their points on the torus,
    \begin{equation*}
      \mathrm{dist}(i,j) := \min_{\boldsymbol k \in \mathbb Z^2} \|\boldsymbol z_i - \boldsymbol z_j - \boldsymbol k\|_2 .
    \end{equation*}
    The $m$ edges join the pairs of vertices with the smallest distances.
    Since edges join nearby vertices, two neighbours of a vertex are often neighbours of each other.
  \item \emph{Preferential attachment.}
    Following~\citet{barabasi1999emergence}, the construction starts from a complete graph and adds the remaining vertices one at a time.
    Each new vertex is joined to a prescribed number of earlier vertices, chosen without replacement with probability proportional to their current degree.
    These numbers are set so that the graph has $m$ edges (\Cref{app:sim-truths}).
    As a result, a few vertices have a much larger degree than the rest.
\end{itemize}

\begin{figure}[tp]
  \begin{center}
    \includegraphics[width=0.9\textwidth]{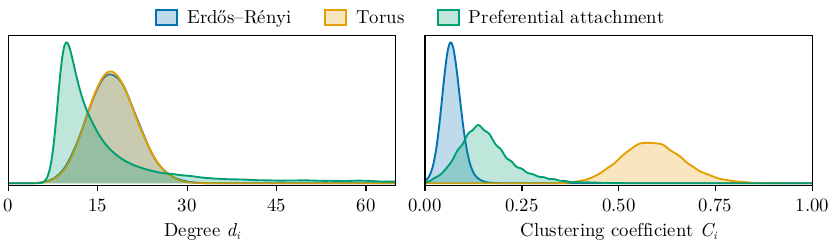}
  \end{center}
  \caption{
    Degree $d_i$ (left) and local clustering coefficient $C_i$ (right) of the vertices of the ground truths with $p = 256$ and $\langle d\rangle = 17.5$, pooled over all ground truths of each model.
    The curves are Gaussian kernel density estimates.
    The degree axis is cut at $65$, which leaves out part of the tail of the preferential attachment model.
    The smallest and largest degrees are $3$ and $36$ for Erd\H{o}s--R\'enyi, $5$ and $35$ for the torus, and $8$ and $105$ for preferential attachment, and the means of the clustering coefficient are $0.07$, $0.59$ and $0.16$, in the same order.
  }
  \label{fig:truths}
\end{figure}

\Cref{fig:truths} compares the three models through the degree $d_i$ of each vertex, its number of neighbours, and its local clustering coefficient $C_i$~\citep{watts1998}, the fraction of pairs of neighbours of $i$ that are themselves neighbours (\Cref{app:sim-truths}).
The Erd\H{o}s--R\'enyi and torus ground truths have almost the same degree distribution and differ mainly in clustering, whereas the preferential attachment ground truths have a much wider degree distribution than the Erd\H{o}s--R\'enyi ones, with clustering coefficients between those of the other two models.

Besides its support, each ground truth depends on the magnitude $r^*$ and on the signs of the nonzero entries, which together must keep $\Omega^*$ positive definite.
Draws of the support and the signs for which the largest eigenvalue $\varphi_{\max}(R^*)$ of $R^*$ is at least $0.9$ are rejected, so that the smallest eigenvalue of every $\Omega^*$ is $\varphi_{\min}(\Omega^*) = 1 - \varphi_{\max}(R^*) > 0.1$.
The magnitude $r^*$ takes five values,
\begin{equation*}
  r^* \in \Big\{ \frac{k}{5}\, r_{\max} : k = 1, 2, \dots, 5 \Big\},
\end{equation*}
where $r_{\max}$ is set separately for each value of $\langle d\rangle$, with one value shared by the Erd\H{o}s--R\'enyi and torus models and another for the preferential attachment model, so that draws with $\varphi_{\max}(R^*) < 0.9$ are found for every $r^* \leq r_{\max}$.

For each model and each value of $\langle d\rangle$ and $r^*$, we draw $16$ ground truths, each with precision matrix $\Omega^*$ and covariance matrix $\Sigma^* = (\Omega^*)^{-1}$.
From each of them we draw samples of size $n \in \{2^5, 2^6, \dots, 2^{12}\}$, from $2^{-3}p$ to $2^4 p$, each consisting of $n$ independent observations from $N_p(\boldsymbol 0, \Sigma^*)$.
We call each combination of model, $\langle d\rangle$, $r^*$ and $n$ a condition, so the simulations comprise $3 \cdot 2 \cdot 5 \cdot 8 = 240$ conditions, each with $16$ ground truths.

\subsection{Recovery of the support}\label{subsec:sim-overview}

We compare the estimators through their estimated support $\widehat{\mathcal S}$, the set of pairs $(i,j)$ joined by an edge of the estimated graph.
For PCSS these are the pairs with $|\widehat r_{ij}| > r_c(\widehat\eta)$, as in \Cref{subsec:edges}, for nodewise regression the pairs given by the OR rule of \Cref{subsec:sim-methods}, and for the other estimators the pairs with a nonzero estimated partial correlation.
Precision is the fraction of estimated edges that are true edges, and recall is the fraction of true edges that are estimated,
\begin{equation*}
  \mathrm{precision} := \frac{|\widehat{\mathcal S} \cap \mathcal S|}{|\widehat{\mathcal S}|},
  \qquad
  \mathrm{recall} := \frac{|\widehat{\mathcal S} \cap \mathcal S|}{|\mathcal S|}.
\end{equation*}
For each estimator and each value of $\langle d\rangle$, $r^*$ and $n$, we summarize precision and recall by their medians over the $16$ ground truths, leaving out of the median of precision the ground truths for which $\widehat{\mathcal S}$ is empty.

The two choices of hyperparameter of PCSS give similar precision and recall in most conditions, and we refer to both as PCSS when the distinction does not matter.
In $192$ of the $240$ conditions, their median recalls differ by at most $0.02$ and their median precisions by at most $0.01$.
In the remaining conditions, PCSS ($\gamma_q$) is the more conservative of the two: it estimates fewer edges, in some conditions none, and wherever both estimate edges its precision is higher or almost the same.

\begin{figure}[tp]
  \begin{center}
    \includegraphics[width=0.95\textwidth]{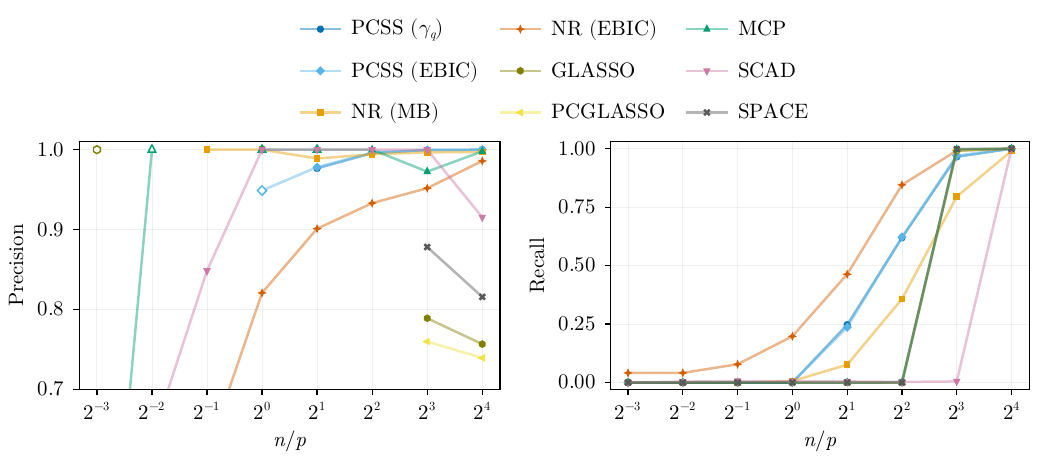}
  \end{center}
  \caption{
    Precision (left) and recall (right) of the estimators as functions of $n/p$, for the Erd\H{o}s--R\'enyi ground truths with $p = 256$, $\langle d\rangle = 17.5$ and $r^* = r_{\max} = 0.1053$.
    Each point is the median over the $16$ ground truths.
    The labels in parentheses give the choice of hyperparameter of PCSS and NR (\Cref{subsec:sim-methods}), and all other estimators use EBIC.
    PCGLASSO is computed only for $n \geq p$.
    Precisions below $0.7$ fall outside the left panel.
    A precision point is hollow when more than half of the $16$ estimates have no edge, so that its median is taken over fewer than $8$ ground truths, and missing when no estimate has an edge.
  }
  \label{fig:overview}
\end{figure}

\Cref{fig:overview} shows precision and recall for the Erd\H{o}s--R\'enyi ground truths with $\langle d\rangle = 17.5$ and $r^* = r_{\max}$.
We describe the behaviour of the estimators in three regimes, which we call small, intermediate and large $n/p$, and which for these ground truths correspond to $n \leq p$, $2p \leq n \leq 2^2 p$ and $n \geq 2^3 p$.
The same regimes appear for the other models and values of $\langle d\rangle$, at values of $n/p$ that depend on the ground truth (\Cref{app:sim-figures}), and the values we quote below are those of \Cref{fig:overview}.

At small $n/p$, PCSS, GLASSO, PCGLASSO, SPACE and NR (MB) estimate almost no edges.
SCAD and MCP estimate a few edges, with a precision that falls as $n$ decreases, to $0.38$ and $0.25$ at $n = 2^{-3}p$.
NR (EBIC) estimates many edges, with a precision that falls from $0.82$ at $n = p$ to $0.17$ at $n = 2^{-3}p$.

At intermediate $n/p$, PCSS recovers a growing fraction of the true edges, with a precision above $0.97$.
Over the same range, NR (MB) recovers fewer of the true edges than PCSS, and GLASSO, PCGLASSO, SCAD, SPACE and MCP almost none.
NR (EBIC) recovers more edges than PCSS, with a lower precision.

At large $n/p$, the recall of GLASSO, PCGLASSO, SPACE and MCP rises abruptly to almost $1$ between $n = 2^2 p$ and $n = 2^3 p$, and that of SCAD between $n = 2^3 p$ and $n = 2^4 p$.
As their recall rises, the precision of GLASSO, PCGLASSO, SPACE and SCAD falls, whereas that of MCP stays above $0.97$.
NR (EBIC) also recovers almost all true edges, with a lower precision than PCSS and NR (MB).
PCSS and NR (MB) keep a precision above $0.99$, and PCSS has the higher recall of the two until both recover almost all true edges at $n = 2^4 p$.

The precision of PCSS therefore stays high in the three regimes, and in all our simulations the median precision of PCSS ($\gamma_q$) is above $0.97$ wherever its median recall exceeds $0.01$.
The estimators that keep a high precision in the three regimes are PCSS, NR (MB) and, apart from the smallest sample sizes, MCP.

\begin{figure}[tp]
  \begin{center}
    \includegraphics[width=0.95\textwidth]{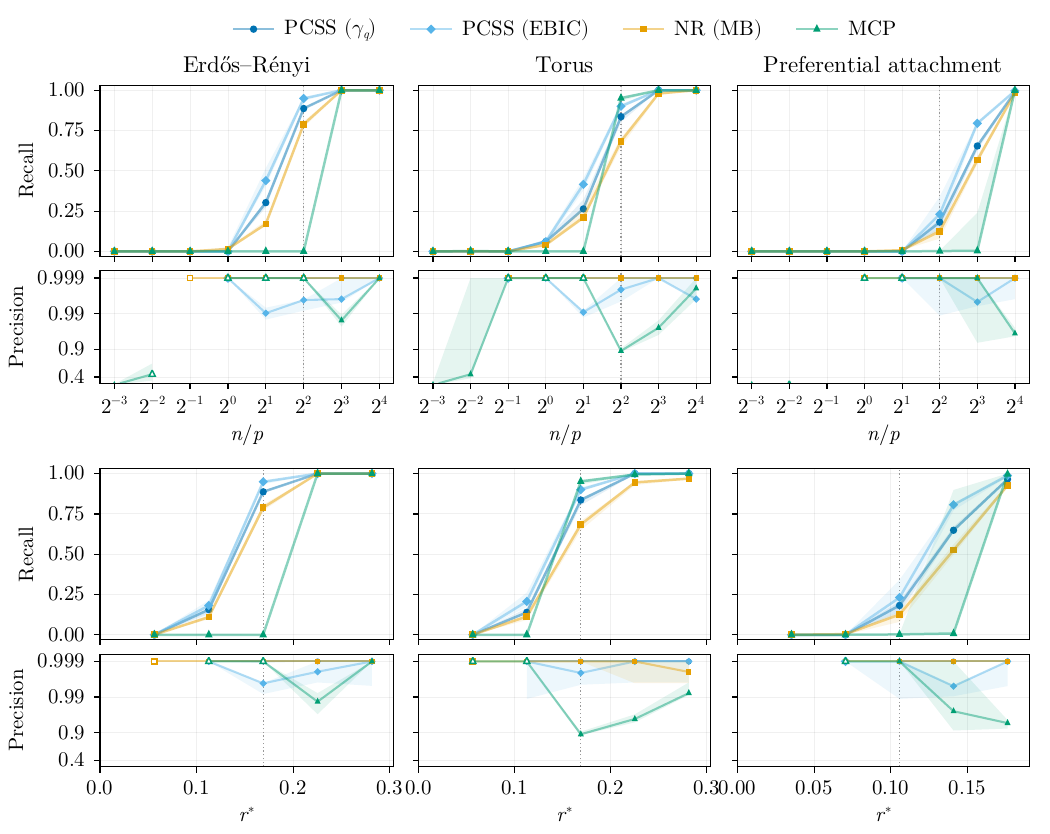}
  \end{center}
  \caption{
    Precision and recall of PCSS, NR (MB) and MCP for the ground truths with $\langle d\rangle = 2$ of the Erd\H{o}s--R\'enyi (left), torus (centre) and preferential attachment (right) models.
    The top row varies $n/p$ at $r^* = 3 r_{\max}/5$, which is $0.1690$ for the Erd\H{o}s--R\'enyi and torus models and $0.1057$ for the preferential attachment model.
    The bottom row varies $r^*$ at $n = 2^2 p$.
    The dotted lines mark the condition shared by the two rows.
    Each point is the median over the $16$ ground truths, and the band extends from the $0.25$- to the $0.75$-quantile.
    The precision axis is linear in $\log(1 - \mathrm{precision})$, precisions above $0.999$ are drawn at $0.999$, and precisions below $0.1$ fall outside the panels.
    A precision point is hollow when more than half of the $16$ estimates have no edge, so that its median is taken over fewer than $8$ ground truths, and missing when no estimate has an edge.
  }
  \label{fig:pr_main_d2}
\end{figure}

\Cref{fig:pr_main_d2} compares these three estimators for each model at $\langle d\rangle = 2$, varying $n$ at a fixed magnitude in the top row and $r^*$ at a fixed sample size in the bottom row.
The comparison of \Cref{fig:overview} holds in every panel.
The precision of PCSS and that of NR (MB) are above $0.99$ throughout.
The recall of PCSS is higher than that of NR (MB), or almost equal where both are close to $0$ or to $1$.
The recall of MCP is close to $0$ over most of the range and, where it rises, it does so within one doubling of $n$ or between two consecutive values of $r^*$.
After this rise its recall can exceed that of PCSS, but only with a lower precision, below $0.97$, against at least $0.997$ for PCSS.
\Cref{fig:pr_main_d17p5} in \Cref{app:sim-figures} shows the same comparison for $\langle d\rangle = 17.5$, with the same conclusions up to small differences in the precisions.

In summary, wherever PCSS recovers part of the support, no other estimator recovers more of it with a higher precision.
In all our simulations, wherever the median recall of PCSS ($\gamma_q$) exceeds $0.01$, no estimator other than PCSS (EBIC) has at the same time a higher median precision and a higher median recall.
In the $157$ conditions in which the median recall of PCSS ($\gamma_q$) is below $0.01$, the other estimators also recover few of the true edges, except NR (EBIC), whose median precision is then below $0.5$ in most of them.
In these conditions, GLASSO, PCGLASSO, SPACE and NR (MB) recover almost none of the true edges, and SCAD and MCP recover few of them, mostly with a precision below $0.9$.

\section{Conclusion}\label{sec:conclusion}

We proposed the PCSS estimator, which places a spike-and-slab prior with a uniform slab on the partial correlations of a Gaussian graphical model.
This combination gives an estimator whose estimates of the partial correlations are scale invariant at every local maximum of the log-posterior, in which large partial correlations are nearly unpenalized, and whose single hyperparameter can be set from $n$ and $p$ alone.
In our simulations, wherever PCSS recovered part of the support, none of the estimators we compared had at the same time a higher precision and a higher recall.
Moreover, a single fit at $\gamma = \gamma_q(\bar m)$ gave almost the same precision and recall as the selection of $\gamma$ by EBIC, although the two criteria need not agree.
Where they differed, the fit at $\gamma_q(\bar m)$ was the more conservative of the two.

The guarantee behind $\gamma_q(\bar m)$ is limited in scope.
Under a diagonal ground truth, \Cref{thm:calibration} shows that, with probability at least $1 - q$, the diagonal model is a local maximum of the log-posterior and is preferred to every estimate supported on a matching with at most $\bar m$ edges.
\Cref{thm:calibration-forest} extends this guarantee to forests with at most $\bar m$ edges and degrees at most $\bar d$, at the cost of a larger hyperparameter.
Neither result guarantees that a diagonal ground truth is recovered, since neither covers estimates supported on other graphs.
Both hold in the limit $p \to \infty$, uniformly in $n$, and we have not quantified how fast this limit is approached.
The restriction to matchings rests on our simulations, in which larger degree bounds left the precision and recall of the estimated graphs almost unchanged (\Cref{app:forest-calibration}), rather than on a theorem.

\Cref{thm:convergence} shows that, from an initial estimate that separates the true edges from the remaining pairs, the algorithm recovers the support in its first iteration and keeps it, and its errors are eventually bounded by multiples of the sampling error that do not depend on $\gamma$.
Its main limitation is this assumption on the initial estimate.
The theorem is also stated in terms of the sampling error $\bar\delta$ of the given sample, and its conditions involve constants that depend on the ground truth.
In the limit of small sampling error, however, only the weak-coupling condition \eqref{eq:coupling-limit} restricts the ground truth.

Our simulations have a limited scope.
All ground truths have $p = 256$ variables and nonzero partial correlations of a common magnitude, and the data are Gaussian.
We compared the estimators only through their recovery of the support, so the simulations do not test whether the estimates of large partial correlations are nearly unbiased.

Our results leave several open problems.
\begin{itemize}
  \item \Cref{thm:convergence} calls for a procedure that provably yields an initial estimate that separates the true edges from the remaining pairs.
    The path of increasing values of $\gamma$, designed for this purpose, is a candidate.
    For $n < p$, where the log-posterior has no global maximum (\Cref{prop:existence}), such a result would also characterize the local maximum that the procedure returns.
  \item The PCSS estimate depends on the data only through the sample correlation matrix $\rho_n$, so it extends to non-Gaussian data by replacing $\rho_n$ with a rank-based correlation matrix, as for Gaussian copula graphical models~\citep{liu2012}.
    \Cref{thm:convergence} does not use the Gaussian assumption, but $\gamma_q(\bar m)$ does, and whether it carries over to this extension remains open.
  \item On the torus ground truths, whose graphs are clustered, the recall of PCSS and of nodewise regression approaches one more slowly than on the Erd\H{o}s--R\'enyi ground truths with the same magnitudes.
    Explaining this requires understanding how clustering affects the recovery of the edges, and the expansion of the Pearson correlations over the paths of the graph~\citep{arenaza2025expansion} may be a tool for this.
\end{itemize}

\bigskip
\begin{center}
{\large\bf CODE AND DATA AVAILABILITY}
\end{center}

The proofs, the implementation of the ECM algorithm, the extension of \Cref{thm:matching,thm:calibration} to forests of bounded degree, and additional simulation figures are given in the appendices.
The Julia package \pkgname{}, which implements the PCSS estimator and the computation of $\gamma_q$, is available at \url{https://github.com/bautistaarenaza/PartialCorrelationSpikeSlab.jl} and archived at \url{https://doi.org/10.5281/zenodo.24663488}.
The scripts that generate the ground truths and the samples, compute the estimators and produce the figures and numbers of \Cref{sec:simulations}, with the versions of \pkgname{} and of the R packages pinned, are available at \url{https://github.com/bautistaarenaza/PCSS-paper} and archived at \url{https://doi.org/10.5281/zenodo.24664242}.
The ground truths and the estimates of every estimator are archived at \url{https://doi.org/10.5281/zenodo.24574158}, so that the figures and numbers can be reproduced without refitting the estimators.

\bibliographystyle{chicago} 
\bibliography{references}

\appendix
\section{Notation}\label{app:notation}

The sample size is $n$ and the number of variables is $p$.
Indices $i$ and $j$ label variables and run over $\{1,\dots,p\}$, and
\begin{equation}
  \ell := \binom{p}{2} = \frac{p(p-1)}{2}
  \label{eq:ell}
\end{equation}
is the number of distinct pairs.
A sum or a product indexed by $i<j$ runs over those $\ell$ pairs.
A hat denotes an estimate and a star denotes the corresponding ground-truth quantity, as in $\widehat{R}$ and $R^*$.
A parenthesised superscript $(t)$ denotes the value at the $t$-th iteration of the algorithm of \Cref{sec:algorithm}, as in $R^{(t)}$.

Vectors are written in boldface, and $\boldsymbol{1}$ is the vector whose entries all equal one.
We write $I_p$ for the $p\times p$ identity matrix.
For a symmetric matrix $M$, we write $M \succ 0$ when $M$ is positive definite and $M \succeq 0$ when it is positive semidefinite.
We write $\varphi_{\max}(M)$ and $\varphi_{\min}(M)$ for its largest and smallest eigenvalues.
The set of partial correlation matrices is
\begin{equation}
  \mathcal{R}_p := \bigl\{ R \in \mathbb{R}^{p\times p} : R = R^\top,\; R_{ii} = 0,\; I_p - R \succ 0 \bigr\}.
  \label{eq:Rp}
\end{equation}
For a vector $\boldsymbol v$, $\diag(\boldsymbol v)$ is the diagonal matrix with diagonal $\boldsymbol v$.
For a diagonal matrix $D$ with positive diagonal, $\log D$ is the diagonal matrix whose entries are the logarithms of those of $D$.
For a matrix $M$, $M^\top$ is its transpose, $\tr M$ its trace, and $\det M$ its determinant.
The symbols $\circ$ and $\otimes$ denote the Hadamard (entrywise) and the Kronecker product.
The norms used throughout are
\begin{equation} \label{eq:norms}
  \|M\|_{\max} := \max_{i,j} |M_{ij}|,
  \qquad
  \|M\|_{\infty} := \max_{i} \sum_{j} |M_{ij}|,
  \qquad
  \|M\|_{F} := \Bigl( \sum_{i,j} M_{ij}^{2} \Bigr)^{1/2},
\end{equation}
and $\|\boldsymbol{v}\|_{\max} := \max_i |v_i|$ for a vector $\boldsymbol{v}$.
We also write $\|M\|_2$ for the spectral norm of $M$, its largest singular value.

The observations are $\boldsymbol{x}_1,\dots,\boldsymbol{x}_n \in \mathbb{R}^p$, collected as the rows of the $n \times p$ data matrix $X$.
Their sample mean and sample covariance are
\begin{align}
  \bar{\boldsymbol{x}} &:= \frac1n\sum_{k=1}^n \boldsymbol{x}_k ,
  \label{eq:sample-mean} \\
  S_n &:= \frac1n\sum_{k=1}^n (\boldsymbol{x}_k-\bar{\boldsymbol{x}})(\boldsymbol{x}_k-\bar{\boldsymbol{x}})^\top ,
  \label{eq:sample-cov}
\end{align}
and the sample correlation matrix $\rho_n$ has entries
\begin{equation}
  (\rho_n)_{ij} := \frac{(S_n)_{ij}}{\sqrt{(S_n)_{ii}\,(S_n)_{jj}}} ,
  \label{eq:sample-corr}
\end{equation}
so that $(\rho_n)_{ij}$ is the sample Pearson correlation between the $i$-th and $j$-th variables.
We write $N_p(\boldsymbol{\mu},\Sigma)$ for the $p$-variate normal distribution, $\Phi_N$ for the cumulative distribution function of the standard normal distribution, $\mathbb{E}$ for expectation, and $\mathds{1}_A$ for the indicator of a set $A$.

\section{Location and scale priors}\label{app:prior}

This appendix derives the dependence of the prior on the location and the scale stated in \Cref{subsec:ignorance}.
Consider the change of variables of \Cref{subsec:ignorance}.
Since this is merely a reparametrization, the prior expressed in the transformed variables satisfies
\begin{equation} \label{eq:change-of-variables}
  p'(\boldsymbol\mu', \Lambda', R') = \det(D)\, p(\boldsymbol{\mu}, \Lambda, R).
\end{equation}
Ignorance of scale and location further requires that the prior keep the same functional form under this transformation,
\begin{equation} \label{eq:consistency}
  p'(\boldsymbol{\mu},\Lambda,R) = p(\boldsymbol{\mu}, \Lambda, R).
\end{equation}
The unique solution of the functional equations \eqref{eq:change-of-variables} and \eqref{eq:consistency} has the form
\begin{equation*}
  p(\boldsymbol{\mu}, \Lambda, R) = \frac{1}{\det(\Lambda)} f(R).
\end{equation*}

\section{Proof of \Cref{prop:existence}}\label{app:existence}

\propexistence*

\begin{proof}
  We call the first three terms of \eqref{eq:map-objective} its smooth part,
  \begin{equation}\label{eq:smooth-part}
    \mathcal L_\mathrm{s}(\Lambda, R) := -\frac n2 \tr\big[S_n \Lambda (I_p - R) \Lambda\big] + (n-1) \log\det\Lambda + \frac n2 \log\det(I_p - R) ,
  \end{equation}
  so that $\mathcal L(\Lambda, R, \eta) = \mathcal L_\mathrm{s}(\Lambda, R) - \sum_{i<j} \mathrm{pen}(r_{ij}, \eta)$.
  The penalty $\mathrm{pen}$ of \eqref{eq:map-objective} is bounded.
  Since $\tilde\gamma\, e^{-\gamma} = \gamma/(e^\gamma - 1) < 1 \le \tilde\gamma$, for every $|r| < 1$ and $\eta \in [0,1]$
  \begin{equation*}
    \tilde\gamma\, e^{-\gamma} \le \eta + (1-\eta)\,\tilde\gamma\, e^{-\gamma|r|} \le \tilde\gamma ,
  \end{equation*}
  so $-\log\tilde\gamma \le \mathrm{pen}(r, \eta) \le \gamma - \log\tilde\gamma$.
  Hence there is a constant $C$, depending only on $\gamma$ and $p$, such that
  \begin{equation}\label{eq:pen-bound}
    \big|\mathcal L(\Lambda, R, \eta) - \mathcal L_\mathrm{s}(\Lambda, R)\big| \le C
  \end{equation}
  for every positive diagonal $\Lambda$, $R \in \mathcal R_p$ and $\eta \in [0,1]$.
  The penalty is also continuous in $(r, \eta)$ on $(-1,1) \times [0,1]$, so $\mathcal L$ is continuous.

  We express $\mathcal L_\mathrm{s}$ in terms of $\Omega = \Lambda(I_p - R)\Lambda$.
  The map $(\Lambda, R) \mapsto \Lambda(I_p - R)\Lambda$ is a bijection from the set of pairs of a positive diagonal $\Lambda$ and an $R \in \mathcal R_p$ onto the set of positive definite matrices, with inverse $\Lambda = \diag(\Omega_{11}, \dots, \Omega_{pp})^{1/2}$ and $R = I_p - \Lambda^{-1}\Omega\Lambda^{-1}$.
  Both the map and its inverse are continuous.
  Since $\Omega_{ii} = \Lambda_{ii}^2$,
  \begin{equation}\label{eq:Ls-omega}
    \mathcal L_\mathrm{s}(\Lambda, R) = \frac n2\Big[\log\det\Omega - \tr\big(S_n\Omega\big)\Big] - \frac12 \sum_{i=1}^p \log\Omega_{ii} .
  \end{equation}
  We now prove the two statements of the proposition in turn.

  \paragraph{First statement.}
  Let $n > p$, and let the observations be independent, each with a distribution that assigns probability zero to every affine hyperplane of $\mathbb R^p$.
  We first show that $S_n$ is then positive definite almost surely.
  The matrix $S_n$ is singular if and only if all the observations lie in an affine hyperplane.
  It therefore suffices to show that, almost surely, no affine hyperplane contains $\boldsymbol x_1, \dots, \boldsymbol x_{p+1}$.
  For $k \le p$, the affine hull of $\{\boldsymbol x_1, \dots, \boldsymbol x_k\}$, the smallest affine subspace of $\mathbb R^p$ that contains it, has dimension at most $k - 1 \le p - 1$, so it is contained in an affine hyperplane.
  Since $\boldsymbol x_{k+1}$ is independent of $\boldsymbol x_1, \dots, \boldsymbol x_k$, the conditional probability that it lies in this affine hull, given $\boldsymbol x_1, \dots, \boldsymbol x_k$, is zero.
  Almost surely, therefore, adding $\boldsymbol x_{k+1}$ raises the dimension of the affine hull by one for every $k \le p$, so the affine hull of $\{\boldsymbol x_1, \dots, \boldsymbol x_{p+1}\}$ has dimension $p$, and no affine hyperplane contains it.

  On the event that $S_n$ is positive definite, Hadamard's inequality, $\prod_{i} \Omega_{ii} \ge \det\Omega$, and $\tr(S_n\Omega) \ge \varphi_{\min}(S_n)\tr\Omega$ give, through \eqref{eq:Ls-omega},
  \begin{equation*}
    \mathcal L_\mathrm{s}(\Lambda, R)
    \le \frac{n-1}{2}\log\det\Omega - \frac n2\,\varphi_{\min}(S_n)\tr\Omega
    = \sum_{j=1}^{p} g\big(\varphi_j(\Omega)\big),
  \end{equation*}
  where $\varphi_1(\Omega), \dots, \varphi_p(\Omega)$ are the eigenvalues of $\Omega$ and
  \begin{equation*}
    g(x) := \frac{n-1}{2}\log x - \frac n2\,\varphi_{\min}(S_n)\,x .
  \end{equation*}
  Since $n \ge 2$ and $\varphi_{\min}(S_n) > 0$, $g$ is bounded above, by $G$ say, and tends to $-\infty$ as $x \to 0^+$ and as $x \to \infty$.
  Since every term of the sum is at most $G$, \eqref{eq:pen-bound} shows that $\mathcal L$ tends to $-\infty$ whenever an eigenvalue of $\Omega$ tends to $0$ or to $\infty$.
  On each superlevel set of $\mathcal L$, the eigenvalues of $\Omega$ therefore lie in a compact interval of $(0, \infty)$, and, since the inverse map above is continuous, each superlevel set is compact.
  The continuous function $\mathcal L$ therefore attains a global maximum.

  \paragraph{Second statement.}
  Let $n < p$.
  Then $S_n$ has rank at most $n - 1$, so its null space has dimension $s \ge p - n + 1$.
  Let $P$ be the orthogonal projection onto it.
  For $h > 0$, the matrix $\Omega(h) := I_p + hP$ is positive definite, and we let $(\Lambda(h), R(h))$ be its preimage under the map above.
  Since $S_n P = 0$, $\tr(S_n\Omega(h)) = \tr S_n$, and since $\Omega(h)$ has $s$ eigenvalues equal to $1 + h$ and the remaining ones equal to $1$, $\det\Omega(h) = (1+h)^s$.
  Moreover, $0 \le P_{ii} \le 1$, so $\log\Omega(h)_{ii} = \log(1 + hP_{ii}) \le \log(1+h)$.
  By \eqref{eq:Ls-omega},
  \begin{equation*}
    \mathcal L_\mathrm{s}\big(\Lambda(h), R(h)\big)
    \ge \frac{ns - p}{2}\log(1+h) - \frac n2\tr S_n
    \ge \frac{(n-1)(p-n)}{2}\log(1+h) - \frac n2\tr S_n ,
  \end{equation*}
  where the second inequality uses $s \ge p - n + 1$ and $n(p - n + 1) - p = (n-1)(p-n)$.
  Since $n \ge 2$ and $p > n$, the right-hand side grows without bound as $h \to \infty$, and so, by \eqref{eq:pen-bound}, does $\mathcal L(\Lambda(h), R(h), \eta)$ for any $\eta \in [0,1]$.
  Along this path, the scales $\Lambda(h)_{ii} = (1 + hP_{ii})^{1/2}$ of the variables with $P_{ii} > 0$ grow without bound.
\end{proof}

\section{Derivation and implementation of the ECM algorithm}\label{app:ecm-details}

This appendix derives the steps of Algorithm~\ref{alg:pcss} (\Cref{app:ecm-monotonicity,app:ecm-estep,app:ecm-mstep}), describes how \pkgname{}~\citep{pcssjl} solves its $R$- and $\Lambda$-updates and stops its iterations (\Cref{app:ecm-admm,app:ecm-lambda,app:ecm-stopping}), and gives the heuristic behind the continuation path of \Cref{subsec:initial_condition} (\Cref{app:ecm-path}).
The package runs the algorithm on the standardized objective \eqref{eq:standardized-objective}, that is, with $S_n$ replaced by $\rho_n$ and $\Lambda$ by the standardized scale $\Lambda_s = D_s\Lambda$ of \Cref{subsec:scale-invariance}, and returns the scales to the data scale at the end.
The relative changes of the scales in the stopping rules below are computed on the standardized scale.

\subsection{Monotonicity}\label{app:ecm-monotonicity}

Given an arbitrary distribution $q(\boldsymbol\sigma)$ over the binary indicators, the EM algorithm exploits the identity
\begin{equation*}
  \log p(\vartheta \mid X) = \mathbb{E}_{q(\boldsymbol\sigma)}\big[\log p(\vartheta, \boldsymbol\sigma \mid X) - \log q(\boldsymbol\sigma)\big] + D_\mathrm{KL}\big(q(\boldsymbol\sigma) \,\|\, p(\boldsymbol\sigma \mid \vartheta, X)\big),
\end{equation*}
where $\mathbb{E}_{q(\boldsymbol\sigma)}[\cdot]$ denotes the expectation with respect to $q(\boldsymbol\sigma)$ and $D_\mathrm{KL}\big(q(\boldsymbol\sigma) \,\|\, p(\boldsymbol\sigma \mid \vartheta, X)\big)$ is the Kullback--Leibler divergence between $q(\boldsymbol\sigma)$ and $p(\boldsymbol\sigma\mid \vartheta, X)$.
Given an estimate $\vartheta^{(t)}$, the E-step sets $q(\boldsymbol\sigma) = p(\boldsymbol\sigma \mid \vartheta^{(t)}, X)$, which leaves
\begin{equation*}
  \log p(\vartheta \mid X) = Q(\vartheta \mid \vartheta^{(t)}) + D_\mathrm{KL}\big(p(\boldsymbol\sigma \mid \vartheta^{(t)}, X) \,\|\, p(\boldsymbol\sigma \mid \vartheta, X)\big),
\end{equation*}
where
\begin{equation} \label{eq:Q_def}
  Q(\vartheta \mid \vartheta^{(t)}) = \mathbb{E}_{p(\boldsymbol\sigma \mid \vartheta^{(t)}, X)}\big[\log p(\vartheta, \boldsymbol\sigma \mid X)\big] + \mathrm{const},
\end{equation}
and the constant depends on $(\vartheta^{(t)}, X)$ but not on $\vartheta$.
\Cref{app:ecm-estep} shows that \eqref{eq:Q_def} coincides with \eqref{eq:Q-function} up to such a constant.
The M-step then proposes a new estimate $\vartheta^{(t+1)}$ such that $Q(\vartheta^{(t+1)} \mid \vartheta^{(t)}) \ge Q(\vartheta^{(t)} \mid \vartheta^{(t)})$.
Since the Kullback--Leibler divergence is nonnegative and vanishes at $\vartheta = \vartheta^{(t)}$, any increase in $Q$ transfers to the log-posterior,
\begin{equation} \label{eq:em_ineq_chain}
  \log p(\vartheta^{(t+1)}\mid X) \ \ge\ Q(\vartheta^{(t+1)}\mid\vartheta^{(t)}) \ \ge\ Q(\vartheta^{(t)}\mid\vartheta^{(t)}) \ =\ \log p(\vartheta^{(t)}\mid X),
\end{equation}
so the log-posterior increases monotonically along the iterates.
The algorithm alternates between the E and the M step until $\log p(\vartheta^{(t+1)}\mid X) \approx \log p(\vartheta^{(t)}\mid X)$ and $\vartheta^{(t+1)} \approx \vartheta^{(t)}$, within some error tolerance.

We stress that the inequality chain \eqref{eq:em_ineq_chain} only requires $Q(\vartheta^{(t+1)} \mid \vartheta^{(t)}) \ge Q(\vartheta^{(t)} \mid \vartheta^{(t)})$.
The new estimate $\vartheta^{(t+1)}$ does not need to be a maximizer of $Q(\cdot \mid \vartheta^{(t)})$ for the argument to hold: any update that increases the expected log-posterior also increases the objective, a variant known as generalized EM~\citep{dempster1977}.
This flexibility is relevant here.
The M-step of \Cref{subsec:mstep} proposes a new estimate that increases $Q(\cdot \mid \vartheta^{(t)})$ without necessarily maximizing it.

\subsection{The E-step}\label{app:ecm-estep}

To compute \eqref{eq:Q_def}, the E-step requires the log-posterior of the full hierarchy, including the binary indicators.
With $\boldsymbol\mu$ fixed at $\boldsymbol{\bar x}$ (\Cref{subsec:objective}), combining the likelihood \eqref{eq:likelihood} with the priors \eqref{eq:prior-mu-lambda} and \eqref{eq:prior-R-sigma-eta} gives, up to an additive constant,
\begin{align} \label{eq:full-log-posterior}
  \log p(\Lambda, R, \boldsymbol\sigma, \eta \mid X)
  &= -\frac n2 \tr\!\big[S_n\Lambda(I_p-R)\Lambda\big]
  + (n-1)\log\det\Lambda + \frac n2\log\det(I_p-R) \nonumber \\
  &\qquad - \gamma \sum_{i<j} (1-\sigma_{ij}) |r_{ij}| + \sum_{i<j} \Big[ \sigma_{ij}\log\eta
  + (1-\sigma_{ij})\log\big((1-\eta)\,\tilde\gamma\big) \Big],
\end{align}
subject to $R\in\mathcal R_p$ and $\eta\in[0,1]$.

Given the current estimate $\vartheta^{(t)} = (\Lambda^{(t)}, R^{(t)}, \eta^{(t)})$, the E-step computes $p(\boldsymbol\sigma \mid \vartheta^{(t)}, X)$.
Since the likelihood \eqref{eq:likelihood} does not involve $\boldsymbol\sigma$, it follows that $p(\boldsymbol\sigma \mid \vartheta, X) = p(\boldsymbol\sigma \mid R, \eta)$, which is the conditional derived in \Cref{subsec:edges}: the indicators are conditionally independent across pairs, and each is on with the slab-membership probability \eqref{eq:responsibility}.
Moreover, the log-posterior \eqref{eq:full-log-posterior} is linear in each $\sigma_{ij}$, so taking its expectation under $p(\boldsymbol\sigma \mid \vartheta^{(t)}, X)$ amounts to replacing every $\sigma_{ij}$ by its conditional mean.
Since $\sigma^{(t)}_{ij}$ in \eqref{eq:sigma_t} is this conditional mean, the expectation of \eqref{eq:full-log-posterior} equals \eqref{eq:Q-function} up to an additive constant independent of $\vartheta$.

\subsection{The M-step}\label{app:ecm-mstep}

The $Q$-function \eqref{eq:Q-function} is concave in each of the blocks $\eta$, $R$ and $\Lambda$ when the other two are held fixed.
With $\Lambda$ and $R$ fixed, the subproblem for $\eta$ is
\begin{equation*}
  \eta^{(t+1)} = \argmax_{\eta \in [0,1]} \ \sum_{i<j} \left[\sigma_{ij}^{(t)} \log\eta + \big(1-\sigma_{ij}^{(t)}\big) \log(1-\eta) \right],
\end{equation*}
whose objective is strictly concave on $(0,1)$ and is maximized in closed form at \eqref{eq:eta_update}.

The objective in \eqref{eq:R_subproblem} is twice the negative of the $R$-dependent part of \eqref{eq:Q-function}.
Notice that here the penalty sum runs over every off-diagonal entry of $\Theta$, so that each pair $(i,j)$ appears twice with the same penalty weight ($\gamma^{(t)}_{ij} = \gamma^{(t)}_{ji}$ and $|\Theta_{ij}| = |\Theta_{ji}|$).
Problem \eqref{eq:R_subproblem} is strictly convex, since the trace term is linear in $\Theta$, $-\log\det\Theta$ is strictly convex on the positive definite matrices, the penalty is a sum of absolute values with nonnegative weights, and the constraints define a convex set.

The objective in \eqref{eq:lambda_subprob} is the negative of the $\Lambda$-dependent part of \eqref{eq:Q-function}, divided by $n$, and \Cref{app:ecm-lambda} shows that it is strictly convex whenever every $(S_n)_{ii}$ is positive.

\subsection{ADMM for the \texorpdfstring{$R$}{R}-update}\label{app:ecm-admm}

We solve the diagonal-constrained, entrywise-weighted graphical lasso \eqref{eq:R_subproblem} by ADMM~\citep{boyd2011}.
The method introduces two copies of the variable, coupled by the constraint $\Phi = \Theta$.
The copy $\Phi$ carries the smooth terms of the objective, while $\Theta$ carries the penalty and the unit-diagonal constraint.
With a symmetric matrix $A$ of dual variables and a parameter $\beta > 0$, the augmented Lagrangian is
\begin{equation} \label{eq:aug-lagrangian}
  L_\beta(\Phi, \Theta, A)
  = n \left[\tr\big(\widehat\Sigma^{(t)} \Phi\big) - \log\det\Phi\right]
  + \sum_{i \neq j} \gamma^{(t)}_{ij} |\Theta_{ij}|
  + \tr\!\big[A (\Phi - \Theta)\big]
  + \frac\beta2 \big\|\Phi - \Theta\big\|_F^2.
\end{equation}
Whenever the constraint $\Phi = \Theta$ is satisfied, the last two terms vanish and $L_\beta$ coincides with the objective of \eqref{eq:R_subproblem}.
ADMM iterates three updates until convergence.

The $\Phi$-update minimizes \eqref{eq:aug-lagrangian} over $\Phi$ exactly.
Its stationarity condition is
\begin{equation*}
  n\Phi^{-1} - \beta\Phi = G,
  \qquad
  G := n\widehat\Sigma^{(t)} + A - \beta\Theta,
\end{equation*}
which diagonalizes in the eigenbasis of the symmetric matrix $G = V \diag(g_1,\dots,g_p) V^\top$.
Each eigenvalue of $\Phi$ is the positive root of a scalar quadratic, giving
\begin{equation*}
  \Phi = V \diag(\varphi_1,\dots,\varphi_p) V^\top,
  \qquad
  \varphi_i = \frac{\sqrt{g_i^2 + 4 n \beta} - g_i}{2\beta} > 0.
\end{equation*}
The update is automatically symmetric and positive definite, and the single symmetric eigendecomposition dominates the cost of each ADMM iteration.
These eigendecompositions also dominate the cost of each iteration of Algorithm~\ref{alg:pcss}.

The $\Theta$-update minimizes \eqref{eq:aug-lagrangian} over $\Theta$, which is separable across entries.
The diagonal is fixed by the constraint $\Theta_{ii} = 1$, and each off-diagonal entry is soft-thresholded,
\begin{equation*}
  \Theta_{ij} = \sgn\!\Big(\Phi_{ij} + \tfrac1\beta A_{ij}\Big)
  \max\Big( \big|\Phi_{ij} + \tfrac1\beta A_{ij}\big|
  - \tfrac1\beta \gamma^{(t)}_{ij},\ 0 \Big),
  \quad \text{for } i \neq j.
\end{equation*}

The dual update is
\begin{equation*}
  A \leftarrow A + \beta(\Phi - \Theta).
\end{equation*}

The iterations stop when the primal and dual residuals,
\begin{equation*}
  \varepsilon_{\mathrm{pri}} = \|\Phi - \Theta\|_F,
  \qquad
  \varepsilon_{\mathrm{dual}} = \beta\,\|\Theta - \Theta_{\mathrm{prev}}\|_F,
\end{equation*}
satisfy
\begin{equation*}
  \varepsilon_{\mathrm{pri}} \le 2^{-7} \max\big(\|\Phi\|_F, \|\Theta\|_F\big),
  \qquad
  \varepsilon_{\mathrm{dual}} \le 2^{-7}\, \|A\|_F,
\end{equation*}
a relative version of the criterion of~\citet[Section~3.3.1]{boyd2011}, or after $2^8$ iterations.
After each iteration, $\beta$ is doubled if $\varepsilon_{\mathrm{pri}} > 10\,\varepsilon_{\mathrm{dual}}$ and halved if $\varepsilon_{\mathrm{dual}} > 10\,\varepsilon_{\mathrm{pri}}$, which keeps the two residuals of comparable size~\citep[Section~3.4.1]{boyd2011}.
At convergence $\Phi = \Theta$, and the common value is taken as $\Theta^{(t+1)}$.
This matrix satisfies the unit-diagonal constraint through the $\Theta$-update and positive definiteness through the $\Phi$-update.

\begin{algorithm}[t]
\caption{$R$-update: ADMM for the subproblem \eqref{eq:R_subproblem}}
\label{alg:R-update}
\begin{algorithmic}[1]
\Function{RUpdate}{$S_n$, $\Lambda$, $R$, $\boldsymbol\sigma$, $\beta$}
  \State $\widehat\Sigma \gets \Lambda S_n \Lambda$;\quad $\Theta \gets I_p - R$;\quad $\Phi \gets \Theta$;\quad $A \gets 0$
  \Repeat
    \State $\Theta_{\mathrm{prev}} \gets \Theta$
    \State $G \gets n\widehat\Sigma + A - \beta\Theta$;\quad compute $G = V \diag(g_1,\dots,g_p) V^\top$
    \State $\Phi \gets V \diag(\varphi_1,\dots,\varphi_p) V^\top$ with $\varphi_i \gets \dfrac{\sqrt{g_i^2 + 4 n \beta} - g_i}{2\beta}$
    \State $\Theta_{ij} \gets \sgn\!\Big(\Phi_{ij} + \tfrac1\beta A_{ij}\Big) \max\Big( \big|\Phi_{ij} + \tfrac1\beta A_{ij}\big| - \tfrac\gamma\beta \big(1-\sigma_{ij}\big),\ 0 \Big)$ for $i \neq j$;\quad $\Theta_{ii} \gets 1$
    \State $A \gets A + \beta(\Phi - \Theta)$
    \State $\varepsilon_{\mathrm{pri}} \gets \|\Phi - \Theta\|_F$;\quad $\varepsilon_{\mathrm{dual}} \gets \beta\|\Theta - \Theta_{\mathrm{prev}}\|_F$
    \State $\beta \gets 2\beta$ if $\varepsilon_{\mathrm{pri}} > 10\,\varepsilon_{\mathrm{dual}}$, and $\beta \gets \beta/2$ if $\varepsilon_{\mathrm{dual}} > 10\,\varepsilon_{\mathrm{pri}}$
  \Until{$\varepsilon_{\mathrm{pri}} \le 2^{-7}\max(\|\Phi\|_F, \|\Theta\|_F)$ and $\varepsilon_{\mathrm{dual}} \le 2^{-7}\|A\|_F$, or after $2^8$ iterations}
  \State \Return $I_p - \Theta$
\EndFunction
\end{algorithmic}
\end{algorithm}

\subsection{Coordinate descent for the \texorpdfstring{$\Lambda$}{Lambda}-update}\label{app:ecm-lambda}

Problem \eqref{eq:lambda_subprob} has a unique minimizer whenever every $(S_n)_{ii}$ is positive.
Indeed, the $R$-update guarantees $R^{(t+1)} \in \mathcal{R}_p$, so $I_p - R^{(t+1)} \succ 0$, and $S_n \succeq 0$ by construction.
By the Schur product theorem, $S_n \circ (I_p - R^{(t+1)}) \succeq 0$, and Oppenheim's inequality bounds its determinant from below,
\begin{equation*}
  \det\bigl[ S_n \circ (I_p - R^{(t+1)}) \bigr] \;\ge\; \det\bigl( I_p - R^{(t+1)} \bigr) \prod_i (S_n)_{ii},
\end{equation*}
which is positive whenever every $(S_n)_{ii} > 0$~\citep{horn2013}.
A positive semidefinite matrix with positive determinant is positive definite, so $S_n \circ (I_p - R^{(t+1)}) \succ 0$ and the quadratic form in \eqref{eq:lambda_subprob} is strictly convex in $\boldsymbol{\lambda}$.
Adding the convex logarithmic barrier, problem~\eqref{eq:lambda_subprob} is strictly convex and coercive on the positive orthant, and therefore has a unique minimizer.

We solve \eqref{eq:lambda_subprob} by coordinate descent.
Sweeping over $i = 1, \dots, p$, each coordinate is set to the exact minimizer of \eqref{eq:lambda_subprob} given the current values of the remaining coordinates, which is the positive root of a scalar quadratic,
\begin{equation*}
  \lambda_i \leftarrow \frac{\sqrt{b_i^2 + 4\left(1-\frac{1}{n}\right) (S_n)_{ii}}-b_i}{2 (S_n)_{ii}},
  \qquad
  b_i := - \sum_{j \neq i} (S_n)_{ij}\, r_{ij}^{(t+1)}\, \lambda_j.
\end{equation*}
Each update uses the most recent values of the other coordinates.
Starting from $\boldsymbol\lambda^{(t)}$, we repeat full sweeps until the scales $\Lambda'$ and $\Lambda''$ obtained in two consecutive sweeps satisfy
\begin{equation*}
  \|\log\Lambda'' - \log\Lambda'\|_F < 2^{-10}\, \|\log\Lambda''\|_F,
\end{equation*}
or for at most $2^{10}$ sweeps, and the result is taken as $\boldsymbol\lambda^{(t+1)}$.
Because the sweeps start from the current estimate and no coordinate update can increase the convex objective, terminating after finitely many sweeps still produces an update that does not decrease $Q$, as required by the inequality chain \eqref{eq:em_ineq_chain}.

\begin{algorithm}[t]
\caption{$\Lambda$-update: coordinate descent for the subproblem \eqref{eq:lambda_subprob}}
\label{alg:lambda-update}
\begin{algorithmic}[1]
\Function{LambdaUpdate}{$S_n$, $R$, $\Lambda$}
  \Repeat
    \State $\Lambda_{\mathrm{prev}} \gets \Lambda$
    \For{$i = 1, \dots, p$}
      \State $b_i \gets - \displaystyle\sum_{j \neq i} (S_n)_{ij}\, r_{ij}\, \Lambda_{jj}$
      \State $\Lambda_{ii} \gets \dfrac{\sqrt{b_i^2 + 4\left(1-\frac{1}{n}\right) (S_n)_{ii}} - b_i}{2 (S_n)_{ii}}$
    \EndFor
  \Until{$\|\log\Lambda - \log\Lambda_{\mathrm{prev}}\|_F < 2^{-10}\|\log\Lambda\|_F$, or after $2^{10}$ sweeps}
  \State \Return $\Lambda$
\EndFunction
\end{algorithmic}
\end{algorithm}

\subsection{Stopping rule of Algorithm~\ref{alg:pcss}}\label{app:ecm-stopping}

Write $\mathcal L_\rho^{(t)}$ for the standardized objective \eqref{eq:standardized-objective} at the $t$-th iterate.
Algorithm~\ref{alg:pcss} stops at the first iteration $t+1$ at which
\begin{gather*}
  \|\log\Lambda^{(t+1)} - \log\Lambda^{(t)}\|_F < 2^{-12}\, \|\log\Lambda^{(t+1)}\|_F,
  \qquad
  \big|\eta^{(t+1)} - \eta^{(t)}\big| < 2^{-12},
  \\
  \big|\mathcal L_\rho^{(t+1)} - \mathcal L_\rho^{(t)}\big| < 2^{-12} \max\big(n, |\mathcal L_\rho^{(t)}|\big),
\end{gather*}
and fewer than $2^{-12}\ell$ pairs enter or leave the support of $R$, or after $\lceil 2^4\sqrt p\,\rceil$ iterations.
The differences $\mathcal L_\rho^{(t+1)} - \mathcal L_\rho^{(t)}$ are those of the log-posterior \eqref{eq:map-objective}.

\subsection{The continuation path}\label{app:ecm-path}

This subsection gives the heuristic behind the continuation path of \Cref{subsec:initial_condition}.
The optimality conditions of the convex problem \eqref{eq:R_subproblem} give
\begin{equation}\label{eq:R-update-residual}
  \Big|\Big[\big(I_p - R^{(t+1)}\big)^{-1} - \widehat\Sigma^{(t)}\Big]_{ij}\Big| \le \frac{\gamma^{(t)}_{ij}}{n},
\end{equation}
with equality when $r^{(t+1)}_{ij} \ne 0$.
We call the left-hand side the residual of $(i,j)$, the difference between the covariance implied by $R^{(t+1)}$ and the sample covariance, both on the scale set by $\Lambda^{(t)}$.
An entry can therefore be nonzero after the update only if its residual reaches the right-hand side.
By \eqref{eq:sigma_t}, the penalty weight on the right-hand side satisfies
\begin{equation}\label{eq:weight-bounds}
  \gamma^{(t)}_{ij} \ge \gamma - 1 \ \ \text{if } |r^{(t)}_{ij}| \le r_c\big(\eta^{(t)}\big) - \frac{\log\gamma}{\gamma},
  \qquad
  \gamma^{(t)}_{ij} \le 1 \ \ \text{if } |r^{(t)}_{ij}| \ge r_c\big(\eta^{(t)}\big) + \frac{\log\gamma}{\gamma}.
\end{equation}
We call the entries in the first case spike entries and those in the second case slab entries, since by \eqref{eq:responsibility} they are assigned to the spike and to the slab, respectively, with probability at least $\gamma/(\gamma+1)$.
Slab entries are almost unpenalized, and spike entries carry a penalty weight close to $\gamma$.
Since $\tilde\gamma > \gamma$, \eqref{eq:rc} gives $r_c(\eta) > \log\gamma/\gamma$ for $\eta \le 1/2$, so for such $\eta^{(t)}$ every zero entry of $R^{(t)}$ is a spike entry.

A simple initial estimate is the diagonal estimate \eqref{eq:diagonal-estimate}, but starting Algorithm~\ref{alg:pcss} from it at the target value of $\gamma$ has two drawbacks.
First, $\eta^{(0)} = 0$ gives $\eta^{(t)} = 0$ for every $t \ge 1$, so the slab plays no role.
Indeed, by \eqref{eq:sigma_t}, $\eta^{(t)} = 0$ gives $\sigma^{(t)}_{ij} = 0$ for every pair, and \eqref{eq:eta_update} then gives $\eta^{(t+1)} = 0$.
Second, for large values of $\gamma$ the algorithm does not leave the diagonal estimate.
At the diagonal estimate,
\begin{equation*}
  \widehat\Sigma^{(0)}_{ij} = \Big(1 - \frac1n\Big)(\rho_n)_{ij}, \qquad i \ne j,
\end{equation*}
so \eqref{eq:R-update-residual} with every penalty weight equal to $\gamma$ gives
\begin{equation*}
  R^{(1)} = 0
  \qquad \iff \qquad
  \gamma \ge (n-1) \max_{i<j} \big|(\rho_n)_{ij}\big| .
\end{equation*}
In that case $\Lambda$ and $\eta$ are unchanged as well, and Algorithm~\ref{alg:pcss} returns the diagonal estimate.

The fit at $\gamma_0$ acts as a screening step.
Its first update starts from the diagonal estimate \eqref{eq:diagonal-estimate}.
By \eqref{eq:R-update-residual}, the optimality condition of a pair $(i,j)$ then holds at $R = 0$ if and only if $|(\rho_n)_{ij}| \le \gamma_0/(n-1) = 1/\sqrt n$.
Since $1/\sqrt n$ is the asymptotic standard deviation of the sample correlation of two independent variables with finite fourth moments~\citep[Theorem~8]{ferguson1996}, the fit leaves the diagonal estimate through the pairs whose sample correlation lies more than about one standard deviation away from its value under independence.
The support of this fit is thus expected to contain the true edges whose marginal correlation stands out from the sampling noise, together with many false ones.
The later fits prune this support.
After each reset of $\eta$, \eqref{eq:weight-bounds} splits the entries into spike entries, with penalty weight at least $\gamma_k - 1$, and slab entries, with penalty weight at most $1$.
As $\gamma_k$ grows, a spike entry needs a residual of at least $(\gamma_k - 1)/n$ to remain nonzero, while slab entries remain almost unpenalized.
Raising $\gamma$ gradually gives each entry several stages, with moderate penalties on spike entries, in which its assignment to the spike or to the slab can change.

The path of \Cref{subsec:initial_condition} requires $\gamma \ge \gamma_0$.
If $\gamma < \gamma_0$, the path is replaced by a single fit at $\gamma$ started from the diagonal estimate, which keeps $\eta = 0$.

\section{Comparison with forests of bounded degree}\label{app:forest}

This appendix extends \Cref{thm:matching,thm:calibration} from matchings to forests with a bounded number of edges and a bounded degree.
\Cref{app:forest-statement} states the extension of \Cref{thm:matching}, and \Cref{app:forest-calibration} the corresponding selected hyperparameter, the extension of \Cref{thm:calibration} and the reason for taking $\bar d = 1$ in the main text.
\Cref{app:supporting,app:proof-forest} prove the extension of \Cref{thm:matching}, and \Cref{app:calibration} proves the extension of \Cref{thm:calibration}.


\subsection{The forest theorem}\label{app:forest-statement}

We use the terminology of \Cref{subsec:gamma_q}.
The edge $(i,j)$ is incident to $i$ and to $j$.
The degree of $k$ in a set of edges $\mathcal E$ is the number of edges of $\mathcal E$ incident to $k$, and a forest is a set of edges that contains no cycle.
The forests with maximum degree at most one are the matchings.
We fix $1 \le \bar m \le p-1$ and $1 \le \bar d \le \bar m$, the largest number of edges and the largest degree of the forests against which the diagonal model is compared.
The bound $p-1$ is the largest number of edges of a forest on $p$ vertices.
The comparison with forests replaces Condition~(\ref{cond:diag-matching}) of \Cref{subsec:gamma_q} by the following condition.
\begin{enumerate}
  \renewcommand{\theenumi}{C\arabic{enumi}}
  \renewcommand{\labelenumi}{(\theenumi)}
  \setcounter{enumi}{2}
  \item\label{cond:diag-forest}
  \emph{Preference over forests.}
  $\mathcal L(\Lambda, R, \eta) < \mathcal L(\widehat\Lambda, 0, 0)$ for every positive diagonal $\Lambda$, every $\eta \in [0,1]$ and every nonzero $R \in \mathcal R_p$ supported on a forest with at most $\bar m$ edges and maximum degree at most $\bar d$.
\end{enumerate}
Since a matching has at most $\lfloor p/2 \rfloor \le p-1$ edges, Condition~(\ref{cond:diag-forest}) with $\bar d = 1$ is Condition~(\ref{cond:diag-matching}).

The lower bound is computed from the ordered absolute sample correlations of \Cref{subsec:gamma_q}, through
\begin{equation*}
  w_{(k)} := \frac{1}{1-|\rho_n|_{(k)}^2},
  \qquad
  \nu(x) := \frac{x}{1-e^{-x}}, \quad \nu(0) := 1 .
\end{equation*}
We write $\gamma_\eta(\bar m)$ for the second term of \eqref{eq:gamma-lower}, which in this notation reads
\begin{equation}\label{eq:gamma2}
  \gamma_\eta(\bar m) := 1 + \frac{1}{\ell} \sum_{k=1}^{\bar m} w_{(k)}^{(n-1)/2} .
\end{equation}
The first term of \eqref{eq:gamma-lower} is replaced by a quantity $\gamma_R(\bar m,\bar d)$.
It is a maximum over the possible arrangements of an edge and of the edges that share an endpoint with it, which we call configurations.

\begin{figure}[t]
  \centering
  \begin{tikzpicture}[
      vertex/.style={circle, fill, inner sep=1.6pt},
      endpoint/.style={circle, draw, line width=0.9pt, fill=white, inner sep=0pt, minimum size=15pt, font=\small},
      typeb/.style={line width=0.9pt},
      types/.style={line width=0.9pt, dash pattern=on 4pt off 2.5pt},
      rank/.style={fill=white, inner sep=1.5pt, font=\small},
    ]
    \node[endpoint] (v1) at (0,0) {$1$};
    \node[endpoint] (v2) at (4.4,0) {$2$};
    \node[vertex] (b1) at (-1.7,1.2) {};
    \node[vertex] (s1) at (-1.7,-1.2) {};
    \node[vertex, gray] (f1) at (-3.6,1.2) {};
    \node[vertex] (b2) at (6.1,1.2) {};
    \node[vertex] (s2) at (6.1,-1.2) {};
    \draw[types, line width=2.4pt] (v1) -- node[rank, above=2pt, fill=none] {$k_0 = 2$} (v2);
    \draw[typeb] (v1) -- node[rank] {$1$} (b1);
    \draw[types] (v1) -- node[rank] {$4$} (s1);
    \draw[types, gray] (b1) -- node[rank, above=2pt, fill=none, text=gray] {$\tau_1 = 1$} (f1);
    \draw[typeb] (v2) -- node[rank] {$3$} (b2);
    \draw[types] (v2) -- node[rank] {$5$} (s2);
  \end{tikzpicture}
  \caption{
    The configuration with $k_0 = 2$, $\mathcal K^\mathrm{b}_1 = \{1\}$, $\mathcal K^\mathrm{s}_1 = \{4\}$, $\tau_1 = 1$, $\mathcal K^\mathrm{b}_2 = \{3\}$, $\mathcal K^\mathrm{s}_2 = \{5\}$ and $\tau_2 = 0$, which belongs to $\mathcal C(\bar m,\bar d)$ whenever $\bar d \ge 3$ and $\bar m \ge 6$.
    The endpoints $1$ and $2$ of the central edge are circled, and the central edge is thick.
    Solid edges are of type $\mathrm b$ and dashed edges of type $\mathrm s$.
    Each edge incident to $1$ or $2$ is labelled with its rank.
    The gray edge is an outer edge, which gives $\tau_1 = 1$ and carries no rank.
  }
  \label{fig:configuration}
\end{figure}
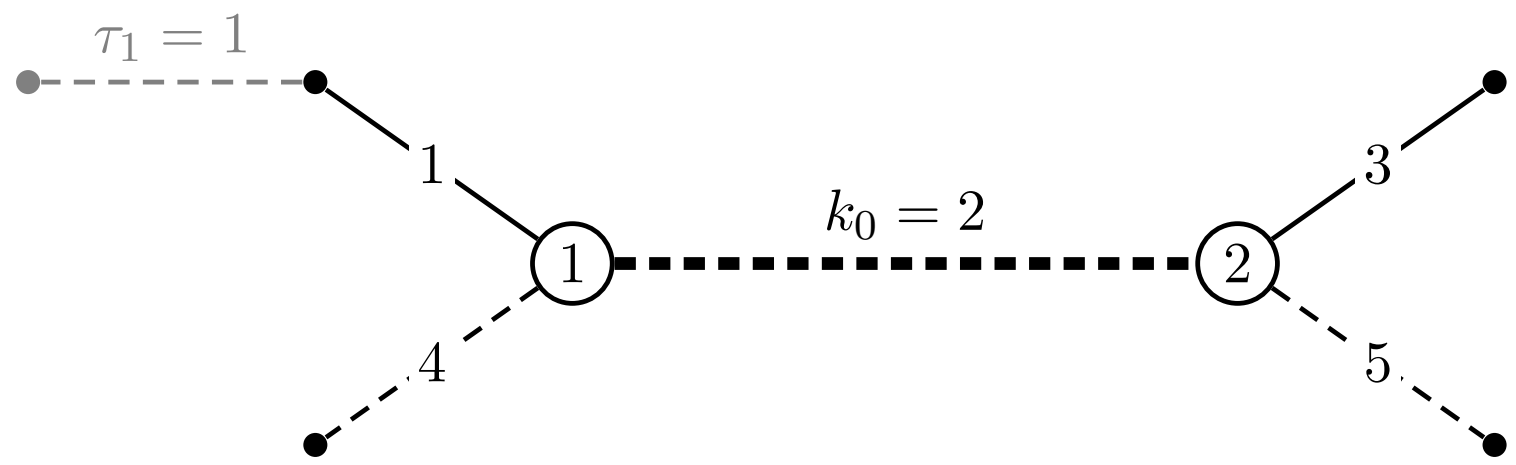

A configuration consists of an edge, which we call the central edge, the edges incident to its endpoints, the sample correlations these edges carry, and one indicator for each endpoint of the central edge.
\Cref{fig:configuration} shows an example.
Since the labels of the variables play no role, we call the endpoints of the central edge $1$ and $2$.

Besides the central edge, other edges may be incident to $1$ or to $2$.
Each of them is of one of two types, labelled $\mathrm b$ and $\mathrm s$.
The letters stand for big and small, and refer to the size of the corresponding entries of $R$ (see the proof of \Cref{thm:forest} in \Cref{app:proof-forest}).
The edges incident to $1$ or $2$ carry the largest sample correlations, in some order, which we record by ranks: the edge of rank $k$ carries $|\rho_n|_{(k)}$, and with it $w_{(k)}$.

The edges incident to neither $1$ nor $2$ carry no rank.
They enter a configuration only through one indicator for each endpoint of the central edge,
\begin{equation*}
  \tau_1 :=
  \begin{cases}
    1 & \text{if there are edges $(1,u)$ of type $\mathrm b$ and $(u,v)$ of type $\mathrm s$ with $v \ne 2$,} \\
    0 & \text{otherwise,}
  \end{cases}
\end{equation*}
and $\tau_2$, defined by exchanging $1$ and $2$.
We call an edge $(u,v)$ as in the first case an outer edge.
It is incident to neither $1$ nor $2$.
Only the existence of an outer edge matters, not their number, and the other edges incident to neither $1$ nor $2$ do not enter at all.
\Cref{lem:scale-bound} and Step 3 of the proof of \Cref{thm:forest} in \Cref{app:proof-forest} show why.

Formally, a configuration is a tuple
\begin{equation*}
  c = \big(k_0,\ \mathcal K^\mathrm{b}_1,\ \mathcal K^\mathrm{s}_1,\ \tau_1,\ \mathcal K^\mathrm{b}_2,\ \mathcal K^\mathrm{s}_2,\ \tau_2\big) .
\end{equation*}
Here $k_0$ is the rank of the central edge, $\mathcal K^\mathrm{b}_1$ and $\mathcal K^\mathrm{s}_1$ are the sets of ranks of the edges of type $\mathrm b$ and of type $\mathrm s$ incident to $1$, other than the central edge, $\mathcal K^\mathrm{b}_2$ and $\mathcal K^\mathrm{s}_2$ are the corresponding sets for $2$, and $\tau_1, \tau_2 \in \{0,1\}$ are the indicators defined above.
We require that $\{k_0\}$, $\mathcal K^\mathrm{b}_1$, $\mathcal K^\mathrm{s}_1$, $\mathcal K^\mathrm{b}_2$ and $\mathcal K^\mathrm{s}_2$ partition $\{1, \dots, |c|\}$, where
\begin{equation*}
  |c| := 1 + |\mathcal K^\mathrm{b}_1| + |\mathcal K^\mathrm{s}_1| + |\mathcal K^\mathrm{b}_2| + |\mathcal K^\mathrm{s}_2|
\end{equation*}
is the number of edges incident to $1$ or $2$, and that
\begin{equation}\label{eq:configuration-constraints}
  \begin{gathered}
    \tau_1 \le |\mathcal K^\mathrm{b}_1|, \qquad \tau_2 \le |\mathcal K^\mathrm{b}_2|, \\
    1 + |\mathcal K^\mathrm{b}_1| + |\mathcal K^\mathrm{s}_1| \le \bar d, \qquad
    1 + |\mathcal K^\mathrm{b}_2| + |\mathcal K^\mathrm{s}_2| \le \bar d, \qquad
    |c| + \tau_1 + \tau_2 \le \bar m .
  \end{gathered}
\end{equation}
The first two constraints say that $\tau_1 = 1$ requires an edge of type $\mathrm b$ incident to $1$, and likewise for $\tau_2$.
The two constraints involving $\bar d$ bound the degrees of $1$ and $2$.
The constraint involving $\bar m$ bounds the number of edges, each indicator equal to $1$ accounting for one outer edge.
We write $\mathcal C(\bar m,\bar d)$ for the set of configurations, which is finite.

For a configuration $c$ set
\begin{equation*}
  \omega_1(c) := \Big(1 + \sum_{k \in \mathcal K^\mathrm{b}_1} \big(w_{(k)} - 1\big)\Big)\,
  \nu\Big[\frac{1+\tau_1}{2}\Big(\log w_{(k_0)} + \sum_{k \in \mathcal K^\mathrm{s}_1} \log w_{(k)}\Big)\Big],
\end{equation*}
and define $\omega_2(c)$ by exchanging the indices $1$ and $2$.
The value $\omega_1(c)$ collects the sample correlations carried by the edges incident to $1$, those of type $\mathrm b$ through the first factor, and those of the central edge and of the edges of type $\mathrm s$ through the second, whose argument is doubled when $\tau_1 = 1$.
We set
\begin{equation}\label{eq:gamma1-forest}
  \gamma_R(\bar m,\bar d) := (n-1) \max_{c \in \mathcal C(\bar m,\bar d)} |\rho_n|_{(k_0)} \sqrt{\omega_1(c)\, \omega_2(c)} .
\end{equation}

The following theorem extends \Cref{thm:matching} to forests.
Its proof is given in \Cref{app:proof-forest}.

\begin{theorem}\label{thm:forest}
  Fix $\bar m$ and $\bar d$ with $1 \le \bar m \le p-1$ and $1 \le \bar d \le \bar m$.
  Suppose
  \begin{equation}\label{eq:gamma_forest_condition}
    \gamma \ge \underline\gamma(\bar m,\bar d) := \max\big\{\gamma_R(\bar m,\bar d),\ \gamma_\eta(\bar m)\big\} ,
  \end{equation}
  where $\gamma_R(\bar m,\bar d)$ and $\gamma_\eta(\bar m)$ are given by \eqref{eq:gamma1-forest} and \eqref{eq:gamma2}.
  Then Conditions~(\ref{cond:diag-local}) and~(\ref{cond:diag-forest}) hold.
\end{theorem}

For $\bar d = 1$, \Cref{thm:forest} reduces to \Cref{thm:matching}.
Indeed, Condition~(\ref{cond:diag-forest}) with $\bar d = 1$ is Condition~(\ref{cond:diag-matching}), and the constraints \eqref{eq:configuration-constraints} leave a single configuration, with $k_0 = 1$, all four sets of ranks empty and $\tau_1 = \tau_2 = 0$.
For it $\omega_1(c) = \omega_2(c) = \nu\big(\tfrac12\log w_{(1)}\big)$, and
\begin{equation*}
  \nu\Big(\frac12 \log \frac{1}{1-r^2}\Big) = \frac{-\log\big(1-r^2\big)}{2\big(1-\sqrt{1-r^2}\big)}, \qquad r \in (0,1),
\end{equation*}
so that $\gamma_R(\bar m,1)$ is the first term of \eqref{eq:gamma-lower} and $\underline\gamma(\bar m,1) = \underline\gamma(\bar m)$.

The set $\mathcal C(\bar m,\bar d)$ grows with $\bar m$ and with $\bar d$, so $\gamma_R(\bar m,\bar d)$ is nondecreasing in both, and comparing the diagonal model with more forests requires a larger lower bound.
For a fixed $\bar d$, however, the growth in $\bar m$ stops.
The degree constraints of \eqref{eq:configuration-constraints} give $|c| \le 2\bar d - 1$, and $\tau_1, \tau_2 \le 1$, so every configuration satisfies $|c| + \tau_1 + \tau_2 \le 2\bar d + 1$.
Hence
\begin{equation*}
  \mathcal C(\bar m,\bar d) \subseteq \mathcal C(2\bar d+1,\bar d) \qquad \text{for every } \bar m ,
\end{equation*}
with equality when $\bar m \ge 2\bar d+1$, so that $\gamma_R(\bar m,\bar d) = \gamma_R(2\bar d+1,\bar d)$ for every $\bar m \ge 2\bar d+1$.
Moreover, every rank of a configuration is at most $|c| \le 2\bar d - 1$, so $\gamma_R(\bar m,\bar d)$ depends on the sample only through the $2\bar d - 1$ largest absolute sample correlations.
Comparing the diagonal model with forests that have more than $2\bar d+1$ edges therefore changes the lower bound only through $\gamma_\eta(\bar m)$.

The maximum in \eqref{eq:gamma1-forest} can be restricted to a smaller set of configurations.
A configuration $c'$ extends a configuration $c$ if it has the same rank $k_0$, each of its four sets of ranks contains the corresponding set of $c$, and each of its indicators is at least the corresponding indicator of $c$.
A configuration of $\mathcal C(\bar m,\bar d)$ is maximal if no other configuration of $\mathcal C(\bar m,\bar d)$ extends it.
The following lemma shows that the maximum is attained at a maximal configuration.

\begin{lemma}\label{lem:maximal-configurations}
  The maximum in \eqref{eq:gamma1-forest} is attained at a maximal configuration.
  If moreover $|\rho_n|_{(\bar m)} > 0$, it is attained only at maximal configurations.
\end{lemma}

\begin{proof}
  Starting from any $c \in \mathcal C(\bar m,\bar d)$, pass repeatedly to an extension other than the current configuration.
  Each such step increases $|c| + \tau_1 + \tau_2$, which is at most $\bar m$, so the process ends, and it ends at a maximal configuration that extends $c$.
  It therefore suffices to show that passing to an extension does not decrease $\omega_1(c)$ and $\omega_2(c)$.
  Passing to an extension keeps $k_0$, and therefore the factor $|\rho_n|_{(k_0)}$ in \eqref{eq:gamma1-forest}.
  Each additional rank $k$ contributes a term $w_{(k)} - 1 \ge 0$ to the first factor of $\omega_1(c)$ or of $\omega_2(c)$, or a term $\log w_{(k)} \ge 0$ to the argument of $\nu$ in the second factor.
  Raising an indicator from $0$ to $1$ doubles the argument of $\nu$.
  Both factors are positive and $\nu$ is increasing, so neither $\omega_1(c)$ nor $\omega_2(c)$ decreases.

  Suppose now that $|\rho_n|_{(\bar m)} > 0$.
  Every rank of a configuration is at most $|c| \le \bar m$, so $|\rho_n|_{(k_0)} > 0$ and $w_{(k)} > 1$ for every rank $k$ that appears.
  The terms added above are then positive, the argument of $\nu$ is positive because it contains $\log w_{(k_0)} > 0$, and $\nu$ is strictly increasing on $[0,\infty)$.
  Hence an extension other than $c$ itself strictly increases $\omega_1(c)$ or $\omega_2(c)$, and with it the quantity maximised in \eqref{eq:gamma1-forest}, so no configuration that is not maximal attains the maximum.
\end{proof}


\subsection{Hyperparameter selection and choice of the degree bound}\label{app:forest-calibration}

Every rank of a configuration is at most $|c| \le \bar m$, so $\underline\gamma(\bar m,\bar d)$, like $\underline\gamma(\bar m)$, depends on the sample only through $|\rho_n|_{(1)}, \dots, |\rho_n|_{(\bar m)}$.
Substituting $|\rho_n|_{(k)} = \bar F_n^{-1}(\Gamma_k/\ell)$ as in \Cref{subsec:gamma_q} gives the selected hyperparameter
\begin{equation} \label{eq:gamma-q-forest}
  \gamma_q(\bar m,\bar d) := \inf\big\{\gamma : \mathrm P\big(\max(\gamma_R^\star(\bar m,\bar d),\gamma_\eta^\star(\bar m)) \le \gamma\big) \ge 1-q\big\},
\end{equation}
where
\begin{equation} \label{eq:gamma-star-forest}
  \gamma_R^\star(\bar m,\bar d) := (n-1) \max_{c \in \mathcal C(\bar m,\bar d)} \varrho_{k_0} \sqrt{\omega_1^\star(c)\, \omega_2^\star(c)} ,
\end{equation}
$\varrho_k$ is given by \eqref{eq:varrho}, $\gamma_\eta^\star(\bar m)$ is $\gamma_\eta(\bar m)$ with $|\rho_n|_{(k)}$ replaced by $\varrho_k$, and $\omega_1^\star(c)$ and $\omega_2^\star(c)$ are defined as $\omega_1(c)$ and $\omega_2(c)$, with $w_{(k)}$ replaced by $(1-\varrho_k^2)^{-1}$.
By the reduction after \Cref{thm:forest}, $\max(\gamma_R^\star(\bar m,1),\gamma_\eta^\star(\bar m)) = \underline\gamma^\star(\bar m)$, so that $\gamma_q(\bar m,1) = \gamma_q(\bar m)$.
The Monte Carlo procedure of \Cref{subsec:gamma_q} computes $\gamma_q(\bar m,\bar d)$ with $\max(\gamma_R^\star(\bar m,\bar d),\gamma_\eta^\star(\bar m))$ in place of $\underline\gamma^\star(\bar m)$.

The following theorem extends \Cref{thm:calibration} to forests.
Its proof is given in \Cref{app:calibration}.

\begin{theorem}\label{thm:calibration-forest}
  Fix $\bar m \ge 1$, $1 \le \bar d \le \bar m$ and $q \in (0,1)$.
  For $n \ge 3$, let $\boldsymbol x_1,\dots,\boldsymbol x_n$ be independent $N_p(\boldsymbol\mu,\Sigma)$ vectors with $\Sigma$ diagonal and positive definite.
  Then
  \begin{equation}\label{eq:calibration-forest}
    \liminf_{p\to\infty}\ \inf_{n \ge 3}\ \mathrm P\big(\underline\gamma(\bar m,\bar d) \le \gamma_q(\bar m,\bar d)\big) \ \ge\ 1-q ,
  \end{equation}
  and the same bound holds for the probability that Conditions~(\ref{cond:diag-local}) and~(\ref{cond:diag-forest}) hold with $\gamma = \gamma_q(\bar m,\bar d)$.
\end{theorem}

For $\bar d = 1$ we have $\underline\gamma(\bar m,1) = \underline\gamma(\bar m)$ and $\gamma_q(\bar m,1) = \gamma_q(\bar m)$, and Condition~(\ref{cond:diag-forest}) is Condition~(\ref{cond:diag-matching}), so \Cref{thm:calibration} is \Cref{thm:calibration-forest} with $\bar d = 1$.

Taking $\bar m = \bar d = p-1$ would rule out every forest-supported alternative, and \Cref{thm:forest} holds for that choice as well.
The restriction to fixed $\bar m$ comes from \Cref{thm:calibration-forest}, which holds for $\bar m$ fixed as $p$ grows.

\begin{figure}[t]
  \centering
  \includegraphics[width=0.8\textwidth]{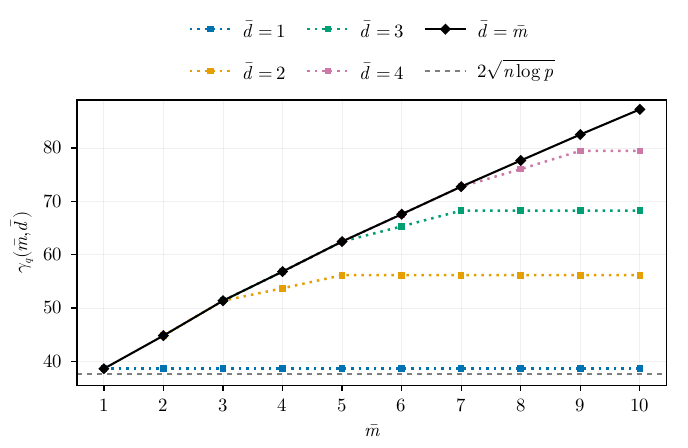}
  \caption{
    Selected hyperparameter $\gamma_q(\bar m,\bar d)$ of \eqref{eq:gamma-q-forest} as a function of $\bar m$, for $\bar d = 1, \dots, 4$ and for $\bar d = \bar m$, with $q = 0.05$, $n = 64$ and $p = 256$.
    The curves are constant from $\bar m = 2\bar d + 1$ on, because $\gamma_R^\star(\bar m,\bar d)$ is constant there and, for these $n$ and $p$, $\gamma_\eta^\star(\bar m)$ rarely exceeds it.
    The gray dashed line marks $2\sqrt{n\log p}$, to which the lower bound \eqref{eq:gamma-q-lower} on $\gamma_q(\bar m) = \gamma_q(\bar m,1)$ is asymptotic.
  }
  \label{fig:gamma-q}
\end{figure}

\Cref{fig:gamma-q} shows $\gamma_q(\bar m,\bar d)$ as a function of $\bar m$ for several values of $\bar d$.
The discussion after \Cref{thm:forest} shows that, for each $\bar d$, $\gamma_R(\bar m,\bar d)$ stops increasing at $\bar m = 2\bar d+1$, and so does $\gamma_R^\star(\bar m,\bar d)$.
For the values of $n$ and $p$ of the figure, $\gamma_\eta^\star(\bar m)$ rarely exceeds $\gamma_R^\star(\bar m,\bar d)$, so $\gamma_q(\bar m,\bar d)$ coincides with the $(1-q)$-quantile of $\gamma_R^\star(\bar m,\bar d)$ and inherits its saturation.
We observed the same saturation at $\bar m = 2\bar d + 1$ for most of the values of $n$ and $p$ we considered, with $p$ moderately large and $n$ not too small.
For much smaller $n$, $\gamma_\eta^\star(\bar m)$ can exceed $\gamma_R^\star(\bar m,\bar d)$, and $\gamma_q(\bar m,\bar d)$ then keeps increasing with $\bar m$ past $2\bar d+1$.

Unlike $\bar m$, the degree bound $\bar d$ has a marked effect on $\gamma_q(\bar m,\bar d)$, as the figure shows.
A larger $\bar d$ widens the class of alternatives covered by \Cref{thm:calibration-forest}, at the cost of a larger hyperparameter.
In our simulations this larger hyperparameter changes the precision and recall of the estimated graphs little.
We fitted the PCSS estimator with $\gamma = \gamma_q(5,1)$ and with $\gamma = \gamma_q(5,2)$ to all the samples of \Cref{sec:simulations}.
In $214$ of the $240$ combinations of ground truth model, mean degree, sample size and magnitude, the number of estimates without edges is the same for both, and their mean precision and recall differ by at most $0.01$.
In $23$ of the remaining $26$, $\bar d = 1$ gives fewer estimated edges, either through more estimates without edges or through a lower recall.
We therefore take $\bar d = 1$ in the main text, for which the selection takes the simpler form of \Cref{subsec:gamma_q}.


\subsection{Supporting results}\label{app:supporting}

For a pair $(i,j)$ write
\begin{equation*}
  w_{ij} := \frac{1}{1-(\rho_n)_{ij}^2} ,
\end{equation*}
so that $w_{(k)}$ is the $k$-th largest of the values $w_{ij}$.
The smooth part $\mathcal L_\mathrm{s}$ of the objective is defined in \eqref{eq:smooth-part}.
For $R \in \mathcal R_p$ we write
\begin{equation*}
  \mathcal L_\mathrm{s}^\star(R) := \max_{\Lambda} \mathcal L_\mathrm{s}(\Lambda, R) ,
\end{equation*}
the maximum running over the positive diagonal matrices $\Lambda$.
The maximum is attained.
Indeed, $\tr[S_n \Lambda (I_p - R) \Lambda]$ is the quadratic form of $S_n \circ (I_p - R)$ evaluated at the diagonal of $\Lambda$, and $S_n \circ (I_p - R) \succ 0$ by the Schur product theorem, because $S_n \succeq 0$ has positive diagonal and $I_p - R \succ 0$.
Hence $\mathcal L_\mathrm{s}(\cdot, R)$ tends to $-\infty$ as any diagonal entry of $\Lambda$ tends to $0$ or to $\infty$.

\begin{lemma}\label{lem:diag-local-max}
  If $\gamma > \gamma_d := (n-1)\max_{i<j}|(\rho_n)_{ij}|$, then the diagonal
  estimate \eqref{eq:diagonal-estimate}
  is a strict local maximum of the objective \eqref{eq:map-objective}. If
  $\gamma < \gamma_d$, then \eqref{eq:map-objective}
  has no local maximum with $R = 0$.
\end{lemma}

\begin{proof}
  Write $\mathcal L_{R=0}(\Lambda, \eta) := \mathcal L(\Lambda, 0, \eta)$ for the restriction of the objective to $R = 0$.
  Since the log-determinant term of \eqref{eq:map-objective} vanishes at $R = 0$,
  \begin{equation*}
    \mathcal{L}_{R=0}(\Lambda, \eta)
    = - \frac{n}{2} \sum_i (S_n)_{ii} \Lambda_{ii}^2 + (n-1) \sum_i \log \Lambda_{ii}
    + \ell \, \log\big[\tilde{\gamma} + \eta (1-\tilde\gamma)\big].
  \end{equation*}
  This function is separable in $\Lambda$ and $\eta$.
  The $\Lambda$-part is strictly concave in each $\Lambda_{ii}$, with derivative vanishing exactly at $\widehat\Lambda_{ii}$ of \eqref{eq:diagonal-estimate}.
  The $\eta$-part is strictly decreasing on $[0,1]$, since $\tilde\gamma > 1$ for every $\gamma > 0$, and is therefore maximized at $\widehat\eta = 0$.
  Hence $(\widehat\Lambda, 0)$ is the unique maximizer of $\mathcal L_{R=0}$, and in particular its unique local maximizer.

  Suppose first that $\gamma > \gamma_d$.
  We show that the objective strictly decreases in a neighborhood of \eqref{eq:diagonal-estimate}, in three steps: we first remove the dependence on $\eta$, then bound the increment of the objective away from the diagonal estimate, and finally choose a neighborhood on which that bound is negative.

  \emph{Step 1: elimination of $\eta$.}
  Differentiating \eqref{eq:map-objective},
  \begin{equation*}
    \frac{\partial \mathcal L}{\partial \eta}
    = \sum_{i<j} \frac{1 - \tilde\gamma\, e^{-\gamma |r_{ij}|}}{\eta + (1-\eta)\, \tilde\gamma\, e^{-\gamma |r_{ij}|}},
  \end{equation*}
  whose denominators are positive, so every summand is negative whenever $|r_{ij}| < \gamma^{-1}\log\tilde\gamma$.
  This condition constrains only $R$, so on the set
  \begin{equation*}
    \mathcal N_1 := \Big\{ (\Lambda, R, \eta) \ : \ \Lambda_{ii} > 0, \ \ R \in \mathcal R_p, \ \ \eta \in [0,1], \ \ \max_{i<j} |r_{ij}| < \gamma^{-1}\log\tilde\gamma \Big\}
  \end{equation*}
  we have $\mathcal L(\Lambda, R, \eta) \le \mathcal L(\Lambda, R, 0)$ for every $\eta \in [0,1]$, with equality only at $\eta = 0$.
  It therefore suffices to show that $(\widehat\Lambda, 0)$ is a strict local maximum of $\mathcal L_{\eta=0}(\Lambda, R) := \mathcal L(\Lambda, R, 0)$.

  \emph{Step 2: an upper bound on the increment.}
  Setting $\eta = 0$ in the penalty of \eqref{eq:map-objective} gives $\mathrm{pen}(r_{ij}, 0) = \gamma|r_{ij}| - \log\tilde\gamma$, so
  \begin{equation*}
    \mathcal L_{\eta=0}(\Lambda, R) = \mathcal{L}_\mathrm{s}(\Lambda, R) - \gamma \sum_{i<j} |r_{ij}| + \ell\log\tilde\gamma,
  \end{equation*}
  where $\mathcal L_\mathrm{s}$ is the smooth part \eqref{eq:smooth-part}.
  Subtracting the value at $(\widehat\Lambda, 0)$ and splitting the variation of $\mathcal L_\mathrm{s}$ into a move in $\Lambda$ and a move in $R$,
  \begin{equation} \label{eq:proof-split}
    \mathcal L_{\eta=0}(\Lambda, R) - \mathcal L_{\eta=0}(\widehat\Lambda, 0)
    = \underbrace{\big[\mathcal{L}_\mathrm{s}(\Lambda, 0) - \mathcal{L}_\mathrm{s}(\widehat\Lambda, 0)\big]}_{=:\ \mathrm{(A)}}
    + \underbrace{\big[\mathcal{L}_\mathrm{s}(\Lambda, R) - \mathcal{L}_\mathrm{s}(\Lambda, 0)\big]}_{=:\ \mathrm{(B)}}
    - \gamma \sum_{i<j} |r_{ij}|.
  \end{equation}
  Term $\mathrm{(A)}$ is the $\Lambda$-part of $\mathcal L_{R=0}$ evaluated at $\Lambda$ and at its maximizer, so $\mathrm{(A)} \le 0$, with equality only at $\Lambda = \widehat\Lambda$.
  For term $\mathrm{(B)}$, the map $\mathcal L_\mathrm{s}(\Lambda, \cdot)$ is concave on $\{R : R \in \mathcal R_p\}$, since the trace term is affine in $R$ and $\log\det(I_p-R)$ is concave, so it lies below its linearization at $R = 0$.
  The derivative of $\mathcal L_\mathrm{s}$ with respect to $r_{ij}$ at $(\Lambda, 0)$ equals $n\Lambda_{ii}\Lambda_{jj}(S_n)_{ij}$, hence
  \begin{equation} \label{eq:proof-linearization}
    \mathrm{(B)} \ \le\ n \sum_{i<j} \Lambda_{ii}\Lambda_{jj} (S_n)_{ij}\, r_{ij}
    \ \le\ n \sum_{i<j} \Lambda_{ii}\Lambda_{jj} \big|(S_n)_{ij}\big|\, |r_{ij}| .
  \end{equation}
  At $\Lambda = \widehat\Lambda$ the coefficient of $|r_{ij}|$ equals $(n-1)|(\rho_n)_{ij}|$.
  The final step bounds it for $\Lambda$ near $\widehat\Lambda$.

  \emph{Step 3: choice of the neighborhood.}
  Set $a := \max_i \widehat\Lambda_{ii}$ and $s := \max_{i<j}|(S_n)_{ij}|$, and define
  \begin{equation*}
    \mathcal N := \Big\{ (\Lambda, R, \eta) \in \mathcal N_1 \ : \ \max_i \big|\Lambda_{ii} - \widehat\Lambda_{ii}\big| \le \delta \Big\},
    \qquad
    \delta := \min\left\{ a, \ \frac{\gamma - \gamma_d}{6\, n\, s\, a} \right\} > 0 .
  \end{equation*}
  On $\mathcal N$ each $\Lambda_{ii} \le a + \delta \le 2a$, so
  \begin{equation*}
    \Lambda_{ii}\Lambda_{jj} - \widehat\Lambda_{ii}\widehat\Lambda_{jj}
    = (\Lambda_{ii} - \widehat\Lambda_{ii})\,\Lambda_{jj} + \widehat\Lambda_{ii}\,(\Lambda_{jj} - \widehat\Lambda_{jj})
    \le \delta\,(\Lambda_{jj} + \widehat\Lambda_{ii}) \le 3 a \delta,
  \end{equation*}
  and therefore, for every pair,
  \begin{equation*}
    n \Lambda_{ii}\Lambda_{jj}\big|(S_n)_{ij}\big|
    \ \le\ n\big|(S_n)_{ij}\big|\, \widehat\Lambda_{ii}\widehat\Lambda_{jj} + 3 n s a\, \delta
    \ \le\ (n-1)\big|(\rho_n)_{ij}\big| + \frac{\gamma - \gamma_d}2 .
  \end{equation*}
  Substituting this and \eqref{eq:proof-linearization} into \eqref{eq:proof-split},
  \begin{equation*}
    \mathcal L(\Lambda, R, 0) - \mathcal L(\widehat\Lambda, 0, 0)
    \ \le\ \mathrm{(A)} + \sum_{i<j} \Big[ (n-1)\big|(\rho_n)_{ij}\big| + \frac{\gamma - \gamma_d}2 - \gamma \Big] |r_{ij}|
    \ \le\ \mathrm{(A)} - \frac{\gamma - \gamma_d}2 \sum_{i<j} |r_{ij}| ,
  \end{equation*}
  where the last inequality uses $(n-1)|(\rho_n)_{ij}| \le \gamma_d$ for every pair.
  The right-hand side is nonpositive and vanishes only at $\Lambda = \widehat\Lambda$, $R = 0$.
  Combined with Step 1, this gives $\mathcal L(\Lambda, R, \eta) < \mathcal L(\widehat\Lambda, 0, 0)$ for every $(\Lambda, R, \eta) \in \mathcal N$ other than \eqref{eq:diagonal-estimate}, which is therefore a strict local maximum.

  Conversely, suppose $\gamma < \gamma_d$, and let $(\Lambda^\star, 0, \eta^\star)$ be a local maximum of \eqref{eq:map-objective}.
  Restricting a local maximum to a subset of its arguments preserves local maximality, so $(\Lambda^\star, \eta^\star)$ is a local maximum of $\mathcal L_{R=0}$.
  By the first paragraph, $(\Lambda^\star, \eta^\star) = (\widehat\Lambda, 0)$.
  It therefore suffices to rule out \eqref{eq:diagonal-estimate} itself.
  Let $(i,j)$ attain $|(\rho_n)_{ij}| = \max_{k<l}|(\rho_n)_{kl}|$, and for $u > 0$ let $R(u)$ be the matrix whose only nonzero entries are $R(u)_{ij} = R(u)_{ji} = u\,\sgn\big((\rho_n)_{ij}\big)$.
  Keeping $\Lambda = \widehat\Lambda$ and $\eta = 0$, the decomposition \eqref{eq:proof-split} with the gradient of Step 2 gives
  \begin{equation*}
    \mathcal L\big(\widehat\Lambda, R(u), 0\big) - \mathcal L(\widehat\Lambda, 0, 0)
    = \Big[ (n-1)\big|(\rho_n)_{ij}\big| - \gamma \Big]\, u + O(u^2) \ >\ 0
  \end{equation*}
  for $u$ small enough, so \eqref{eq:diagonal-estimate} is not a local maximum either, and \eqref{eq:map-objective} has no local maximum with $R = 0$.
\end{proof}

\begin{lemma}\label{lem:sign-gauge}
  Let $R \in \mathcal R_p$ be supported on a forest $\mathcal E$.
  There is a diagonal $D$ with entries in $\{-1,1\}$ such that $DRD \in \mathcal R_p$ is supported on $\mathcal E$, satisfies $(\rho_n)_{ij} (DRD)_{ij} \ge 0$ for every $(i,j) \in \mathcal E$, and
  \begin{equation}\label{eq:sign-gauge}
    \mathcal L_\mathrm{s}(\Lambda, DRD) \ge \mathcal L_\mathrm{s}(\Lambda, R)
  \end{equation}
  for every positive diagonal $\Lambda$.
\end{lemma}

\begin{proof}
  Let $D$ be diagonal with $D_{kk} \in \{-1,1\}$.
  Then $(DRD)_{ij} = D_{ii} D_{jj} R_{ij}$, so $DRD$ is symmetric, has null diagonal and is supported on $\mathcal E$.
  Moreover $I_p - DRD = D(I_p - R)D \succ 0$, so $DRD \in \mathcal R_p$.

  We now construct a $D$ with the two remaining properties.
  In each tree of $\mathcal E$ choose one vertex, which we call its root, and for every other vertex $k$ of the tree write $\mathrm{pa}(k)$ for its parent, the vertex that follows $k$ on the path from $k$ to the root.
  Set $D_{kk} := 1$ for every root and for every vertex to which no edge of $\mathcal E$ is incident, and define the remaining entries in order of increasing distance to the root by
  \begin{equation}\label{eq:sign-recursion}
    D_{kk} :=
    \begin{cases}
      D_{\mathrm{pa}(k)\mathrm{pa}(k)} & \text{if } (\rho_n)_{k\mathrm{pa}(k)} R_{k\mathrm{pa}(k)} \ge 0, \\
      -D_{\mathrm{pa}(k)\mathrm{pa}(k)} & \text{otherwise.}
    \end{cases}
  \end{equation}
  Since $\mathcal E$ has no cycle, every vertex other than a root has exactly one parent, which is closer to the root.
  Hence \eqref{eq:sign-recursion} defines each entry exactly once, and $D_{\mathrm{pa}(k)\mathrm{pa}(k)}$ is already defined when $D_{kk}$ is.
  Every edge of $\mathcal E$ joins some vertex $k$ to $\mathrm{pa}(k)$, and \eqref{eq:sign-recursion} gives $D_{ii} D_{jj} (\rho_n)_{ij} R_{ij} = |(\rho_n)_{ij} R_{ij}|$ for every $(i,j) \in \mathcal E$.
  Hence
  \begin{equation}\label{eq:sign-aligned}
    (\rho_n)_{ij} (DRD)_{ij} = \big|(\rho_n)_{ij} R_{ij}\big| \ge 0
    \qquad \text{for every } (i,j) \in \mathcal E .
  \end{equation}

  It remains to prove \eqref{eq:sign-gauge} for this $D$.
  Since $D^2 = I_p$, the last term of \eqref{eq:smooth-part} is unchanged,
  \begin{equation*}
    \log\det(I_p - DRD) = \log\det\big(D (I_p - R) D\big) = \log\det(I_p - R) .
  \end{equation*}
  Expanding the trace term of \eqref{eq:smooth-part}, and using that $R$ has null diagonal and is supported on $\mathcal E$,
  \begin{equation*}
    \tr\big[S_n \Lambda (I_p - R) \Lambda\big]
    = \sum_{k=1}^{p} (S_n)_{kk} \Lambda_{kk}^2
    - 2 \sum_{(i,j) \in \mathcal E} (S_n)_{ij} \Lambda_{ii} \Lambda_{jj} R_{ij} .
  \end{equation*}
  The same identity holds for $DRD$, which is supported on $\mathcal E$ as well, and only the second sum depends on the matrix.
  Therefore
  \begin{equation*}
    \mathcal L_\mathrm{s}(\Lambda, DRD) - \mathcal L_\mathrm{s}(\Lambda, R)
    = n \sum_{(i,j) \in \mathcal E} (S_n)_{ij} \Lambda_{ii} \Lambda_{jj} \big[(DRD)_{ij} - R_{ij}\big] .
  \end{equation*}
  By \eqref{eq:sample-corr} we have $(S_n)_{ij} = \sqrt{(S_n)_{ii} (S_n)_{jj}}\, (\rho_n)_{ij}$, so \eqref{eq:sign-aligned} turns each summand into
  \begin{equation*}
    \sqrt{(S_n)_{ii} (S_n)_{jj}}\; \Lambda_{ii} \Lambda_{jj}
    \Big[\big|(\rho_n)_{ij} R_{ij}\big| - (\rho_n)_{ij} R_{ij}\Big] \ge 0 ,
  \end{equation*}
  because $\Lambda$ is positive.
  Summing over $\mathcal E$ gives \eqref{eq:sign-gauge}.
\end{proof}

\begin{lemma}\label{lem:edge-removal}
  Let $R \in \mathcal R_p$ be supported on a forest $\mathcal E$, and let $\mathcal E' \subseteq \mathcal E$.
  Let $R'$ be the matrix obtained from $R$ by setting $R_{ij}$ and $R_{ji}$ to zero for every $(i,j) \in \mathcal E'$.
  Then $R' \in \mathcal R_p$ and
  \begin{equation*}
    \log\det(I_p - R) \le \sum_{(i,j) \in \mathcal E'} \log\big(1-R_{ij}^2\big) + \log\det(I_p - R') .
  \end{equation*}
\end{lemma}

\begin{proof}
  We first write $\det(I_p - R)$ as a sum over matchings, that is, over sets of edges no two of which share an endpoint.
  By the Leibniz formula,
  \begin{equation*}
    \det(I_p - R) = \sum_{\pi} \operatorname{sgn}(\pi) \prod_{k=1}^{p} (I_p - R)_{k\pi(k)} ,
  \end{equation*}
  the sum running over the permutations $\pi$ of $\{1, \dots, p\}$.
  The term of $\pi$ vanishes unless, for every $k$, either $\pi(k) = k$ or $(k,\pi(k)) \in \mathcal E$.
  Write $\pi$ as a product of disjoint cyclic permutations.
  If one of them is $k_1 \mapsto k_2 \mapsto \dots \mapsto k_r \mapsto k_1$ with $r \ge 3$ distinct indices, a nonvanishing term would require the edges $(k_1,k_2), \dots, (k_{r-1},k_r), (k_r,k_1)$ to belong to $\mathcal E$.
  These edges form a cycle, which $\mathcal E$ does not contain because it is a forest.
  Hence, in every permutation with a nonvanishing term, each cyclic permutation either fixes a single index or exchanges two indices $i$ and $j$ with $(i,j) \in \mathcal E$.
  The exchanged pairs are disjoint, so they form a matching $\mathcal M \subseteq \mathcal E$, and conversely each matching arises from exactly one such permutation.
  That permutation has sign $(-1)^{|\mathcal M|}$, and in its term each exchanged pair $(i,j)$ contributes $(-R_{ij})(-R_{ji}) = R_{ij}^2$, and each fixed index contributes a diagonal entry equal to $1$.
  Therefore
  \begin{equation}\label{eq:matching-expansion}
    \det(I_p - R) = \sum_{\mathcal M} (-1)^{|\mathcal M|} \prod_{(i,j) \in \mathcal M} R_{ij}^2 ,
  \end{equation}
  the sum running over the matchings $\mathcal M \subseteq \mathcal E$, the empty matching included.
  The same argument applies to every principal submatrix $(I_p - R)_{\mathcal U}$, $\mathcal U \subseteq \{1, \dots, p\}$, with $\mathcal E$ replaced by the set of edges of $\mathcal E$ with both endpoints in $\mathcal U$.
  Here and below, the determinant of an empty matrix is $1$.

  Consider first $\mathcal E' = \{(k,l)\}$, and write $\mathcal V := \{1, \dots, p\}$.
  We split the matchings of \eqref{eq:matching-expansion} according to whether they contain $(k,l)$.
  Those that do not are the matchings of the forest obtained from $\mathcal E$ by removing $(k,l)$, on which $R'$ is supported, so their terms add up to $\det(I_p - R')$.
  Those that do are the sets $\mathcal M \cup \{(k,l)\}$ with $\mathcal M$ a matching of the edges of $\mathcal E$ with both endpoints in $\mathcal V \setminus \{k,l\}$, so their terms add up to $-R_{kl}^2 \det[(I_p - R)_{\mathcal V \setminus \{k,l\}}]$.
  Hence
  \begin{equation}\label{eq:edge-expansion}
    \det(I_p - R) = \det(I_p - R') - R_{kl}^2 \det\big[(I_p - R)_{\mathcal V \setminus \{k,l\}}\big] .
  \end{equation}

  We now show that $R' \in \mathcal R_p$.
  Let $\mathcal U \subseteq \mathcal V$.
  If $\mathcal U$ does not contain both $k$ and $l$, then $(I_p - R')_{\mathcal U} = (I_p - R)_{\mathcal U}$.
  Otherwise \eqref{eq:edge-expansion} applied to $(I_p - R)_{\mathcal U}$ gives
  \begin{equation*}
    \det\big[(I_p - R)_{\mathcal U}\big] = \det\big[(I_p - R')_{\mathcal U}\big] - R_{kl}^2 \det\big[(I_p - R)_{\mathcal U \setminus \{k,l\}}\big] \le \det\big[(I_p - R')_{\mathcal U}\big] ,
  \end{equation*}
  because the last determinant is that of a principal submatrix of $I_p - R \succ 0$, hence positive.
  In both cases $\det[(I_p - R')_{\mathcal U}] \ge \det[(I_p - R)_{\mathcal U}] > 0$, so every principal minor of $I_p - R'$ is positive and $R' \in \mathcal R_p$.

  Next, let $A \succ 0$ have rows and columns indexed by a set $\mathcal U$, let $k \in \mathcal U$, and let $\boldsymbol a_k$ be the $k$-th column of $A$ with its $k$-th entry removed.
  The Schur complement formula and $A_{\mathcal U \setminus \{k\}}^{-1} \succ 0$ give
  \begin{equation}\label{eq:schur-bound}
    \det A = \det A_{\mathcal U \setminus \{k\}} \big(A_{kk} - \boldsymbol a_k^\top A_{\mathcal U \setminus \{k\}}^{-1} \boldsymbol a_k\big) \le A_{kk} \det A_{\mathcal U \setminus \{k\}} .
  \end{equation}
  We apply \eqref{eq:schur-bound} to $I_p - R'$ with index $k$, and then to $(I_p - R')_{\mathcal V \setminus \{k\}}$ with index $l$.
  Both matrices are positive definite with unit diagonal, and deleting the rows and columns $k$ and $l$ removes the only entries in which $R$ and $R'$ differ, so
  \begin{equation*}
    \det(I_p - R') \le \det\big[(I_p - R')_{\mathcal V \setminus \{k,l\}}\big] = \det\big[(I_p - R)_{\mathcal V \setminus \{k,l\}}\big] .
  \end{equation*}
  Substituting this into \eqref{eq:edge-expansion} gives $\det(I_p - R) \le (1 - R_{kl}^2)\det(I_p - R')$, which is the claim for one edge.
  Since $R'$ is again in $\mathcal R_p$ and supported on a forest, the general case follows by removing the edges of $\mathcal E'$ one at a time.
\end{proof}

\begin{lemma}\label{lem:forest-likelihood}
  Let $\mathcal E$ be a forest and let $\rho$ be symmetric with unit diagonal and $|\rho_{ij}| < 1$ for every $(i,j) \in \mathcal E$.
  Then every $\Omega \succ 0$ supported on $\mathcal E$ satisfies
  \begin{equation*}
    \log\det\Omega - \tr(\rho\,\Omega) + p \le \sum_{(i,j) \in \mathcal E} \log \frac{1}{1 - \rho_{ij}^2} .
  \end{equation*}
  For $\rho = \rho_n$ the right-hand side is $\sum_{(i,j) \in \mathcal E} \log w_{ij}$.
\end{lemma}

\begin{proof}
  The matrix $\rho$ need not be positive definite.
  The idea is to replace it by a positive definite matrix $\Sigma$ that agrees with $\rho$ on the diagonal and on $\mathcal E$, which are the only entries that $\tr(\rho\,\Omega)$ involves, and whose determinant is explicit.

  In each tree of $\mathcal E$ choose a root and define parents as in the proof of \Cref{lem:sign-gauge}.
  Let $A$ be the $p \times p$ matrix with unit diagonal, $A_{l\mathrm{pa}(l)} := -\rho_{l\mathrm{pa}(l)}$ for every vertex $l$ that has a parent $\mathrm{pa}(l)$, and all other entries zero.
  Let $D$ be the diagonal matrix with $D_{ll} := 1 - \rho_{l\mathrm{pa}(l)}^2$ for every vertex $l$ that has a parent, and $D_{ll} := 1$ otherwise.
  Listing the vertices so that every vertex comes after its parent, for instance by increasing distance to the root, makes $A$ lower triangular with unit diagonal, so $A$ is invertible and $\det A = 1$.
  Let $\boldsymbol Z = (Z_1, \dots, Z_p)$ be a standard normal vector in $\mathbb R^p$ and let $\boldsymbol Y = (Y_1, \dots, Y_p) := A^{-1} D^{1/2} \boldsymbol Z$, whose covariance matrix is
  \begin{equation*}
    \Sigma := A^{-1} D \big(A^{-1}\big)^\top .
  \end{equation*}
  Since $D \succ 0$ and $\det A = 1$,
  \begin{equation}\label{eq:tree-determinant}
    \Sigma \succ 0 ,
    \qquad
    \det\Sigma = \det D = \prod_{(i,j) \in \mathcal E} \big(1 - \rho_{ij}^2\big) ,
  \end{equation}
  where the last equality holds because every edge of $\mathcal E$ joins a vertex to its parent.

  We now show that $\Sigma$ agrees with $\rho$ on the diagonal and on $\mathcal E$.
  The equation $A\boldsymbol Y = D^{1/2}\boldsymbol Z$ reads $Y_k = Z_k$ for every vertex $k$ without a parent, and
  \begin{equation*}
    Y_l = \rho_{l\mathrm{pa}(l)}\, Y_{\mathrm{pa}(l)} + \sqrt{1 - \rho_{l\mathrm{pa}(l)}^2}\; Z_l
  \end{equation*}
  for every vertex $l$ with a parent.
  In the latter case $Y_{\mathrm{pa}(l)}$ depends only on the coordinates of $\boldsymbol Z$ at $\mathrm{pa}(l)$ and at the vertices on the path from $\mathrm{pa}(l)$ to the root, so it is independent of $Z_l$.
  By induction on the distance to the root, every $Y_k$ therefore has unit variance, and $\Sigma_{l\mathrm{pa}(l)} = \rho_{l\mathrm{pa}(l)}$ for every vertex $l$ with a parent.
  Since every edge of $\mathcal E$ joins a vertex to its parent, $\Sigma$ agrees with $\rho$ on the diagonal and on $\mathcal E$.
  As $\Omega$ is supported on $\mathcal E$, this gives $\tr(\rho\,\Omega) = \tr(\Sigma\Omega)$.

  Let $\psi_1, \dots, \psi_p$ be the eigenvalues of $\Sigma^{1/2}\Omega\Sigma^{1/2}$, which are positive.
  Then $\log\det\Omega = \sum_k \log\psi_k - \log\det\Sigma$ and $\tr(\Sigma\Omega) = \sum_k \psi_k$, and $\log\psi - \psi + 1 \le 0$ for $\psi > 0$, so by \eqref{eq:tree-determinant}
  \begin{equation*}
    \log\det\Omega - \tr(\rho\,\Omega) + p
    = \sum_{k=1}^{p} \big(\log\psi_k - \psi_k + 1\big) - \log\det\Sigma
    \le \sum_{(i,j) \in \mathcal E} \log \frac{1}{1 - \rho_{ij}^2} .
  \end{equation*}
\end{proof}

\begin{lemma}\label{lem:scale-bound}
  Let $\mathcal E$ be a forest, partitioned into $\mathcal E_\mathrm{b}$ and $\mathcal E_\mathrm{s}$.
  For $k \in \{1, \dots, p\}$ set
  \begin{equation}\label{eq:tau-vertex}
    \tau_k :=
    \begin{cases}
      1 & \text{if there are $l$ and $l'$ with $(k,l) \in \mathcal E_\mathrm{b}$ and $(l,l') \in \mathcal E_\mathrm{s}$,} \\
      0 & \text{otherwise,}
    \end{cases}
  \end{equation}
  and
  \begin{equation}\label{eq:omega-vertex}
    \omega_k := \Big(1 + \sum_{l \,:\, (k,l) \in \mathcal E_\mathrm{b}} \big(w_{kl} - 1\big)\Big)\,
    \nu\Big[\frac{1+\tau_k}{2} \sum_{l \,:\, (k,l) \in \mathcal E_\mathrm{s}} \log w_{kl}\Big] .
  \end{equation}
  Suppose $\gamma \ge (n-1) |(\rho_n)_{ij}| \sqrt{\omega_i \omega_j}$ for every $(i,j) \in \mathcal E_\mathrm{s}$.
  Then every nonzero $R \in \mathcal R_p$ supported on $\mathcal E$ with $(\rho_n)_{ij} R_{ij} \ge 0$ on $\mathcal E$ and $\gamma |R_{ij}| \le \tfrac{n-1}{2} \log w_{ij}$ on $\mathcal E_\mathrm{s}$ satisfies
  \begin{equation}\label{eq:scale-bound}
    \mathcal L_\mathrm{s}^\star(R) - \mathcal L_\mathrm{s}^\star(0) < \frac{n-1}{2} \sum_{(i,j) \in \mathcal E_\mathrm{b}} \log w_{ij} + \gamma \sum_{(i,j) \in \mathcal E_\mathrm{s}} |R_{ij}| .
  \end{equation}
\end{lemma}

\begin{proof}
  We call the edges of $\mathcal E_\mathrm{b}$ big and those of $\mathcal E_\mathrm{s}$ small, and we write $g_{kl} := (\rho_n)_{kl}R_{kl}$, which is nonnegative on $\mathcal E$ by hypothesis.
  The proof has six steps.
  Step 1 writes $\mathcal L_\mathrm{s}^\star(R) - \mathcal L_\mathrm{s}^\star(0)$ as the maximum of a function $v$ of the rescaled diagonal entries of $\Lambda$, in which the edges enter only through the numbers $g_{kl}$.
  Step 2 assigns to each small edge two positive weights, one for each endpoint, and shows that at every vertex the weighted sum of the numbers $g_{kl}$ over the incident small edges is smaller than $1$.
  Step 3 uses these weights and the inequality between the arithmetic and geometric means to bound $v$ by a function in which only the big edges appear.
  Step 4 bounds that function with \Cref{lem:forest-likelihood}, which gives the first sum in \eqref{eq:scale-bound} plus a remainder $h$.
  Step 5 bounds $h$ by a sum of terms, one for each vertex.
  Step 6 bounds each vertex term, and the hypothesis on $\gamma$ then bounds $\tfrac{n-1}{2}h$ by the second sum in \eqref{eq:scale-bound}.

  For a vertex $k$ let $\mathcal E_\mathrm{b}(k)$ and $\mathcal E_\mathrm{s}(k)$ be the sets of big and of small edges incident to $k$, and set
  \begin{equation*}
    b_k := \sum_{(k,l) \in \mathcal E_\mathrm{b}(k)} \big(w_{kl}-1\big),
    \qquad
    t_k := \frac12 \sum_{(k,l) \in \mathcal E_\mathrm{s}(k)} \log w_{kl},
    \qquad
    q_k := 1 + \tau_k ,
  \end{equation*}
  so that $\omega_k = (1+b_k)\,\nu(q_kt_k)$ by \eqref{eq:omega-vertex}.

  \emph{Step 1: rewriting the objective in terms of the scales.}
  By \eqref{eq:diagonal-estimate} we have $(S_n)_{kk}\widehat\Lambda_{kk}^2 = (n-1)/n$, hence $(S_n)_{kl}\widehat\Lambda_{kk}\widehat\Lambda_{ll} = \tfrac{n-1}{n}(\rho_n)_{kl}$.
  Substituting into \eqref{eq:smooth-part} and writing $\boldsymbol\zeta = (\zeta_1, \dots, \zeta_p)$ with $\zeta_k := \Lambda_{kk}/\widehat\Lambda_{kk}$,
  \begin{equation}\label{eq:profiled}
    \mathcal L_\mathrm{s}^\star(R) - \mathcal L_\mathrm{s}^\star(0)
    = \max_{\boldsymbol\zeta > 0} \frac{n-1}{2}\, v(\boldsymbol\zeta) + \frac n2 \log\det(I_p-R),
  \end{equation}
  where
  \begin{equation*}
    v(\boldsymbol\zeta) := 2\sum_k \log\zeta_k - \sum_k \zeta_k^2 + 2\sum_{(k,l) \in \mathcal E} g_{kl}\zeta_k\zeta_l + p ,
  \end{equation*}
  the constant $p$ making $v(\boldsymbol 1) = 0$ at $R = 0$.

  \emph{Step 2: weighting the small edges.}
  For $(k,l) \in \mathcal E_\mathrm{s}$ set $\theta_{kl} := \sqrt{\omega_l/\omega_k}$, so that $\theta_{kl}\theta_{lk} = 1$, and set
  \begin{equation*}
    u_k := \sum_{(k,l) \in \mathcal E_\mathrm{s}(k)} g_{kl}\theta_{kl},
    \qquad
    \bar u_k := \frac{t_k}{\omega_k},
    \qquad
    z_k := 1 - e^{-q_kt_k} .
  \end{equation*}
  For every $(k,l) \in \mathcal E_\mathrm{s}$, the hypothesis on $\gamma$ and the condition $\gamma|R_{kl}| \le \tfrac{n-1}{2}\log w_{kl}$ give
  \begin{equation*}
    g_{kl}\theta_{kl}
    = \frac{|(\rho_n)_{kl}|\sqrt{\omega_k\omega_l}}{\omega_k}\,|R_{kl}|
    \le \frac{\gamma |R_{kl}|}{(n-1)\,\omega_k}
    \le \frac{\log w_{kl}}{2\,\omega_k} .
  \end{equation*}
  Summing over $\mathcal E_\mathrm{s}(k)$ and using $\nu(x) = x/(1-e^{-x})$,
  \begin{equation}\label{eq:weight-bound}
    0 \le u_k \le \bar u_k = \frac{t_k}{(1+b_k)\, \nu(q_kt_k)} = \frac{z_k}{q_k(1+b_k)} < \frac{1}{q_k(1+b_k)} \le 1 .
  \end{equation}

  \emph{Step 3: removing the small edges.}
  This step bounds $v$ by a function in which only the big edges appear, namely the right-hand side of \eqref{eq:after-decoupling} below.
  Set $y_k := 1 - u_k$, which lies in $(0,1]$ by \eqref{eq:weight-bound}, and fix $\boldsymbol\zeta > 0$.
  For every small edge $(k,l)$, expanding $(\sqrt{\theta_{kl}}\,\zeta_k - \sqrt{\theta_{lk}}\,\zeta_l)^2 \ge 0$ and using $\theta_{kl}\theta_{lk} = 1$ gives $2\zeta_k\zeta_l \le \theta_{kl}\zeta_k^2 + \theta_{lk}\zeta_l^2$.
  Multiplying by $g_{kl} \ge 0$ and summing over the small edges,
  \begin{equation*}
    2\sum_{(k,l) \in \mathcal E_\mathrm{s}} g_{kl}\zeta_k\zeta_l
    \le \sum_{(k,l) \in \mathcal E_\mathrm{s}} g_{kl}\big(\theta_{kl}\zeta_k^2 + \theta_{lk}\zeta_l^2\big)
    = \sum_k u_k\zeta_k^2 .
  \end{equation*}
  Substituting into the definition of $v$,
  \begin{equation*}
    v(\boldsymbol\zeta) \le 2\sum_k \log\zeta_k - \sum_k y_k\zeta_k^2 + 2\sum_{(k,l) \in \mathcal E_\mathrm{b}} g_{kl}\zeta_k\zeta_l + p .
  \end{equation*}
  Since $y_k > 0$ we may write the right-hand side in terms of $\zeta'_k := \sqrt{y_k}\,\zeta_k$, which gives
  \begin{equation}\label{eq:after-decoupling}
    v(\boldsymbol\zeta) \le -\sum_k \log y_k + 2\sum_k \log \zeta'_k - \sum_k \zeta_k'^2 + 2 \sum_{(k,l) \in \mathcal E_\mathrm{b}} \frac{g_{kl}}{\sqrt{y_ky_l}}\, \zeta'_k\zeta'_l + p .
  \end{equation}

  \emph{Step 4: bounding the big edges.}
  This step bounds the right-hand side of \eqref{eq:after-decoupling} with \Cref{lem:forest-likelihood}, and thereby proves \eqref{eq:after-forest-likelihood} below.
  Let $\rho$ be the symmetric matrix with unit diagonal, $\rho_{kl} := (\rho_n)_{kl}/\sqrt{y_ky_l}$ for $(k,l) \in \mathcal E_\mathrm{b}$, and all other entries zero.
  Let $R'$ be $R$ with the small edges zeroed, which lies in $\mathcal R_p$ by \Cref{lem:edge-removal}.
  Set $\Lambda' := \diag \boldsymbol\zeta'$ and $\Omega := \Lambda'(I_p - R')\Lambda'$, which is positive definite and supported on $\mathcal E_\mathrm{b}$.
  Since $\rho_{kl}R_{kl} = g_{kl}/\sqrt{y_ky_l}$ on $\mathcal E_\mathrm{b}$,
  \begin{align*}
    \log\det\Omega &= 2\sum_k\log\zeta'_k + \log\det(I_p-R'), \\
    \tr(\rho\,\Omega) &= \sum_k \zeta_k'^2 - 2\sum_{(k,l) \in \mathcal E_\mathrm{b}} \frac{g_{kl}}{\sqrt{y_ky_l}}\, \zeta'_k\zeta'_l .
  \end{align*}
  Therefore, adding $\log\det(I_p-R')$ to both sides, \eqref{eq:after-decoupling} can be written as
  \begin{equation}\label{eq:omega-form}
    v(\boldsymbol\zeta) + \log\det(I_p-R') \le -\sum_k \log y_k + \log\det\Omega - \tr(\rho\,\Omega) + p .
  \end{equation}

  To apply \Cref{lem:forest-likelihood} to $\mathcal E_\mathrm{b}$ and $\rho$ it remains to check that $|\rho_{kl}| < 1$ for every $(k,l) \in \mathcal E_\mathrm{b}$.
  Since $w_{kl} \ge 1$ and
  \begin{equation}\label{eq:rho-identity}
    1 - \rho_{kl}^2 = \frac{1}{w_{kl}}\Big[1 - \big(w_{kl}-1\big)\Big(\frac{1}{y_ky_l}-1\Big)\Big] ,
  \end{equation}
  this amounts to $(w_{kl}-1)(1/(y_ky_l)-1) < 1$.
  Set $\chi_k := y_k^{-q_k} - 1 \ge 0$.
  Since $q_k \in \{1,2\}$, we have $(1-\bar u_k)^{q_k} \ge 1 - q_k\bar u_k$, with equality for $q_k = 1$, and $q_k\bar u_k = z_k/(1+b_k)$ by \eqref{eq:weight-bound}, so
  \begin{equation}\label{eq:key-vertex}
    (1-\bar u_k)^{-q_k} \le \frac{1+b_k}{1+b_k - z_k} ,
  \end{equation}
  the denominator being positive because $z_k < 1$.
  As $y_k \ge 1 - \bar u_k$ by \eqref{eq:weight-bound}, it follows that
  \begin{equation}\label{eq:positivity}
    b_k\chi_k \le b_k\big[(1-\bar u_k)^{-q_k}-1\big] \le \frac{b_k z_k}{1 + b_k - z_k} < 1 ,
  \end{equation}
  the last step because $b_kz_k < 1 + b_k - z_k$ is equivalent to $z_k < 1$.
  For $(k,l) \in \mathcal E_\mathrm{b}$ we have $w_{kl}-1 \le b_k$ and $w_{kl}-1 \le b_l$, so $(w_{kl}-1)\chi_k < 1$ and $(w_{kl}-1)\chi_l < 1$.
  Note that $y_k < 1$ only if $u_k > 0$, that is, only if some small edge is incident to $k$.
  We now distinguish two cases, which will also serve in Step 5.
  In the first case, at least one of $y_k$ and $y_l$ equals $1$, say $y_l = 1$ after exchanging $k$ and $l$.
  Then $\chi_l = 0$ and, since $q_k \ge 1$ and $y_k \le 1$,
  \begin{equation*}
    \big(w_{kl}-1\big)\Big(\frac{1}{y_ky_l} - 1\Big) = \big(w_{kl}-1\big)\Big(\frac{1}{y_k} - 1\Big) \le \big(w_{kl}-1\big) \chi_k < 1 .
  \end{equation*}
  In the second case, $y_k < 1$ and $y_l < 1$, so small edges are incident to both $k$ and $l$.
  As $(k,l)$ is a big edge, \eqref{eq:tau-vertex} gives $\tau_k = \tau_l = 1$, that is, $q_k = q_l = 2$.
  Then $1/(y_ky_l) \le \tfrac12(y_k^{-2} + y_l^{-2})$ gives
  \begin{equation}\label{eq:midpoint}
    \big(w_{kl}-1\big)\Big(\frac{1}{y_ky_l} - 1\Big) \le \tfrac12 \big(w_{kl}-1\big) \chi_k + \tfrac12 \big(w_{kl}-1\big) \chi_l < 1 .
  \end{equation}
  In both cases $|\rho_{kl}| < 1$, and \Cref{lem:forest-likelihood} with $\mathcal E_\mathrm{b}$ in place of $\mathcal E$, together with \eqref{eq:rho-identity}, gives
  \begin{equation}\label{eq:big-edge-bound}
    \log\det\Omega - \tr(\rho\,\Omega) + p
    \le \sum_{(k,l) \in \mathcal E_\mathrm{b}} \log w_{kl} - \sum_{(k,l) \in \mathcal E_\mathrm{b}} \log\Big[1 - \big(w_{kl}-1\big)\Big(\frac{1}{y_ky_l}-1\Big)\Big] .
  \end{equation}

  It remains to insert these bounds into \eqref{eq:profiled}.
  By \Cref{lem:edge-removal} with $\mathcal E' = \mathcal E_\mathrm{s}$, and since $\log(1-R_{ij}^2) \le 0$, we have $\log\det(I_p-R) \le \log\det(I_p-R')$.
  We apply this bound to the part $\tfrac{n-1}{2}\log\det(I_p-R)$ of the last term of \eqref{eq:profiled}, and then \eqref{eq:omega-form} and \eqref{eq:big-edge-bound}.
  The resulting bound on the expression maximised in \eqref{eq:profiled} does not depend on $\boldsymbol\zeta$, so it bounds the maximum, and
  \begin{equation}\label{eq:after-forest-likelihood}
    \mathcal L_\mathrm{s}^\star(R) - \mathcal L_\mathrm{s}^\star(0)
    \le \frac{n-1}{2} \sum_{(k,l) \in \mathcal E_\mathrm{b}} \log w_{kl} + \frac{n-1}{2}\, h + \frac12 \log\det(I_p-R) ,
  \end{equation}
  where
  \begin{equation*}
    h := -\sum_k \log y_k - \sum_{(k,l) \in \mathcal E_\mathrm{b}} \log\Big[1 - \big(w_{kl}-1\big)\Big(\frac{1}{y_ky_l}-1\Big)\Big] .
  \end{equation*}

  \emph{Step 5: splitting the remainder over the vertices.}
  We claim that every $(k,l) \in \mathcal E_\mathrm{b}$ satisfies
  \begin{equation}\label{eq:edge-split}
    - \log\big[1 - (w_{kl}-1)\big(\tfrac{1}{y_ky_l}-1\big)\big]
    \le - \frac{1}{q_k} \log\big[1 - (w_{kl}-1)\chi_k\big] - \frac{1}{q_l} \log\big[1 - (w_{kl}-1)\chi_l\big] .
  \end{equation}
  In the first case of Step 4, with $y_l = 1$ after exchanging $k$ and $l$, we have $\chi_l = 0$, so the second term on the right-hand side of \eqref{eq:edge-split} vanishes and the claim reads
  \begin{equation*}
    -\log\Big[1-\big(w_{kl}-1\big)\big(y_k^{-1}-1\big)\Big] \le -\frac{1}{q_k}\log\Big[1-\big(w_{kl}-1\big)\big(y_k^{-q_k}-1\big)\Big] .
  \end{equation*}
  For $q_k = 1$ both sides are equal, and for $q_k = 2$ the claim is equivalent to $(w_{kl}-1)(y_k^{-1}-1)^2 w_{kl} \ge 0$, which holds.
  In the second case, $q_k = q_l = 2$.
  The first inequality in \eqref{eq:midpoint}, and the fact that $x \mapsto -\log(1-(w_{kl}-1)x)$ is nondecreasing and convex on $[0, (w_{kl}-1)^{-1})$, give
  \begin{align*}
    - \log\Big[1 - \big(w_{kl}-1\big)\Big(\frac{1}{y_ky_l}-1\Big)\Big]
    &\le - \log\Big[1 - \big(w_{kl}-1\big)\frac{\chi_k + \chi_l}{2}\Big] \\
    &\le - \tfrac12 \log\big[1 - (w_{kl}-1)\chi_k\big] - \tfrac12 \log\big[1 - (w_{kl}-1)\chi_l\big] ,
  \end{align*}
  which is \eqref{eq:edge-split}.
  Hence the claim is true in both cases.

  We now sum \eqref{eq:edge-split} over the big edges.
  Each big edge $(k,l)$ contributes one term to its endpoint $k$ and one to its endpoint $l$, so
  \begin{equation*}
    - \sum_{(k,l) \in \mathcal E_\mathrm{b}} \log\Big[1 - \big(w_{kl}-1\big)\Big(\frac{1}{y_ky_l}-1\Big)\Big]
    \le - \sum_k \frac{1}{q_k} \sum_{(k,l) \in \mathcal E_\mathrm{b}(k)} \log\big[1 - (w_{kl}-1)\chi_k\big] .
  \end{equation*}
  For numbers $a_1, \dots, a_r \in [0,1]$ we have $\prod_{i \le r}(1-a_i) \ge 1 - \sum_{i \le r} a_i$, by induction on $r$, since $(1 - \sum_{i<r} a_i)(1 - a_r) = 1 - \sum_{i \le r} a_i + a_r\sum_{i<r} a_i$.
  The numbers $(w_{kl}-1)\chi_k$, for $(k,l) \in \mathcal E_\mathrm{b}(k)$, lie in $[0,1]$ because they are nonnegative and at most $b_k\chi_k$, which is smaller than $1$ by \eqref{eq:positivity}, so
  \begin{equation*}
    \prod_{(k,l) \in \mathcal E_\mathrm{b}(k)} \big[1-(w_{kl}-1)\chi_k\big]
    \ge 1 - \sum_{(k,l) \in \mathcal E_\mathrm{b}(k)} \big(w_{kl}-1\big)\chi_k
    = 1 - b_k\chi_k > 0 .
  \end{equation*}
  Taking logarithms, the inner sum on the right-hand side above is at least $\log(1 - b_k\chi_k)$.
  Since $y_k = 1 - u_k$ and $\chi_k = (1-u_k)^{-q_k} - 1$, this gives
  \begin{equation}\label{eq:h-split}
    h \le \sum_k h_k(u_k),
    \qquad
    h_k(u) := -\log(1-u) - \frac{1}{q_k}\log\Big[1 - b_k\big((1-u)^{-q_k}-1\big)\Big] .
  \end{equation}

  \emph{Step 6: bounding each vertex term.}
  The function $h_k$ is defined and finite on $[0,\bar u_k]$ by \eqref{eq:positivity}, and it is convex and nondecreasing there, with $h_k(0) = 0$.
  If $\bar u_k = 0$ then $u_k = 0$ and $h_k(u_k) = 0 = u_k\omega_k$.
  Otherwise we claim that
  \begin{equation*}
    h_k(u_k) \le \frac{u_k}{\bar u_k}\, h_k(\bar u_k) \le u_k\, \frac{t_k}{\bar u_k} = u_k \omega_k .
  \end{equation*}
  The first inequality holds because $h_k$ is convex with $h_k(0) = 0$ and $0 \le u_k \le \bar u_k$.
  The equality is the definition of $\bar u_k$.
  It remains to prove the second inequality, that is, $h_k(\bar u_k) \le t_k$.
  Inequality \eqref{eq:key-vertex} reads $(1+b_k)(1-\bar u_k)^{q_k} \ge 1 + b_k - z_k$, and $1 - z_k = e^{-q_kt_k}$, so
  \begin{align*}
    h_k(\bar u_k)
    &= -\frac{1}{q_k} \log\Big[(1-\bar u_k)^{q_k}\Big(1 + b_k - b_k(1-\bar u_k)^{-q_k}\Big)\Big] \\
    &= -\frac{1}{q_k} \log\Big[(1+b_k)(1-\bar u_k)^{q_k} - b_k\Big] \\
    &\le -\frac{1}{q_k} \log\big(1 - z_k\big)
    = t_k .
  \end{align*}

  Combining \eqref{eq:h-split} with $h_k(u_k) \le u_k\omega_k$,
  \begin{align*}
    \frac{n-1}{2}\, h
    &\le \frac{n-1}{2} \sum_k u_k\omega_k
    = \frac{n-1}{2} \sum_k \sum_{(k,l) \in \mathcal E_\mathrm{s}(k)} g_{kl}\sqrt{\omega_k\omega_l} \\
    &= (n-1) \sum_{(k,l) \in \mathcal E_\mathrm{s}} g_{kl}\sqrt{\omega_k\omega_l}
    \le \gamma \sum_{(k,l) \in \mathcal E_\mathrm{s}} |R_{kl}| .
  \end{align*}
  The first equality uses $u_k\omega_k = \sum_{(k,l) \in \mathcal E_\mathrm{s}(k)} g_{kl}\theta_{kl}\omega_k$ and $\theta_{kl}\omega_k = \sqrt{\omega_k\omega_l}$.
  The second holds because each small edge appears at the two vertices it joins.
  The last inequality follows from $g_{kl} = |(\rho_n)_{kl}||R_{kl}|$ and the hypothesis $(n-1)|(\rho_n)_{kl}|\sqrt{\omega_k\omega_l} \le \gamma$.
  Substituting into \eqref{eq:after-forest-likelihood} and using $\log\det(I_p-R) < 0$ for $R \ne 0$, which holds by \Cref{lem:edge-removal} with $\mathcal E' = \mathcal E$, gives \eqref{eq:scale-bound}.
\end{proof}

\begin{lemma}\label{lem:eta-elimination}
  Fix $1 \le \bar m \le p-1$ and $1 \le \bar d \le \bar m$, and suppose $\gamma \ge \gamma_\eta(\bar m)$.
  Suppose moreover that
  \begin{equation}\label{eq:eta-free}
    \mathcal L_\mathrm{s}^\star(R) - \mathcal L_\mathrm{s}^\star(0) < \sum_{(i,j) \in \mathcal E} \min\Big\{\gamma |R_{ij}|,\ \log\big(1+w_{ij}^{(n-1)/2}\big)\Big\}
  \end{equation}
  for every forest $\mathcal E$ with at most $\bar m$ edges and maximum degree at most $\bar d$ and every nonzero $R \in \mathcal R_p$ supported on $\mathcal E$.
  Then $\mathcal L(\Lambda, R, \eta) < \mathcal L(\widehat\Lambda, 0, 0)$ for every positive diagonal $\Lambda$, every $\eta \in [0,1]$ and every such $R$.
\end{lemma}

\begin{proof}
  Fix such an $\mathcal E$ and $R$, a positive diagonal $\Lambda$ and $\eta \in [0,1]$, and write $K_{ij} := w_{ij}^{(n-1)/2}$ and $K_{(k)} := w_{(k)}^{(n-1)/2}$.
  Set
  \begin{equation*}
    a := \frac{\eta}{\eta + (1-\eta)\tilde\gamma} \in [0,1],
    \qquad \text{so that} \qquad
    \eta + (1-\eta)\tilde\gamma = \frac{\tilde\gamma}{1 + (\tilde\gamma-1)a} .
  \end{equation*}
  Factoring $\eta + (1-\eta)\tilde\gamma$ out of the logarithm in each $\mathrm{pen}(R_{ij}, \eta)$, and using $R_{ij} = 0$ off $\mathcal E$,
  \begin{equation*}
    \mathcal L(\Lambda, R, \eta)
    \le \mathcal L_\mathrm{s}^\star(R) + \ell\log\tilde\gamma - \ell \log\big[1+(\tilde\gamma-1)a\big] + \sum_{(i,j) \in \mathcal E} \log\big[a + (1-a)e^{-\gamma|R_{ij}|}\big] .
  \end{equation*}
  Since $\mathcal L(\widehat\Lambda, 0, 0) = \mathcal L_\mathrm{s}^\star(0) + \ell\log\tilde\gamma$, it suffices to prove
  \begin{equation}\label{eq:eta-target}
    \mathcal L_\mathrm{s}^\star(R) - \mathcal L_\mathrm{s}^\star(0) + \sum_{(i,j) \in \mathcal E} \log\big[a + (1-a)e^{-\gamma|R_{ij}|}\big] < \ell \log\big[1+(\tilde\gamma-1)a\big] .
  \end{equation}

  We first claim that every $x \ge 0$ and $y \ge 1$ satisfy
  \begin{equation*}
    \min\big\{x, \log(1+y)\big\} + \log\big[a + (1-a)e^{-x}\big] \le \log(1+ay) .
  \end{equation*}
  If $x \le \log(1+y)$ the left-hand side equals $\log[1 + a(e^x-1)]$, and $e^x - 1 \le y$.
  If $x > \log(1+y)$ then $e^{-x} < (1+y)^{-1}$, so the left-hand side is at most $\log[(1+y)a + 1 - a]$, which is the right-hand side.

  Applying the claim with $x = \gamma|R_{ij}|$ and $y = K_{ij}$, and then \eqref{eq:eta-free},
  \begin{equation*}
    \mathcal L_\mathrm{s}^\star(R) - \mathcal L_\mathrm{s}^\star(0) + \sum_{(i,j) \in \mathcal E} \log\big[a + (1-a)e^{-\gamma|R_{ij}|}\big]
    < \sum_{(i,j) \in \mathcal E} \log\big(1 + aK_{ij}\big) .
  \end{equation*}
  Order the edges of $\mathcal E$ by decreasing $|\rho_n|$, and let $(i,j)$ be the $k$-th of them.
  The first $k$ edges are $k$ distinct pairs, all with a sample correlation at least $|(\rho_n)_{ij}|$ in absolute value.
  Hence at least $k$ of the $\ell$ sample correlations are at least $|(\rho_n)_{ij}|$ in absolute value, so $|(\rho_n)_{ij}| \le |\rho_n|_{(k)}$ and $K_{ij} \le K_{(k)}$.
  Write $m := |\mathcal E| \le \bar m$.
  Since $t \mapsto \log(1+at)$ is nonnegative, nondecreasing and concave on $[0,\infty)$,
  \begin{align*}
    \sum_{(i,j) \in \mathcal E} \log\big(1 + aK_{ij}\big)
    &\le \sum_{k=1}^{m} \log\big(1 + aK_{(k)}\big)
    \le \sum_{k=1}^{\bar m} \log\big(1 + aK_{(k)}\big) \\
    &\le \bar m \log\Big(1 + \frac{a}{\bar m}\sum_{k=1}^{\bar m} K_{(k)}\Big) .
  \end{align*}
  Now $\sum_{k \le \bar m} K_{(k)} = \ell\,[\gamma_\eta(\bar m)-1]$ by \eqref{eq:gamma2}, and $\tilde\gamma = \gamma/(1-e^{-\gamma}) > \gamma \ge \gamma_\eta(\bar m)$.
  Moreover $\ell/\bar m \ge 1$ and $\log(1+xz) \le x\log(1+z)$ for every $x \ge 1$ and $z \ge 0$.
  Therefore
  \begin{equation*}
    \bar m \log\Big(1 + \frac{a}{\bar m}\sum_{k=1}^{\bar m} K_{(k)}\Big)
    \le \bar m \log\Big[1 + \frac{\ell}{\bar m}\, a(\tilde\gamma-1)\Big]
    \le \ell \log\big[1 + a(\tilde\gamma-1)\big] ,
  \end{equation*}
  which, combined with the two previous displays, is \eqref{eq:eta-target}.
\end{proof}


\subsection{Proof of \Cref{thm:forest}}\label{app:proof-forest}

\begin{proof}[Proof of \Cref{thm:forest}]
  The proof has three steps.
  Step 1 shows that $\gamma > \gamma_d$, so that the diagonal estimate is a strict local maximum by \Cref{lem:diag-local-max}.
  Step 2 reduces the comparison with the forest models to a condition on $\gamma$ alone.
  It uses \Cref{lem:eta-elimination} to remove $\eta$ and $\Lambda$ from the comparison, \Cref{lem:sign-gauge} to fix the signs of $R$, and \Cref{lem:scale-bound} to bound the smooth part.
  The condition on $\gamma$ is the hypothesis of \Cref{lem:scale-bound}, and Step 3 shows that it follows from $\gamma \ge \gamma_R(\bar m,\bar d)$.

  \emph{Step 1: the diagonal estimate is a strict local maximum.}
  The tuple $c$ with $k_0 = 1$, all four sets of ranks empty and $\tau_1 = \tau_2 = 0$ belongs to $\mathcal C(\bar m,\bar d)$ for every $\bar m \ge 1$ and $\bar d \ge 1$.
  It satisfies $\omega_1(c) = \omega_2(c) = \nu\big(\tfrac12 \log w_{(1)}\big)$.
  Hence
  \begin{equation*}
    \gamma_R(\bar m,\bar d) \ge (n-1)\,|\rho_n|_{(1)}\, \nu\big(\tfrac12 \log w_{(1)}\big) \ge (n-1)\,|\rho_n|_{(1)} = \gamma_d ,
  \end{equation*}
  where the second inequality uses $\nu(x) \ge 1$ for $x \ge 0$.
  Since $1 - e^{-x} < x$ for $x > 0$, that inequality is strict when $|\rho_n|_{(1)} > 0$.
  If $|\rho_n|_{(1)} = 0$, then $\gamma \ge \gamma_\eta(\bar m) > 1 > 0 = \gamma_d$.
  In both cases $\gamma > \gamma_d$, and \Cref{lem:diag-local-max} shows that \eqref{eq:diagonal-estimate} is a strict local maximum of \eqref{eq:map-objective}, which is Condition~(\ref{cond:diag-local}).

  \emph{Step 2: the comparison with the forest models reduces to a condition on $\gamma$.}
  It remains to compare the diagonal model with the forest models.
  Since $\gamma \ge \gamma_\eta(\bar m)$, by \Cref{lem:eta-elimination} it suffices to prove \eqref{eq:eta-free}.
  Fix a forest $\mathcal E$ with at most $\bar m$ edges and maximum degree at most $\bar d$, and a nonzero $R \in \mathcal R_p$ supported on $\mathcal E$.

  We first reduce to the case $(\rho_n)_{ij} R_{ij} \ge 0$ on $\mathcal E$.
  Let $D$ be the matrix of \Cref{lem:sign-gauge} and put $R' := DRD$.
  By \Cref{lem:sign-gauge}, $R' \in \mathcal R_p$ is supported on $\mathcal E$, and it is nonzero because $|R'_{ij}| = |R_{ij}|$ for every pair.
  Maximising \eqref{eq:sign-gauge} over $\Lambda$ gives $\mathcal L_\mathrm{s}^\star(R') \ge \mathcal L_\mathrm{s}^\star(R)$.
  For the same reason, the right-hand side of \eqref{eq:eta-free} takes the same value at $R'$ and at $R$.
  It therefore suffices to prove \eqref{eq:eta-free} for $R'$, and from now on we assume $(\rho_n)_{ij} R_{ij} \ge 0$ on $\mathcal E$.

  Split the edges according to
  \begin{align*}
    \mathcal E_\mathrm{s} &:= \Big\{ (i,j) \in \mathcal E : \gamma |R_{ij}| < \tfrac{n-1}{2} \log w_{ij} \Big\}, \\
    \mathcal E_\mathrm{b} &:= \Big\{ (i,j) \in \mathcal E : \gamma |R_{ij}| \ge \tfrac{n-1}{2} \log w_{ij} \Big\} .
  \end{align*}
  Every pair satisfies
  \begin{equation*}
    \tfrac{n-1}{2} \log w_{ij} = \log w_{ij}^{(n-1)/2} < \log\big(1+w_{ij}^{(n-1)/2}\big) .
  \end{equation*}
  On $\mathcal E_\mathrm{s}$ the minimum in \eqref{eq:eta-free} therefore equals $\gamma |R_{ij}|$.
  On $\mathcal E_\mathrm{b}$ we have $\tfrac{n-1}{2} \log w_{ij} \le \gamma |R_{ij}|$, so the minimum is at least $\tfrac{n-1}{2} \log w_{ij}$.
  Therefore
  \begin{equation}\label{eq:split-bound}
    \frac{n-1}{2} \sum_{(i,j) \in \mathcal E_\mathrm{b}} \log w_{ij} + \gamma \sum_{(i,j) \in \mathcal E_\mathrm{s}} |R_{ij}|
    \le \sum_{(i,j) \in \mathcal E} \min\Big\{\gamma |R_{ij}|,\ \log\big(1+w_{ij}^{(n-1)/2}\big)\Big\} .
  \end{equation}
  The matrix $R$ is nonzero and satisfies the sign and size conditions of \Cref{lem:scale-bound} for this partition.
  The only hypothesis of that lemma left to check is the one on $\gamma$, namely
  \begin{equation}\label{eq:gamma-condition}
    \gamma \ge (n-1)\, |(\rho_n)_{ij}| \sqrt{\omega_i \omega_j}
    \qquad \text{for every } (i,j) \in \mathcal E_\mathrm{s} ,
  \end{equation}
  with $\omega_i$ and $\omega_j$ as in \eqref{eq:omega-vertex}.
  Once \eqref{eq:gamma-condition} holds, the lemma gives \eqref{eq:scale-bound}, and combining it with \eqref{eq:split-bound} gives \eqref{eq:eta-free}.
  The rest of the proof establishes \eqref{eq:gamma-condition}.

  \emph{Step 3: the condition on $\gamma$ follows from $\gamma \ge \gamma_R(\bar m,\bar d)$.}
  Since $\gamma \ge \gamma_R(\bar m,\bar d)$, condition \eqref{eq:gamma-condition} follows once we show that
  \begin{equation}\label{eq:domination}
    (n-1)\, |(\rho_n)_{ij}| \sqrt{\omega_i \omega_j} \le \gamma_R(\bar m,\bar d)
    \qquad \text{for every } (i,j) \in \mathcal E_\mathrm{s} .
  \end{equation}
  If $\mathcal E_\mathrm{s}$ is empty there is nothing to prove, so fix a small edge.
  Relabelling the vertices, we may assume that this edge is $(1,2)$.
  Let $\mathcal E_{12}$ be the set of edges of $\mathcal E$ incident to $1$ or to $2$, of which only $(1,2)$ is incident to both.
  Order the edges of $\mathcal E_{12}$ by decreasing $|\rho_n|$, breaking ties arbitrarily, and give the rank $k$ to the $k$-th edge of that order.
  Let $k_0$ be the rank of $(1,2)$, let $\mathcal K^\mathrm{b}_1$ and $\mathcal K^\mathrm{s}_1$ be the sets of ranks of the edges of $\mathcal E_\mathrm{b}$ and of the edges of $\mathcal E_\mathrm{s}$ other than $(1,2)$ incident to $1$, define $\mathcal K^\mathrm{b}_2$ and $\mathcal K^\mathrm{s}_2$ likewise, and let $\tau_1$ and $\tau_2$ be given by \eqref{eq:tau-vertex}.
  We claim that $c := (k_0, \mathcal K^\mathrm{b}_1, \mathcal K^\mathrm{s}_1, \tau_1, \mathcal K^\mathrm{b}_2, \mathcal K^\mathrm{s}_2, \tau_2)$ belongs to $\mathcal C(\bar m,\bar d)$.

  By construction $\{k_0\}$ and the four sets of ranks partition $\{1, \dots, |\mathcal E_{12}|\}$, so $|c| = |\mathcal E_{12}|$.
  If $\tau_1 = 1$ then some edge of $\mathcal E_\mathrm{b}$ is incident to $1$, so $\tau_1 \le |\mathcal K^\mathrm{b}_1|$, and likewise $\tau_2 \le |\mathcal K^\mathrm{b}_2|$.
  The degrees of $1$ and $2$ in $\mathcal E$ are $1 + |\mathcal K^\mathrm{b}_1| + |\mathcal K^\mathrm{s}_1|$ and $1 + |\mathcal K^\mathrm{b}_2| + |\mathcal K^\mathrm{s}_2|$, so both are at most $\bar d$.
  For the constraint of \eqref{eq:configuration-constraints} involving $\bar m$ we exhibit $|c| + \tau_1 + \tau_2$ distinct edges of $\mathcal E$.
  Suppose $\tau_1 = 1$, so that there are $(1,u) \in \mathcal E_\mathrm{b}$ and $(u,v) \in \mathcal E_\mathrm{s}$.
  Then $u \ne 2$ because $(1,u) \in \mathcal E_\mathrm{b}$ and $(1,2) \in \mathcal E_\mathrm{s}$, and $v \ne 1$ because $(u,v) \in \mathcal E_\mathrm{s}$ and $(1,u) \in \mathcal E_\mathrm{b}$.
  Moreover $v \ne 2$, since otherwise $1$, $u$, $2$ would form a cycle in $\mathcal E$.
  Hence $(u,v)$ is incident to neither $1$ nor $2$, and $(u,v) \notin \mathcal E_{12}$.
  If $\tau_2 = 1$, the same argument gives $(2,u') \in \mathcal E_\mathrm{b}$ and $(u',v') \in \mathcal E_\mathrm{s}$ with $(u',v') \notin \mathcal E_{12}$.
  If $\tau_1 = \tau_2 = 1$, the edges $(u,v)$ and $(u',v')$ are distinct.
  Indeed, $u = u'$ would make $1$, $u$, $2$ a cycle, and $u = v'$ together with $v = u'$ would make $1$, $u$, $u'$, $2$ a cycle.
  Therefore $\mathcal E$ contains $|c| + \tau_1 + \tau_2$ distinct edges, and $|c| + \tau_1 + \tau_2 \le \bar m$.
  This proves the claim.

  These ranks order the edges of $\mathcal E_{12}$ among themselves, whereas $|\rho_n|_{(k)}$ is the $k$-th largest of all $\ell$ sample correlations.
  The edges of ranks $1, \dots, k$ are $k$ distinct pairs whose sample correlations are no smaller in absolute value than that of the edge of rank $k$, so at least $k$ of the $\ell$ sample correlations are no smaller in absolute value than it.
  Hence the sample correlation of the edge of rank $k$ is at most $|\rho_n|_{(k)}$ in absolute value, and its value of $w$ is at most $w_{(k)}$.
  In particular $|(\rho_n)_{12}| \le |\rho_n|_{(k_0)}$.

  The function $\nu$ is nondecreasing on $[0,\infty)$ and satisfies $\nu \ge 1$.
  Since $w_{1l} \ge 1$ for every $l$, the two factors of $\omega_1$ in \eqref{eq:omega-vertex} are positive and nondecreasing in each $w_{1l}$.
  The first factor runs over the edges with ranks in $\mathcal K^\mathrm{b}_1$, and the second over $(1,2)$ and the edges with ranks in $\mathcal K^\mathrm{s}_1$.
  Replacing each $w_{1l}$ in \eqref{eq:omega-vertex} by the value $w_{(k)}$ attached to the rank $k$ of that edge therefore gives $\omega_1 \le \omega_1(c)$, and likewise $\omega_2 \le \omega_2(c)$.
  Consequently
  \begin{equation*}
    (n-1)\, |(\rho_n)_{12}| \sqrt{\omega_1 \omega_2}
    \le (n-1)\, |\rho_n|_{(k_0)} \sqrt{\omega_1(c)\, \omega_2(c)}
    \le \gamma_R(\bar m,\bar d) .
  \end{equation*}
  Since the small edge was arbitrary, this proves \eqref{eq:domination}.
  Hence \eqref{eq:gamma-condition} holds for every forest $\mathcal E$ and every nonzero $R$ fixed in Step 2.
  This completes the proof of \eqref{eq:eta-free}, and \Cref{lem:eta-elimination} gives Condition~(\ref{cond:diag-forest}).
\end{proof}

\section{Proof of \Cref{thm:calibration-forest}}\label{app:calibration}

The proof rests on a description of the joint law of the $\bar m$ largest absolute sample correlations under a diagonal ground truth.
\citet{jiang2004} obtained the Gumbel limit of $|\rho_n|_{(1)}$, and \citet{heiny2021point} obtained point process convergence for the off-diagonal entries of sample covariance and correlation matrices, from which the joint behaviour of the extremes of the signed entries follows.
The lower bound $\underline\gamma(\bar m,\bar d)$ depends instead on the largest entries in absolute value, through a function that changes with $p$.
We therefore prove a bound in total variation for the absolute entries, uniform in $n$, using the Chen--Stein method in the form of Theorem 2 of \citet{arratia1989}.

\begin{lemma}\label{lem:joint-order-stats}
  Fix $\bar m \ge 1$.
  For $n \ge 3$, let $\boldsymbol x_1,\dots,\boldsymbol x_n$ be independent $N_p(\boldsymbol\mu,\Sigma)$ vectors with $\Sigma$ diagonal and positive definite, and let $\bar F_n$ and $\Gamma_1, \Gamma_2, \dots$ be as in \Cref{subsec:gamma_q}.
  For $\ell \ge \bar m$ set
  \begin{equation*}
    V_k := \ell\,\bar F_n\big(|\rho_n|_{(k)}\big),
    \qquad
    \boldsymbol V := (V_1, \dots, V_{\bar m}),
    \qquad
    \boldsymbol\Gamma := (\Gamma_1, \dots, \Gamma_{\bar m}).
  \end{equation*}
  Then
  \begin{equation}\label{eq:tv-convergence}
    \sup_{n \ge 3}\ \sup_{\mathcal A}\big|\mathrm P(\boldsymbol V \in \mathcal A) - \mathrm P(\boldsymbol\Gamma \in \mathcal A)\big| \longrightarrow 0,
    \qquad p\to\infty,
  \end{equation}
  where the inner supremum runs over the Borel sets $\mathcal A \subseteq \mathbb R^{\bar m}$.
\end{lemma}

\begin{proof}
  Since $\Sigma$ is diagonal, each $(\rho_n)_{ij}$ is invariant under the map $\boldsymbol x \mapsto \Sigma^{-1/2}(\boldsymbol x - \boldsymbol\mu)$ applied to every observation, so we may assume that $\boldsymbol\mu = \boldsymbol 0$ and $\Sigma = I_p$.
  Let $\boldsymbol y_i \in \mathbb R^n$ be the $i$-th column of the data matrix $X$, which collects the $n$ observations of variable $i$, so that $\boldsymbol y_1,\dots,\boldsymbol y_p$ are independent $N_n(\boldsymbol 0,I_n)$ vectors.
  Define the centered vectors
  \begin{equation*}
    \boldsymbol y_i^c = \left(I_n - \frac{1}{n} \boldsymbol{1}\boldsymbol{1}^\top\right) \boldsymbol y_i, \qquad i = 1, \dots, p,
  \end{equation*}
  which live in the subspace $\boldsymbol 1^\bot \subset \mathbb R^n$ of vectors orthogonal to $\boldsymbol 1 = (1, \dots, 1)^\top$.
  Writing each $\boldsymbol y_i^c$ in the coordinates of an orthonormal basis of $\boldsymbol 1^\bot$ produces independent $N_{n-1}(\boldsymbol 0, I_{n-1})$ vectors $\{\boldsymbol z_1, \dots, \boldsymbol z_p\} \subset \mathbb R^{n-1}$.
  The sample correlation can then be written as
  \begin{equation} \label{eq:rho_as_inner_prod}
    (\rho_n)_{ij} = \boldsymbol u_i^\top \boldsymbol u_j, \qquad \boldsymbol u_i := \boldsymbol z_i/\|\boldsymbol z_i\|.
  \end{equation}
  The vectors $\boldsymbol z_1, \dots, \boldsymbol z_p$ are independent and almost surely nonzero, so each $\boldsymbol u_i$ is well defined and $\boldsymbol u_1, \dots, \boldsymbol u_p$ are independent.
  Each $\boldsymbol u_i$ is uniform on the unit sphere $S^{n-2} \subset \mathbb R^{n-1}$, since the distribution of $\boldsymbol z_i$ is invariant under rotations.
  Moreover, for $n \ge 3$ the event $\{\boldsymbol u_i = \pm \boldsymbol u_j\}$ has probability zero, so $|(\rho_n)_{ij}| < 1$ almost surely.

  For $1 \le i < j \le p$ set $f_{ij} := \bar F_n(|(\rho_n)_{ij}|)$, and $f_{ji} := f_{ij}$.
  Write $f_{(1)} \le \dots \le f_{(\ell)}$ for the ordered values of the $f_{ij}$, $i<j$.
  Since $\bar F_n$ is strictly decreasing, $f_{(k)} = \bar F_n(|\rho_n|_{(k)})$, so that $V_k = \ell f_{(k)}$.
  The proof uses three properties of the $f_{ij}$.

  First, each $f_{ij}$ is uniformly distributed on $(0,1]$.
  Indeed, $\bar F_n$ is a continuous strictly decreasing bijection from $[0,1)$ onto $(0,1]$, and $\bar F_n(r) = \mathrm P(|(\rho_n)_{ij}| > r)$, so this is the probability integral transform.
  Hence
  \begin{equation}\label{eq:uniform-cells}
    \mathrm P\big(\ell f_{ij} \in (t, t']\big) = \frac{t' - t}{\ell}, \qquad 0 \le t < t' \le \ell .
  \end{equation}

  Second, for every $i$ and every two distinct indices $j$ and $j'$ different from $i$, $f_{ij}$ and $f_{ij'}$ are independent.
  Conditional on $\boldsymbol u_i = \boldsymbol v$, the variables $(\rho_n)_{ij} = \boldsymbol u_j^\top \boldsymbol v$ and $(\rho_n)_{ij'} = \boldsymbol u_{j'}^\top \boldsymbol v$ from \eqref{eq:rho_as_inner_prod} are functions of $\boldsymbol u_j$ and $\boldsymbol u_{j'}$, which are independent of each other and of $\boldsymbol u_i$, so they are conditionally independent.
  Moreover, since the distributions of $\boldsymbol u_j$ and $\boldsymbol u_{j'}$ are invariant under rotations,
  \begin{equation*}
    (\boldsymbol u_j^\top \boldsymbol v, \boldsymbol u_{j'}^\top \boldsymbol v) \overset{d}= (\boldsymbol u_j^\top \boldsymbol e_1, \boldsymbol u_{j'}^\top \boldsymbol e_1),
  \end{equation*}
  so the conditional distributions do not depend on $\boldsymbol v$, and $(\rho_n)_{ij}$ and $(\rho_n)_{ij'}$ are independent.

  Third, $f_{ij}$ is a function of $\boldsymbol y_i$ and $\boldsymbol y_j$, so it is independent of the family of the $f_{kl}$ with $\{k,l\} \cap \{i,j\} = \emptyset$, which is a function of the other columns of the data matrix $X$.

  The proof has three steps.
  Step 1 bounds the difference between the laws of the numbers of values $\ell f_{ij}$ in finitely many intervals and of the corresponding counts of a Poisson process.
  Step 2 extends this bound to every event about $V_1, \dots, V_{\bar m}$ truncated at a fixed level.
  Step 3 removes the truncation and chooses the level.

  \emph{Step 1: the counts in finitely many intervals are close to Poisson counts.}
  Fix $H \ge 1$ and $0 = t_0 < t_1 < \dots < t_H \le \ell$, and for $1 \le i < j \le p$ and $s \le H$ set
  \begin{equation*}
    \mathds 1_{ijs} := \mathds 1_{(t_{s-1}, t_s]}(\ell f_{ij}),
    \qquad
    N_s := \sum_{i<j} \mathds 1_{ijs} ,
  \end{equation*}
  so that $N_s$ is the number of pairs with $\ell f_{ij} \in (t_{s-1}, t_s]$.
  The $\Gamma_k$ are the ordered points of a Poisson process $\Pi$ of unit rate on $(0,\infty)$, whose counts $\Pi(t_{s-1}, t_s]$, $s \le H$, are independent Poisson variables with means $t_s - t_{s-1}$.
  We show that, for every $\mathcal C \subseteq \mathbb Z_{\ge 0}^H$,
  \begin{equation}\label{eq:count-tv}
    \Big|\mathrm P\big((N_s)_{s \le H} \in \mathcal C\big) - \mathrm P\big((\Pi(t_{s-1},t_s])_{s \le H} \in \mathcal C\big)\Big| \le \frac{16\, t_H^2}{p-1} .
  \end{equation}

  We use Theorem 2 of \citet{arratia1989}.
  It applies to a finite family of indicators $(\mathds 1_\alpha)_{\alpha \in \mathcal I}$, in which each $\alpha$ is assigned a set $\mathcal B_\alpha \subseteq \mathcal I$ that contains it.
  Set
  \begin{gather*}
    b_1 := \sum_{\alpha \in \mathcal I} \sum_{\beta \in \mathcal B_\alpha} \mathbb E[\mathds 1_\alpha]\, \mathbb E[\mathds 1_\beta],
    \qquad
    b_2 := \sum_{\alpha \in \mathcal I} \sum_{\beta \in \mathcal B_\alpha \setminus \{\alpha\}} \mathbb E[\mathds 1_\alpha \mathds 1_\beta],
    \\
    b_3 := \sum_{\alpha \in \mathcal I} \mathbb E\Big[\Big|\mathbb E\big[\mathds 1_\alpha - \mathbb E[\mathds 1_\alpha] \,\big|\, \mathds 1_\beta,\ \beta \notin \mathcal B_\alpha\big]\Big|\Big] .
  \end{gather*}
  The theorem states that, for every event about the family, its probability and that of the same event about independent Poisson variables with means $\mathbb E[\mathds 1_\alpha]$ differ by at most $2b_1 + 2b_2 + b_3$.
  We apply it with $\mathcal I := \{(i,j,s) : 1 \le i < j \le p,\ s \le H\}$, and with $\mathcal B_{ijs}$ the set of $(k,l,s') \in \mathcal I$ such that $\{k,l\} \cap \{i,j\} \ne \emptyset$.

  Each $(i,j)$ pair shares an index with $2p-3$ pairs, itself included, and by \eqref{eq:uniform-cells} the means of $\mathds 1_{ij1}, \dots, \mathds 1_{ijH}$ add up to $t_H/\ell$, so
  \begin{equation*}
    b_1 = \ell\,(2p-3)\,\frac{t_H^2}{\ell^2} = \frac{(2p-3)\, t_H^2}{\ell} .
  \end{equation*}
  In $b_2$, the terms with $(k,l) = (i,j)$ and $s' \ne s$ vanish, because the intervals $(t_{s-1}, t_s]$ are disjoint.
  The remaining terms involve two distinct pairs that share an index, for which the second property of the $f_{ij}$ gives $\mathbb E[\mathds 1_{ijs}\mathds 1_{kls'}] = \mathbb E[\mathds 1_{ijs}]\,\mathbb E[\mathds 1_{kls'}]$, so
  \begin{equation*}
    b_2 = \ell\,(2p-4)\,\frac{t_H^2}{\ell^2} = \frac{(2p-4)\, t_H^2}{\ell} .
  \end{equation*}
  The indicator $\mathds 1_{ijs}$ is a function of $f_{ij}$, and the indicators outside $\mathcal B_{ijs}$ are functions of the $f_{kl}$ with $\{k,l\} \cap \{i,j\} = \emptyset$, so $b_3 = 0$ by the third property of the $f_{ij}$.
  Since $\ell = p(p-1)/2$,
  \begin{equation*}
    2b_1 + 2b_2 + b_3 = \frac{(8p-14)\, t_H^2}{\ell} = \frac{2(8p-14)\, t_H^2}{p(p-1)} \le \frac{16\, t_H^2}{p-1} .
  \end{equation*}
  Each $N_s$ is the sum of the $\mathds 1_{ijs}$ over the pairs, so every event about $(N_s)_{s \le H}$ is an event about the indicators.
  The corresponding sums of the independent Poisson variables are independent Poisson variables with means $t_s - t_{s-1}$, which have the joint law of the counts $\Pi(t_{s-1}, t_s]$, and \eqref{eq:count-tv} follows.

  \emph{Step 2: the bound extends to every event about the truncated order statistics.}
  The counts of Step 1 describe only the values $\ell f_{ij}$ that lie in $(0, t_H]$, and the bound \eqref{eq:count-tv} grows with $t_H$, so $t_H$ must stay finite.
  An order statistic $V_k$ larger than $t_H$ is therefore not determined by the counts, and we first truncate the order statistics at $t_H$.
  Fix $t_H \in (0,\ell]$, set $V'_k := \min(V_k, t_H)$ and $\Gamma'_k := \min(\Gamma_k, t_H)$ for $k \le \bar m$, and write $\boldsymbol V'$ and $\boldsymbol\Gamma'$ for the corresponding vectors.
  We show that, for every Borel $\mathcal A \subseteq \mathbb R^{\bar m}$,
  \begin{equation}\label{eq:trunc-tv}
    \big|\mathrm P(\boldsymbol V' \in \mathcal A) - \mathrm P(\boldsymbol\Gamma' \in \mathcal A)\big| \le \frac{16\, t_H^2}{p-1} .
  \end{equation}

  We first prove \eqref{eq:trunc-tv} when $\mathcal A$ is a finite union of rectangles, that is, of sets of the form
  \begin{equation*}
    \mathcal T = (\zeta_1, \zeta'_1] \times \dots \times (\zeta_{\bar m}, \zeta'_{\bar m}],
    \qquad -\infty \le \zeta_k \le \zeta'_k \le \infty,
  \end{equation*}
  where an interval $(\zeta_k, \infty]$ stands for $(\zeta_k, \infty)$.
  Let $\mathcal F$ be a finite set of rectangles and $\mathcal A := \bigcup_{\mathcal T \in \mathcal F} \mathcal T$.
  Given $\mathcal A$, let $t_1 < \dots < t_{H-1}$ be the endpoints $\zeta_k$ and $\zeta'_k$ of the rectangles of $\mathcal F$ that lie in $(0, t_H)$, so that $H$ and these points depend on $\mathcal A$, while $t_H$ stays fixed.
  Every interval $(\zeta_k, \zeta'_k]$ of these rectangles satisfies
  \begin{equation*}
    (\zeta_k, \zeta'_k] \cap (0, t_H] = \big(\max\{\zeta_k, 0\},\ \min\{\zeta'_k, t_H\}\big] .
  \end{equation*}
  Unless this intersection is empty, $\max\{\zeta_k, 0\}$ is either $0 = t_0$ or an endpoint in $(0,t_H)$, and $\min\{\zeta'_k, t_H\}$ is either $t_H$ or an endpoint in $(0,t_H)$, so both are among $t_0, t_1, \dots, t_H$.
  Hence the intersection is a union of some of the intervals $(t_{h-1}, t_h]$ with $h \in \{1, \dots, H\}$.
  Since $V'_k$ and $\Gamma'_k$ lie in $(0, t_H]$, whether $\boldsymbol V' \in \mathcal A$ therefore depends only on which of these intervals contain $V'_1, \dots, V'_{\bar m}$, and likewise for $\boldsymbol\Gamma'$.

  These intervals are determined by the counts of Step 1.
  Indeed, $V_k$ is the $k$-th smallest of the values $\ell f_{ij}$, and $(0, t_h]$ is the union of the intervals $(t_{g-1}, t_g]$ with $g \le h$, so for $h < H$
  \begin{equation*}
    V'_k \le t_h \iff N_1 + \dots + N_h \ge k,
    \qquad
    \Gamma'_k \le t_h \iff \Pi(0, t_h] \ge k,
  \end{equation*}
  where the truncation plays no role because $t_h < t_H$.
  Since moreover $0 < V'_k \le t_H$, the interval that contains $V'_k$ is $(t_{h-1}, t_h]$ for the smallest $h < H$ with $N_1 + \dots + N_h \ge k$, and $(t_{H-1}, t_H]$ if there is no such $h$, and likewise for $\Gamma'_k$ with the counts of $\Pi$.
  Hence there is a set $\mathcal C \subseteq \mathbb Z_{\ge 0}^H$ such that
  \begin{equation*}
    \{\boldsymbol V' \in \mathcal A\} = \big\{(N_s)_{s\le H} \in \mathcal C\big\},
    \qquad
    \{\boldsymbol \Gamma' \in \mathcal A\} = \big\{(\Pi(t_{s-1},t_s])_{s\le H} \in \mathcal C\big\} .
  \end{equation*}
  The counts $N_s$ depend on $\mathcal A$ through $t_1, \dots, t_{H-1}$, but \eqref{eq:count-tv} holds for every choice of these points, with a bound that depends only on $t_H$.
  It therefore gives \eqref{eq:trunc-tv} for $\mathcal A$.

  Now let $\mathcal A$ be any Borel set, and for every Borel set $\mathcal B \subseteq \mathbb R^{\bar m}$ set
  \begin{equation*}
    \mathrm Q(\mathcal B) := \mathrm P(\boldsymbol V' \in \mathcal B) + \mathrm P(\boldsymbol\Gamma' \in \mathcal B),
  \end{equation*}
  which defines a measure on the Borel sets of $\mathbb R^{\bar m}$ with $\mathrm Q(\mathbb R^{\bar m}) = 2$.
  We approximate $\mathcal A$ by a finite union of rectangles using Theorem 11.4(ii) of \citet{billingsley1995}, whose hypotheses hold for the rectangles and $\mathrm Q$.
  The rectangles form a semiring, that is, the empty set is a rectangle, the intersection of two rectangles is a rectangle, and the difference of two rectangles, one contained in the other, is a finite union of disjoint rectangles.
  They generate the Borel sets of $\mathbb R^{\bar m}$.
  Finally, $\mathrm Q$ is $\sigma$-finite on the rectangles, because $\mathbb R^{\bar m}$ is itself a rectangle and $\mathrm Q(\mathbb R^{\bar m})$ is finite.
  For every $\delta > 0$, the theorem therefore gives a finite union of rectangles $\mathcal A_\delta$ such that
  \begin{equation*}
    \mathrm Q(\mathcal D_\delta) \le \delta,
    \qquad
    \mathcal D_\delta := (\mathcal A \setminus \mathcal A_\delta) \cup (\mathcal A_\delta \setminus \mathcal A),
  \end{equation*}
  where $\mathcal D_\delta$ is the set of points that lie in exactly one of $\mathcal A$ and $\mathcal A_\delta$.
  Since $\mathrm P(\boldsymbol V' \in \mathcal A) - \mathrm P(\boldsymbol V' \in \mathcal A_\delta) = \mathrm P(\boldsymbol V' \in \mathcal A \setminus \mathcal A_\delta) - \mathrm P(\boldsymbol V' \in \mathcal A_\delta \setminus \mathcal A)$, and likewise for $\boldsymbol\Gamma'$,
  \begin{equation*}
    \big|\mathrm P(\boldsymbol V' \in \mathcal A) - \mathrm P(\boldsymbol V' \in \mathcal A_\delta)\big| \le \mathrm P(\boldsymbol V' \in \mathcal D_\delta),
    \qquad
    \big|\mathrm P(\boldsymbol\Gamma' \in \mathcal A) - \mathrm P(\boldsymbol\Gamma' \in \mathcal A_\delta)\big| \le \mathrm P(\boldsymbol\Gamma' \in \mathcal D_\delta) .
  \end{equation*}
  The two right-hand sides add up to $\mathrm Q(\mathcal D_\delta) \le \delta$, so the triangle inequality and \eqref{eq:trunc-tv} for $\mathcal A_\delta$ give
  \begin{equation*}
    \big|\mathrm P(\boldsymbol V' \in \mathcal A) - \mathrm P(\boldsymbol\Gamma' \in \mathcal A)\big|
    \le \big|\mathrm P(\boldsymbol V' \in \mathcal A_\delta) - \mathrm P(\boldsymbol\Gamma' \in \mathcal A_\delta)\big| + \delta
    \le \frac{16\, t_H^2}{p-1} + \delta .
  \end{equation*}
  Since $\delta$ is arbitrary, this proves \eqref{eq:trunc-tv}.

  \emph{Step 3: removing the truncation gives \eqref{eq:tv-convergence}.}
  Let $\mathcal A \subseteq \mathbb R^{\bar m}$ be Borel, and set $\mathcal H := \mathbb R^{\bar m - 1} \times (-\infty, t_H)$, so that $\boldsymbol V' \in \mathcal H$ if and only if $V_{\bar m} < t_H$.
  Since $V_1 \le \dots \le V_{\bar m}$, on the event $\{\boldsymbol V' \in \mathcal H\}$ every $V_k$ is smaller than $t_H$ and $\boldsymbol V' = \boldsymbol V$, and likewise $\boldsymbol\Gamma' = \boldsymbol\Gamma$ on $\{\boldsymbol\Gamma' \in \mathcal H\}$.
  Splitting each probability according to whether the truncated vector lies in $\mathcal H$ therefore gives
  \begin{gather*}
    \mathrm P(\boldsymbol V \in \mathcal A) = \mathrm P(\boldsymbol V' \in \mathcal A \cap \mathcal H) + \mathrm P(\boldsymbol V \in \mathcal A,\ \boldsymbol V' \notin \mathcal H),
    \\
    \mathrm P(\boldsymbol\Gamma \in \mathcal A) = \mathrm P(\boldsymbol\Gamma' \in \mathcal A \cap \mathcal H) + \mathrm P(\boldsymbol\Gamma \in \mathcal A,\ \boldsymbol\Gamma' \notin \mathcal H) .
  \end{gather*}
  The last terms lie in $[0, \mathrm P(\boldsymbol V' \notin \mathcal H)]$ and in $[0, \mathrm P(\boldsymbol\Gamma' \notin \mathcal H)]$, so their difference is at most $\max\{\mathrm P(\boldsymbol V' \notin \mathcal H), \mathrm P(\boldsymbol\Gamma' \notin \mathcal H)\}$ in absolute value, and
  \begin{equation*}
    \begin{aligned}
      \big|\mathrm P(\boldsymbol V \in \mathcal A) - \mathrm P(\boldsymbol\Gamma \in \mathcal A)\big|
      &\le \big|\mathrm P(\boldsymbol V' \in \mathcal A \cap \mathcal H) - \mathrm P(\boldsymbol\Gamma' \in \mathcal A \cap \mathcal H)\big| \\
      &\qquad + \max\big\{\mathrm P(\boldsymbol V' \notin \mathcal H), \mathrm P(\boldsymbol\Gamma' \notin \mathcal H)\big\} \\
      &\le \frac{32\, t_H^2}{p-1} + \mathrm P(\Gamma_{\bar m} \ge t_H) .
    \end{aligned}
  \end{equation*}
  The second inequality applies \eqref{eq:trunc-tv} twice.
  Applied to $\mathcal A \cap \mathcal H$, it bounds the first term by $16\, t_H^2/(p-1)$.
  Applied to the complement of $\mathcal H$, it gives $\mathrm P(\boldsymbol V' \notin \mathcal H) \le \mathrm P(\boldsymbol\Gamma' \notin \mathcal H) + 16\, t_H^2/(p-1)$, so the maximum is at most $\mathrm P(\boldsymbol\Gamma' \notin \mathcal H) + 16\, t_H^2/(p-1)$.
  Finally, $\{\boldsymbol\Gamma' \notin \mathcal H\} = \{\Gamma_{\bar m} \ge t_H\}$.
  This event is the event that $\Pi$ has fewer than $\bar m$ points in $(0, t_H)$, so
  \begin{equation*}
    \mathrm P(\Gamma_{\bar m} \ge t_H) = \sum_{k=0}^{\bar m - 1} e^{-t_H}\,\frac{t_H^k}{k!} .
  \end{equation*}
  Choosing $t_H = \log p$, which satisfies $t_H \le \ell$, we obtain
  \begin{equation*}
    \big|\mathrm P(\boldsymbol V \in \mathcal A) - \mathrm P(\boldsymbol\Gamma \in \mathcal A)\big|
    \le \frac{32(\log p)^2}{p-1} + \frac{1}{p}\sum_{k=0}^{\bar m - 1}\frac{(\log p)^k}{k!} .
  \end{equation*}
  The right-hand side does not depend on $n$ or on $\mathcal A$, and it tends to zero as $p \to \infty$, which proves \eqref{eq:tv-convergence}.
\end{proof}

\begin{proof}[Proof of \Cref{thm:calibration-forest}]
  Let $p$ be large enough that $\bar m \le p-1$, so that $\bar m \le \ell$, and fix $n \ge 3$.
  By \eqref{eq:configuration-constraints}, every rank of a configuration of $\mathcal C(\bar m,\bar d)$ is at most $\bar m$, so $\underline\gamma(\bar m,\bar d)$ depends on the sample only through $|\rho_n|_{(1)}, \dots, |\rho_n|_{(\bar m)}$.
  These lie in $[0,1)$ almost surely, and since $\bar F_n$ is a bijection from $[0,1)$ onto $(0,1]$, with the notation of \Cref{lem:joint-order-stats},
  \begin{equation*}
    |\rho_n|_{(k)} = \bar F_n^{-1}\big(\min\{V_k/\ell, 1\}\big), \qquad k \le \bar m .
  \end{equation*}
  Let $g : (0,\infty)^{\bar m} \to \mathbb R$ be the function obtained by substituting $\bar F_n^{-1}(\min\{v_k/\ell, 1\})$ for $|\rho_n|_{(k)}$ in $\underline\gamma(\bar m,\bar d)$.
  It is continuous, and by the display above and \eqref{eq:gamma-star-forest},
  \begin{equation*}
    \underline\gamma(\bar m,\bar d) = g(\boldsymbol V),
    \qquad
    \max\big(\gamma_R^\star(\bar m,\bar d), \gamma_\eta^\star(\bar m)\big) = g(\boldsymbol\Gamma) .
  \end{equation*}
  Set $\mathcal A_q := \{\boldsymbol v \in (0,\infty)^{\bar m} : g(\boldsymbol v) \le \gamma_q(\bar m,\bar d)\}$, which is a Borel set.
  Then
  \begin{align*}
    \mathrm P\big(\underline\gamma(\bar m,\bar d) \le \gamma_q(\bar m,\bar d)\big)
    &= \mathrm P(\boldsymbol V \in \mathcal A_q) \\
    &\ge \mathrm P(\boldsymbol\Gamma \in \mathcal A_q) - \sup_{\mathcal A}\big|\mathrm P(\boldsymbol V \in \mathcal A) - \mathrm P(\boldsymbol\Gamma \in \mathcal A)\big| \\
    &\ge 1 - q - \sup_{\mathcal A}\big|\mathrm P(\boldsymbol V \in \mathcal A) - \mathrm P(\boldsymbol\Gamma \in \mathcal A)\big| .
  \end{align*}
  The first inequality holds by the definition of the supremum.
  The second holds because the cumulative distribution function of $g(\boldsymbol\Gamma)$ is right-continuous, so the infimum in \eqref{eq:gamma-q-forest} is attained and $\mathrm P(\boldsymbol\Gamma \in \mathcal A_q) \ge 1-q$.
  Taking the infimum over $n \ge 3$ on the left and the supremum over $n \ge 3$ on the right, \Cref{lem:joint-order-stats} gives \eqref{eq:calibration-forest}.
  Finally, on the event $\{\underline\gamma(\bar m,\bar d) \le \gamma_q(\bar m,\bar d)\}$, \Cref{thm:forest} applies with $\gamma = \gamma_q(\bar m,\bar d)$.
  The probability that Conditions~(\ref{cond:diag-local}) and~(\ref{cond:diag-forest}) hold is therefore at least $\mathrm P(\underline\gamma(\bar m,\bar d) \le \gamma_q(\bar m,\bar d))$, and the same bound follows.
\end{proof}

\section{Proof of \Cref{thm:convergence}}\label{app:convergence}

\subsection{Notation}

The notation is that of \Cref{sec:theory}, in particular of \Cref{def:delta,def:errors,def:separated,def:radii,def:recursion}, and the norms are those of \eqref{eq:norms}.
For a ground-truth matrix $M^*$ we write $M^*_\infty := \|M^*\|_\infty$, which gives $\Theta^*_\infty$ and $\widetilde\Sigma^*_\infty$.

The proofs below use the submultiplicativity relations
\begin{gather*}
  \|MN\|_{\max} \le \|M\|_{\infty}\|N\|_{\max},
  \qquad
  \|MN\|_{\max} \le \|M\|_{\max}\|N^\top\|_{\infty},
  \\
  \|MN\|_{\infty} \le \|M\|_{\infty}\|N\|_{\infty},
\end{gather*}
the specialization $\|M\boldsymbol{v}\|_{\max} \le \|M\|_{\infty}\|\boldsymbol{v}\|_{\max}$ to a vector $\boldsymbol{v}$, and the bound
\begin{equation*}
  \|D M D' \boldsymbol{1}\|_{\max} \le \|D\|_{\max}\|M\|_{\infty}\|D'\|_{\max}
\end{equation*}
for diagonal $D$ and $D'$.

\subsection{Supporting results}

\begin{lemma}\label{lem:restr}
  Let $\gamma > 1$, $\xi > 0$ and $c_0 \in [0,1)$, and let $(R^{(t)}, \eta^{(t)})$ be $(\xi, c_0)$-separated.
  Then $\eta^{(t+1)}$, given by \eqref{eq:eta_update}, satisfies
  \begin{equation*}
    \eta_{\min} \leq \eta^{(t+1)} \leq \eta_{\max}.
  \end{equation*}
\end{lemma}

\begin{proof}
  Let $\sigma^{(t)}_{ij}$ denote the summands in \eqref{eq:eta_update}.
  For $(i,j) \notin \supp(R^*)$, the second condition in \eqref{eq:sep_estimate} and $\tilde\gamma \geq \gamma$ give $\tilde\gamma e^{-\gamma |r^{(t)}_{ij}|} \geq \gamma^{1-c_0}$.
  Writing $\sigma^{(t)}_{ij}$ in the form of \eqref{eq:sigma_t} then yields
  \begin{equation}\label{eq:sigma_off}
    \sigma^{(t)}_{ij} \leq \frac{\eta^{(t)}}{\eta^{(t)} + \gamma^{1-c_0}(1-\eta^{(t)})} \leq \frac{\eta_{\max}}{\eta_{\max} + \gamma^{1-c_0}(1-\eta_{\max})},
  \end{equation}
  where the second inequality holds because the middle term is increasing in $\eta^{(t)}$.
  Since $R^*$ is symmetric, $(i,j) \in \supp(R^*)$ if and only if $(j,i) \in \supp(R^*)$, so exactly $\eta^*\ell$ of its elements satisfy $i<j$.
  Bounding $\sigma^{(t)}_{ij} \leq 1$ on these pairs and using \eqref{eq:sigma_off} on the remaining $(1-\eta^*)\ell$ pairs with $i<j$ gives
  \begin{equation*}
    \eta^{(t+1)} \leq \eta^* + (1-\eta^*)\frac{\eta_{\max}}{\eta_{\max} + \gamma^{1-c_0}(1-\eta_{\max})} = \eta_{\max}.
  \end{equation*}

  For $(i,j) \in \supp(R^*)$, since $r_c$ is decreasing and $\eta^{(t)} \geq \eta_{\min}$, we have $r_c(\eta^{(t)}) \leq r_c(\eta_{\min})$.
  The first condition in \eqref{eq:sep_estimate} then yields $|r^{(t)}_{ij}| - r_c(\eta^{(t)}) \geq \xi$, and hence
  \begin{equation} \label{eq:sigma_on}
    1-\sigma^{(t)}_{ij} = \frac{1}{1 + e^{\gamma(|r^{(t)}_{ij}| - r_c(\eta^{(t)}))}} \leq e^{-\gamma\xi}.
  \end{equation}
  Dropping the nonnegative summands outside $\supp(R^*)$ gives
  \begin{equation*}
    \eta^{(t+1)} \geq \frac{1}{\ell} \sum_{\substack{(i,j) \in \supp(R^*) \\ i<j}} \sigma^{(t)}_{ij} \geq \eta^* \big(1-e^{-\gamma\xi}\big) = \eta_{\min}.
  \end{equation*}
\end{proof}

\begin{corollary}\label{cor:cor1}
  Under the conditions of \Cref{lem:restr}, suppose in addition that $\gamma^{1-c_0} > 1/(1-\eta^*)$.
  Then the penalty weights $\gamma^{(t)}_{ij}$ of the $R$-update \eqref{eq:R_subproblem} satisfy
  \begin{equation}\label{eq:boundsgamma}
    \gamma^{(t)}_{ij} \leq \gamma e^{-\gamma\xi} \quad \text{for } (i,j) \in \supp(R^*), \qquad \gamma^{(t)}_{ij} \geq \gamma\,(1-\eta_{\max}) \quad \text{for } (i,j) \notin \supp(R^*).
  \end{equation}
\end{corollary}

\begin{proof}
  The first bound follows from \eqref{eq:sigma_on}.
  By \eqref{eq:boundseta2}, the condition on $\gamma$ is equivalent to $\eta_{\max} < 1$.
  For $(i,j) \notin \supp(R^*)$, this and $\gamma^{1-c_0} > 1$ make the right-hand side of \eqref{eq:sigma_off} at most $\eta_{\max}$, so $\sigma^{(t)}_{ij} \leq \eta_{\max}$.
  The definition of $\gamma^{(t)}_{ij}$ then gives the second bound.
\end{proof}

\begin{lemma}\label{lem:updateR}
  Let $b \geq 0$ satisfy $\|\log\Lambda^{(t)} - \log\Lambda^*\|_{\max} \leq b\bar\delta$, and let the penalty weights $\gamma^{(t)}_{ij}$ satisfy \eqref{eq:boundsgamma}.
  Let $\bar\varepsilon_R > 0$ satisfy the first two conditions in \eqref{eq:thm_eps}, let $\kappa_R$ be given by \eqref{eq:thm_kappas}, and define
  \begin{equation}\label{eq:epsR}
    \varepsilon_R(b) := \kappa_R\, C_\Gamma\Big(e^{2b\bar\delta}\big(1 + 2b\,\widetilde\Sigma^*_{\max,\mathcal S}\big)\bar\delta + \gamma e^{-\gamma\xi}/n\Big).
  \end{equation}
  If
  \begin{equation}\label{eq:updateR_cond}
    \begin{gathered}
      \varepsilon_R(b) \leq \bar\varepsilon_R,
      \\
      \begin{aligned}
        \frac{\gamma\,(1-\eta_{\max})}{n} &\geq e^{2b\bar\delta}\big(1 + 2b\,\widetilde\Sigma^*_{\max,\mathcal S^c}\big)\bar\delta + C_{\mathrm{irr}}\Big(e^{2b\bar\delta}\big(1 + 2b\,\widetilde\Sigma^*_{\max,\mathcal S}\big)\bar\delta + \frac{\gamma e^{-\gamma\xi}}{n}\Big) \\
        &\quad + 2\big(1+C_{\mathrm{irr}}\big)\,d\,(\widetilde\Sigma^*_\infty)^3\varepsilon_R(b)^2,
      \end{aligned}
    \end{gathered}
  \end{equation}
  then $R^{(t+1)} = I_p - \Theta^{(t+1)}$, with $\Theta^{(t+1)}$ given by \eqref{eq:R_subproblem}, satisfies $\supp(R^{(t+1)}) \subseteq \supp(R^*)$ and
  \begin{equation}\label{eq:deltaR}
    \|R^{(t+1)} - R^*\|_{\max} \leq \varepsilon_R(b).
  \end{equation}
\end{lemma}

\begin{proof}
  We follow the primal-dual witness construction of~\citet{ravikumar2011}, and within this proof write $\varepsilon_R$ for $\varepsilon_R(b)$, which differs from the $\varepsilon_R$ of \Cref{def:recursion}.
  Problem~\eqref{eq:R_subproblem} is strictly convex, and a feasible $\Theta$ solves it if and only if there is a symmetric $Z$ such that, for every $i \neq j$,
  \begin{equation}\label{eq:KKT}
    \big[\Theta^{-1} - \widehat\Sigma^{(t)}\big]_{ij} = \frac{\gamma^{(t)}_{ij}}{n} Z_{ij}, \qquad Z_{ij} = \sgn(\Theta_{ij}) \ \text{if } \Theta_{ij} \neq 0, \qquad |Z_{ij}| \leq 1 \ \text{if } \Theta_{ij} = 0.
  \end{equation}
  The diagonal is absent from \eqref{eq:KKT} because it is fixed by the constraint $\Theta_{ii} = 1$.
  Let $\Theta'$ be the minimizer of the restricted problem obtained by adding the constraints $\Theta_{ij} = 0$ for $(i,j) \notin \mathcal S$, $i \neq j$, to problem~\eqref{eq:R_subproblem}.
  It exists because the unit diagonal bounds the feasible set and the log-determinant term diverges at its boundary, and it is unique by strict convexity.
  It satisfies \eqref{eq:KKT} for $(i,j) \in \mathcal S$ with some $Z'$.
  The matrix $\Theta' - \Theta^*$ is symmetric, has zero diagonal, and vanishes outside $\mathcal S$.
  We show that $\|\Theta' - \Theta^*\|_{\max} \leq \varepsilon_R$ and that $\Theta'$ also satisfies \eqref{eq:KKT} for $(i,j) \notin \mathcal S$.
  Then $\Theta' = \Theta^{(t+1)}$ by uniqueness, and since $\Theta = I_p - R$, we have $R^{(t+1)} - R^* = \Theta^* - \Theta'$, which gives both claims.

  \emph{Preliminary bounds.}
  Let $\mathcal B_{\varepsilon_R}$ be the set of symmetric matrices $\Delta$ with zero diagonal that vanish outside $\mathcal S$ and satisfy $\|\Delta\|_{\max} \leq \varepsilon_R$.
  For $\Delta \in \mathcal B_{\varepsilon_R}$, each row of $\Delta$ has at most $d$ nonzero entries, so $\|\widetilde\Sigma^*\Delta\|_\infty \leq d\,\widetilde\Sigma^*_\infty \varepsilon_R \leq 1/2$ by $\varepsilon_R \leq \bar\varepsilon_R$ and the first condition in \eqref{eq:thm_eps}.
  Since $\widetilde\Sigma^*\Delta$ is similar to the symmetric matrix $(\Theta^*)^{-1/2}\Delta(\Theta^*)^{-1/2}$, the eigenvalues of the latter lie in $[-1/2, 1/2]$.
  Hence $\Theta^* + \Delta = (\Theta^*)^{1/2}\big(I_p + (\Theta^*)^{-1/2}\Delta(\Theta^*)^{-1/2}\big)(\Theta^*)^{1/2}$ is positive definite and feasible for the restricted problem.
  The identity $(\Theta^*+\Delta)^{-1} = \widetilde\Sigma^* - \widetilde\Sigma^*\Delta(\Theta^*+\Delta)^{-1}$ then gives $\|(\Theta^*+\Delta)^{-1}\|_\infty \leq 2\widetilde\Sigma^*_\infty$.
  Applying the identity twice,
  \begin{equation}\label{eq:inverse}
    (\Theta^*+\Delta)^{-1} = \widetilde\Sigma^* - \widetilde\Sigma^*\Delta\widetilde\Sigma^* + \Psi(\Delta), \qquad \Psi(\Delta) := \big(\widetilde\Sigma^*\Delta\big)^2(\Theta^*+\Delta)^{-1},
  \end{equation}
  where
  \begin{equation}\label{eq:bound1}
    \begin{gathered}
      \|\widetilde\Sigma^*\Delta\widetilde\Sigma^*\|_{\max} \leq (\widetilde\Sigma^*_\infty)^2 \varepsilon_R,
      \\
      \|\Psi(\Delta)\|_{\max} \leq \|\widetilde\Sigma^*\Delta\|_\infty\|\widetilde\Sigma^*\Delta\|_{\max}\|(\Theta^*+\Delta)^{-1}\|_\infty \leq 2d(\widetilde\Sigma^*_\infty)^3 \varepsilon_R^2.
    \end{gathered}
  \end{equation}
  Finally, let $\boldsymbol\psi \in \mathbb R^p$ be the vector of diagonal entries of $\log\Lambda^{(t)} - \log\Lambda^*$, so that $\|\boldsymbol\psi\|_{\max} \leq b\bar\delta$, and write $\Lambda^{(t)} = \Lambda^* E_{\boldsymbol\psi}$ with $E_{\boldsymbol\psi} := \diag\big(e^{\psi_1}, \dots, e^{\psi_p}\big)$.
  Diagonal matrices commute, so
  \begin{equation*}
    \widehat\Sigma^{(t)} = E_{\boldsymbol\psi}\,\Lambda^*\big(\Sigma^* + W\big)\Lambda^*\,E_{\boldsymbol\psi} = E_{\boldsymbol\psi}\big(\widetilde\Sigma^* + \widetilde W\big)E_{\boldsymbol\psi},
    \qquad
    \widehat\Sigma^{(t)}_{ij} - \widetilde\Sigma^*_{ij} = \big(e^{\psi_i+\psi_j}-1\big)\widetilde\Sigma^*_{ij} + e^{\psi_i+\psi_j}\,\widetilde W_{ij}.
  \end{equation*}
  Since $|e^x - 1| \leq |x|e^{|x|}$, $|\psi_i + \psi_j| \leq 2b\bar\delta$ and $\|\widetilde W\|_{\max} \leq \bar\delta$ by \Cref{def:delta}, we have $|\widehat\Sigma^{(t)}_{ij} - \widetilde\Sigma^*_{ij}| \leq e^{2b\bar\delta}\big(1 + 2b\,|\widetilde\Sigma^*_{ij}|\big)\bar\delta$, and hence, by \eqref{eq:epsR},
  \begin{equation}\label{eq:bound3}
    \begin{gathered}
      \big\|[\widehat\Sigma^{(t)} - \widetilde\Sigma^*]_{\mathcal S}\big\|_{\max} \leq e^{2b\bar\delta}\big(1 + 2b\,\widetilde\Sigma^*_{\max,\mathcal S}\big)\bar\delta = \frac{\varepsilon_R}{\kappa_R C_\Gamma} - \frac{\gamma e^{-\gamma\xi}}{n},
      \\
      \big\|[\widehat\Sigma^{(t)} - \widetilde\Sigma^*]_{\mathcal S^c}\big\|_{\max} \leq e^{2b\bar\delta}\big(1 + 2b\,\widetilde\Sigma^*_{\max,\mathcal S^c}\big)\bar\delta.
    \end{gathered}
  \end{equation}

  \emph{First step: $\|\Theta' - \Theta^*\|_{\max} \leq \varepsilon_R$.}
  For a $p \times p$ matrix $M$, let $[M]_{\mathcal S} \in \mathbb R^{|\mathcal S|}$ collect the entries of $M$ indexed by $\mathcal S$, and identify $\mathbb R^{|\mathcal S|}$ with the matrices that vanish outside $\mathcal S$.
  For symmetric $\Delta$ with zero diagonal that vanishes outside $\mathcal S$ and makes $\Theta^* + \Delta$ positive definite, define
  \begin{equation*}
    G(\Delta) := \big[\widehat\Sigma^{(t)} - (\Theta^*+\Delta)^{-1} + n^{-1}\gamma^{(t)} \circ Z'\big]_{\mathcal S}, \qquad F(\Delta) := [\Delta]_{\mathcal S} - (\Gamma^*_{\mathcal S\mathcal S})^{-1} G(\Delta).
  \end{equation*}
  By the preliminary bounds, every $\Delta \in \mathcal B_{\varepsilon_R}$ belongs to that set.
  Since $\Delta$ vanishes outside $\mathcal S$, $[\widetilde\Sigma^*\Delta\widetilde\Sigma^*]_{\mathcal S} = \Gamma^*_{\mathcal S\mathcal S}[\Delta]_{\mathcal S}$, and \eqref{eq:inverse} gives
  \begin{equation*}
    F(\Delta) = -(\Gamma^*_{\mathcal S\mathcal S})^{-1}\big[\widehat\Sigma^{(t)} - \widetilde\Sigma^* - \Psi(\Delta) + n^{-1}\gamma^{(t)} \circ Z'\big]_{\mathcal S}.
  \end{equation*}
  Equation~\eqref{eq:bound1}, $|Z'_{ij}| \leq 1$, and the first bound in \eqref{eq:boundsgamma} give, for $\Delta \in \mathcal B_{\varepsilon_R}$,
  \begin{equation*}
    \|F(\Delta)\|_{\max} \leq C_\Gamma\big(\|[\widehat\Sigma^{(t)} - \widetilde\Sigma^*]_{\mathcal S}\|_{\max} + \gamma e^{-\gamma\xi}/n + 2d(\widetilde\Sigma^*_\infty)^3 \varepsilon_R^2\big) \leq \frac{\varepsilon_R}{\kappa_R} + 2dC_\Gamma (\widetilde\Sigma^*_\infty)^3 \varepsilon_R^2 \leq \varepsilon_R,
  \end{equation*}
  where the second inequality follows from the first bound in \eqref{eq:bound3} and the last from $\varepsilon_R \leq \bar\varepsilon_R$ and the definition of $\kappa_R$ in \eqref{eq:thm_kappas}.
  By construction, $F(\Delta)$ has zero diagonal and vanishes outside $\mathcal S$.
  It is also symmetric, since $G(\Delta)$ is symmetric and $\Gamma^*_{\mathcal S\mathcal S}$, and hence its inverse, commutes with the transposition $(i,j) \mapsto (j,i)$ of its row and column indices.
  Hence $F$ maps $\mathcal B_{\varepsilon_R}$ into itself.
  Since $F$ is continuous and $\mathcal B_{\varepsilon_R}$ is compact and convex, Brouwer's theorem gives a fixed point $\Delta \in \mathcal B_{\varepsilon_R}$, that is, $G(\Delta) = 0$.
  Since also $G(\Theta' - \Theta^*) = 0$ by \eqref{eq:KKT}, subtracting the two equations eliminates $Z'$ and gives $[(\Theta^*+\Delta)^{-1}]_{\mathcal S} = [\Theta'^{-1}]_{\mathcal S}$.
  On the feasible set of the restricted problem, $\Theta \mapsto [\Theta^{-1}]_{\mathcal S}$ is injective.
  Hence $\Theta' - \Theta^* = \Delta \in \mathcal B_{\varepsilon_R}$.

  \emph{Second step: \eqref{eq:KKT} outside $\mathcal S$.}
  Let $\Delta := \Theta' - \Theta^*$, which is the fixed point of $F$ found in the first step.
  Since $\Delta$ vanishes outside $\mathcal S$, $[\widetilde\Sigma^*\Delta\widetilde\Sigma^*]_{\mathcal S^c} = \Gamma^*_{\mathcal S^c\mathcal S}[\Delta]_{\mathcal S}$, and $[\Delta]_{\mathcal S} = F(\Delta)$ then gives
  \begin{equation*}
    [\widetilde\Sigma^*\Delta\widetilde\Sigma^*]_{\mathcal S^c} = -\Gamma^*_{\mathcal S^c\mathcal S}(\Gamma^*_{\mathcal S\mathcal S})^{-1}\big[\widehat\Sigma^{(t)} - \widetilde\Sigma^* - \Psi(\Delta) + n^{-1}\gamma^{(t)} \circ Z'\big]_{\mathcal S}.
  \end{equation*}
  For $(i,j) \in \mathcal S^c$, \eqref{eq:inverse}, this identity, \eqref{eq:bound1} and~\eqref{eq:bound3}, $|Z'_{kl}| \leq 1$ and the first bound in \eqref{eq:boundsgamma} give
  \begin{equation*}
    \begin{aligned}
      \big|[\Theta'^{-1} - \widehat\Sigma^{(t)}]_{ij}\big|
      &\leq e^{2b\bar\delta}\big(1 + 2b\,\widetilde\Sigma^*_{\max,\mathcal S^c}\big)\bar\delta + C_{\mathrm{irr}}\Big(e^{2b\bar\delta}\big(1 + 2b\,\widetilde\Sigma^*_{\max,\mathcal S}\big)\bar\delta + \frac{\gamma e^{-\gamma\xi}}{n}\Big) \\
      &\quad + 2\big(1+C_{\mathrm{irr}}\big)\,d\,(\widetilde\Sigma^*_\infty)^3\varepsilon_R^2 \\
      &\leq \frac{\gamma\,(1-\eta_{\max})}{n} \leq \frac{\gamma^{(t)}_{ij}}{n},
    \end{aligned}
  \end{equation*}
  where the second inequality is the second condition in \eqref{eq:updateR_cond} and the last follows from the second bound in \eqref{eq:boundsgamma}.
  Hence $\Theta'$ satisfies \eqref{eq:KKT} for $(i,j) \notin \mathcal S$ with $Z'_{ij} := n[\Theta'^{-1} - \widehat\Sigma^{(t)}]_{ij}/\gamma^{(t)}_{ij}$.
\end{proof}

\begin{lemma}\label{lem:updateLambda}
  Let $a \geq 0$ satisfy $\|R^{(t+1)} - R^*\|_{\max} \leq a\bar\delta$, let $R^{(t+1)} \in \mathcal{R}_p$ satisfy $\supp(R^{(t+1)}) \subseteq \supp(R^*)$, and let $\Lambda^{(t+1)}$ be the minimizer in \eqref{eq:lambda_subprob}.
  Let $\bar\varepsilon_\Lambda > 0$ satisfy the third condition in \eqref{eq:thm_eps}, let $\kappa_\Lambda$ be given by \eqref{eq:thm_kappas}, and define
  \begin{equation}\label{eq:epsLambda}
    \varepsilon_\Lambda(a) := \kappa_\Lambda \, C_J \big[\Theta^*_\infty + a\big(\widetilde\Sigma^*_{\infty,\mathcal S} + d\,\bar\delta\big)\big]\,\bar\delta .
  \end{equation}
  If $\varepsilon_\Lambda(a) \leq \bar\varepsilon_\Lambda$, then
  \begin{equation}\label{eq:deltalambda}
    \|\log \Lambda^{(t+1)} - \log \Lambda^*\|_{\max} \leq \varepsilon_\Lambda(a) + \frac{1}{2(n-1)} .
  \end{equation}
\end{lemma}

\begin{proof}
  Within this proof we write $\varepsilon_\Lambda$ for $\varepsilon_\Lambda(a)$, which differs from the $\varepsilon_\Lambda$ of \Cref{def:recursion}.
  Since $\Theta^*$ is symmetric, the row sums of $\widetilde\Sigma^* \circ \Theta^* = (\Theta^*)^{-1} \circ \Theta^*$ are the diagonal entries of $(\Theta^*)^{-1}\Theta^*$, so
  \begin{equation}\label{eq:lambda_star}
    (\widetilde\Sigma^* \circ \Theta^*) \boldsymbol 1 = \boldsymbol 1.
  \end{equation}
  By the Schur product theorem $\widetilde\Sigma^* \circ \Theta^* \succeq 0$, so $J^*$ is invertible.

  The stationarity condition of problem~\eqref{eq:lambda_subprob} is $\Lambda \Xi \Lambda \boldsymbol 1 = (1-1/n)\boldsymbol 1$ with $\Xi := S_n \circ (I_p - R^{(t+1)})$, so $\Lambda' := \Lambda^{(t+1)}/\sqrt{1-1/n}$ satisfies
  \begin{equation}\label{eq:lambda_prime}
    \Lambda' \, \Xi \, \Lambda' \, \boldsymbol 1 = \boldsymbol 1,
  \end{equation}
  and by the strict convexity established in \Cref{app:ecm-lambda} it is its only positive diagonal solution.

  Writing $S_n = \Sigma^* + W$ and $I_p - R^{(t+1)} = \Theta^* - \Delta_R$ with $\Delta_R := R^{(t+1)} - R^*$, we get $\Lambda^* \Xi \Lambda^* = \widetilde\Sigma^* \circ \Theta^* - B$ with
  \begin{equation*}
    B := \widetilde\Sigma^* \circ \Delta_R + \widetilde W \circ \Delta_R - \widetilde W \circ \Theta^* .
  \end{equation*}
  Both $\Delta_R$ and $\Theta^* - I_p$ are supported in $\mathcal S$, whose rows carry at most $d$ entries, and $\|\widetilde W\|_{\max} \leq \bar\delta$ by \Cref{def:delta}, so bounding the three terms in turn gives
  \begin{equation}\label{eq:boundB}
    \|B\|_\infty \leq \big(\widetilde\Sigma^*_{\infty,\mathcal S} + d\,\bar\delta\big)\|\Delta_R\|_{\max} + \Theta^*_\infty\,\bar\delta \leq \frac{\varepsilon_\Lambda}{\kappa_\Lambda\, C_J},
  \end{equation}
  where the last inequality uses $\|\Delta_R\|_{\max} \leq a\bar\delta$ and \eqref{eq:epsLambda}.

  We now construct a map whose fixed points are the solutions of \eqref{eq:lambda_prime} in a ball of radius $\varepsilon_\Lambda$ around $\Lambda^*$ on the logarithmic scale, and show it maps that ball into itself, so that Brouwer's theorem applies.
  For $\boldsymbol\psi \in \mathbb{R}^p$ with $\|\boldsymbol\psi\|_{\max} \leq \varepsilon_\Lambda$, write $E_{\boldsymbol\psi} := \diag\big(e^{\psi_1}, \dots, e^{\psi_p}\big)$ and define
  \begin{equation*}
    G(\boldsymbol\psi) := \Lambda^* E_{\boldsymbol\psi}\, \Xi\, E_{\boldsymbol\psi} \Lambda^* \boldsymbol 1 - \boldsymbol 1,
    \qquad F(\boldsymbol\psi) := \boldsymbol\psi - (J^*)^{-1}G(\boldsymbol\psi).
  \end{equation*}
  Since $J^*$ is invertible, $F(\boldsymbol\psi) = \boldsymbol\psi$ if and only if $G(\boldsymbol\psi) = \boldsymbol 0$.
  Diagonal matrices commute, so $\Lambda^* E_{\boldsymbol\psi} \Xi E_{\boldsymbol\psi} \Lambda^* = E_{\boldsymbol\psi}(\widetilde\Sigma^* \circ \Theta^* - B)E_{\boldsymbol\psi}$, and \eqref{eq:lambda_star} gives
  \begin{equation*}
    G_i(\boldsymbol\psi) = \sum_j \widetilde\Sigma^*_{ij}\Theta^*_{ij}\big(e^{\psi_i + \psi_j} - 1\big) - \big(E_{\boldsymbol\psi} B E_{\boldsymbol\psi} \boldsymbol 1\big)_i .
  \end{equation*}
  The same identity gives $\sum_j \widetilde\Sigma^*_{ij}\Theta^*_{ij}(\psi_i + \psi_j) = (J^* \boldsymbol\psi)_i$, so
  \begin{equation*}
    F(\boldsymbol\psi) = -(J^*)^{-1}\big[\boldsymbol f(\boldsymbol\psi) - E_{\boldsymbol\psi} B E_{\boldsymbol\psi} \boldsymbol 1\big],
    \qquad f_i(\boldsymbol\psi) := \sum_j \widetilde\Sigma^*_{ij}\Theta^*_{ij}\big[e^{\psi_i+\psi_j} - 1 - (\psi_i+\psi_j)\big].
  \end{equation*}
  Since $|e^x - 1 - x| \leq \tfrac{1}{2}x^2 e^{|x|}$ and $|\psi_i + \psi_j| \leq 2\varepsilon_\Lambda$, we have $\|\boldsymbol f(\boldsymbol\psi)\|_{\max} \leq 2\|\widetilde\Sigma^* \circ \Theta^*\|_\infty \varepsilon_\Lambda^2 e^{2\varepsilon_\Lambda}$, and therefore
  \begin{equation*}
    \|F(\boldsymbol\psi)\|_{\max} \leq C_J e^{2\varepsilon_\Lambda}\Big(2\|\widetilde\Sigma^* \circ \Theta^*\|_\infty \varepsilon_\Lambda^2 + \|B\|_\infty\Big) \leq \varepsilon_\Lambda .
  \end{equation*}
  The last inequality follows from \eqref{eq:boundB}, since $\varepsilon_\Lambda \leq \bar\varepsilon_\Lambda$ gives
  \begin{equation*}
    \left(2 C_J \|\widetilde\Sigma^* \circ \Theta^*\|_\infty \varepsilon_\Lambda e^{2\varepsilon_\Lambda} + \frac{e^{2\varepsilon_\Lambda}}{\kappa_\Lambda}\right) \varepsilon_\Lambda \leq \left(2 C_J \|\widetilde\Sigma^* \circ \Theta^*\|_\infty \bar\varepsilon_\Lambda\, e^{2\bar\varepsilon_\Lambda} + \frac{e^{2\bar\varepsilon_\Lambda}}{\kappa_\Lambda}\right) \varepsilon_\Lambda = \varepsilon_\Lambda
  \end{equation*}
  by the definition of $\kappa_\Lambda$ in \eqref{eq:thm_kappas}.

  Since $F$ is continuous and the ball is compact and convex, Brouwer's theorem gives a fixed point $\boldsymbol\psi$.
  The matrix $\Lambda^* E_{\boldsymbol\psi}$ is positive diagonal, hence equals $\Lambda'$ by the uniqueness noted above, giving $\|\log\Lambda' - \log\Lambda^*\|_{\max} \leq \varepsilon_\Lambda$.
  Finally $\log \Lambda^{(t+1)} = \log\Lambda' + \tfrac{1}{2}\log(1-1/n)\,I_p$ and $-\tfrac{1}{2}\log(1-1/n) \leq \tfrac{1}{2(n-1)}$ give \eqref{eq:deltalambda}.
\end{proof}

\subsection{Proof of the theorem}

\thmconvergence*

\begin{proof}
  Throughout, we use that a sweep updates $\eta$, then $R$, then $\Lambda$, so that $\eta^{(t+1)}$ is computed from $R^{(t)}$ and $\eta^{(t)}$, while $R^{(t+1)}$ uses $\Lambda^{(t)}$ and the penalty weights $\gamma^{(t)}_{ij}$ built from $R^{(t)}$ and $\eta^{(t)}$, and $\Lambda^{(t+1)}$ uses $R^{(t+1)}$.
  Since $\bar\varepsilon_R$ and $\bar\varepsilon_\Lambda$ are admissible, they satisfy the conditions that \Cref{lem:updateR} and \Cref{lem:updateLambda} impose on them.
  We first record the two one-sweep bounds
  \begin{equation}\label{eq:one_sweep}
    a_{t+1} \leq \alpha_R + \beta_R b_t, \qquad b_{t+1} \leq \alpha_\Lambda + \beta_\Lambda a_{t+1},
  \end{equation}
  valid whenever $b_t \leq B_0$ and the penalty weights $\gamma^{(t)}_{ij}$ satisfy \eqref{eq:boundsgamma}, and then show by induction that
  \begin{equation}\label{eq:induction}
    a_t \leq A_0, \qquad b_t \leq B_0, \qquad \supp(R^{(t)}) = \supp(R^*), \qquad \eta^{(t)} \in [\eta_{\min}, \eta_{\max}]
  \end{equation}
  holds for every $t \geq 1$.

  \emph{The two bounds.}
  Suppose $b_t \leq B_0$ and that \eqref{eq:boundsgamma} holds, so that, together with the definition of $b_t$, the hypotheses of \Cref{lem:updateR} on $\Lambda^{(t)}$ and the penalty weights are met.
  From $b_t \leq B_0$ and the first inequality of Assumption~(\ref{cond:localization}) we get $b_t\bar\delta \leq \bar\varepsilon_\Lambda$ and hence $e^{2b_t\bar\delta} \leq e^{2\bar\varepsilon_\Lambda}$, while the third inequality of Assumption~(\ref{cond:weights}) bounds the last term of \eqref{eq:epsR} by $\kappa_R C_\Gamma\bar\delta$.
  Therefore
  \begin{equation}\label{eq:epsR_linear}
    \varepsilon_R(b_t) \leq \kappa_R C_\Gamma\big[e^{2\bar\varepsilon_\Lambda}\big(1 + 2\widetilde\Sigma^*_{\max,\mathcal S}b_t\big) + 1\big]\bar\delta = \big(\alpha_R + \beta_R b_t\big)\bar\delta.
  \end{equation}
  At $b_t = B_0$ the right-hand side is the $\varepsilon_R$ of \Cref{def:recursion}, so $\varepsilon_R(b_t) \leq \varepsilon_R$.
  The first condition of \eqref{eq:updateR_cond} at $b = b_t$ therefore follows from the second inequality of Assumption~(\ref{cond:localization}).
  For the second, replacing $b_t$ by $B_0$, $e^{2b_t\bar\delta}$ by $e^{2\bar\varepsilon_\Lambda}$, $\gamma e^{-\gamma\xi}/n$ by $\bar\delta$ and $\varepsilon_R(b_t)$ by $\varepsilon_R$ can only increase its right-hand side, which then becomes the right-hand side of the fourth inequality of Assumption~(\ref{cond:weights}) divided by $n$.
  \Cref{lem:updateR} applies and gives $\supp(R^{(t+1)}) \subseteq \supp(R^*)$ together with $a_{t+1}\bar\delta \leq \varepsilon_R(b_t)$, which by \eqref{eq:epsR_linear} is the first bound in \eqref{eq:one_sweep}.
  For the second, the support inclusion just obtained is the remaining hypothesis of \Cref{lem:updateLambda}.
  Moreover $a_{t+1} \leq \alpha_R + \beta_R B_0 \leq A_0$, and the $\varepsilon_\Lambda$ of \Cref{def:recursion} is $\varepsilon_\Lambda(A_0)$ in the notation of \eqref{eq:epsLambda}, so $\varepsilon_\Lambda(a_{t+1}) \leq \varepsilon_\Lambda \leq \bar\varepsilon_\Lambda$ because $\varepsilon_\Lambda(\cdot)$ is increasing and by the third inequality of Assumption~(\ref{cond:localization}).
  \Cref{lem:updateLambda} applies, and using $1/\big(2(n-1)\big) \leq \bar\delta$, from \Cref{def:delta}, \eqref{eq:deltalambda} gives
  \begin{equation*}
    b_{t+1}\bar\delta \leq \kappa_\Lambda C_J\big[\Theta^*_\infty + a_{t+1}\big(\widetilde\Sigma^*_{\infty,\mathcal S} + d\bar\delta\big)\big]\bar\delta + \frac{1}{2(n-1)} \leq \big(\alpha_\Lambda + \beta_\Lambda a_{t+1}\big)\bar\delta.
  \end{equation*}

  \emph{Base case.}
  The hypotheses of \Cref{lem:restr} at $t = 0$ are that $(R^{(0)}, \eta^{(0)})$ is $(\xi, c_0)$-separated, which is Assumption~(\ref{cond:init}), and $\gamma > 1$, which follows from the first inequality of Assumption~(\ref{cond:weights}), since $1/(1-\eta^*) \geq 1$ and $c_0 < 1$.
  \Cref{lem:restr} then gives $\eta^{(1)} \in [\eta_{\min}, \eta_{\max}]$, and \Cref{cor:cor1}, whose additional hypothesis is the first inequality of Assumption~(\ref{cond:weights}), gives the bounds \eqref{eq:boundsgamma} on the penalty weights at $t = 0$.
  Since $b_0 \leq B_0$, the previous paragraph applies and yields $\supp(R^{(1)}) \subseteq \supp(R^*)$, $a_1 \leq \alpha_R + \beta_R B_0 \leq A_0$ and $b_1 \leq \alpha_\Lambda + \beta_\Lambda a_1$.
  The last is at most $\alpha_\Lambda + \beta_\Lambda\alpha_R + \beta_R\beta_\Lambda B_0$, which is at most $B_0$ because $B_0 \geq b_\infty$ gives $\alpha_\Lambda + \beta_\Lambda\alpha_R = (1-\beta_R\beta_\Lambda)b_\infty \leq (1-\beta_R\beta_\Lambda)B_0$.
  For the reverse support inclusion, $a_1 \leq A_0$ gives, for every $(i,j) \in \supp(R^*)$,
  \begin{equation}\label{eq:separation}
    |r^{(1)}_{ij}| \geq |r^*_{ij}| - a_1\bar\delta \geq R^*_{\min} - A_0\bar\delta \geq r_c(\eta_{\min}) + \xi > 0,
  \end{equation}
  where the third inequality is Assumption~(\ref{cond:signal}), and the last holds because the second inequality of Assumption~(\ref{cond:weights}) makes $r_c(\eta_{\min})$ positive by \eqref{eq:rc}.
  Hence $\supp(R^*) \subseteq \supp(R^{(1)})$, and \eqref{eq:induction} holds at $t = 1$.

  \emph{Induction step.}
  Assume \eqref{eq:induction} at some $t \geq 1$.
  Off $\supp(R^*)$ the entries of $R^{(t)}$ vanish, so the second condition in \eqref{eq:sep_estimate} holds trivially, and on it \eqref{eq:separation} with $1$ replaced by $t$ gives $|r^{(t)}_{ij}| \geq r_c(\eta_{\min}) + \xi$, which is the first condition.
  Since also $\eta^{(t)} \in [\eta_{\min}, \eta_{\max}]$, the pair $(R^{(t)}, \eta^{(t)})$ is $(\xi, c_0)$-separated.
  \Cref{lem:restr} then gives $\eta^{(t+1)} \in [\eta_{\min}, \eta_{\max}]$ and \Cref{cor:cor1} the bounds \eqref{eq:boundsgamma} on the penalty weights.
  Since $b_t \leq B_0$, the paragraph on the two bounds applies and yields $\supp(R^{(t+1)}) \subseteq \supp(R^*)$, $a_{t+1} \leq \alpha_R + \beta_R B_0 \leq A_0$ and $b_{t+1} \leq \alpha_\Lambda + \beta_\Lambda a_{t+1} \leq B_0$, the last as in the base case.
  The reverse support inclusion follows as in the base case, from \eqref{eq:separation} with $1$ replaced by $t+1$.
  Hence \eqref{eq:induction} holds at $t+1$, and by induction at every $t \geq 1$, which proves the first claim of the theorem.

  \emph{Recursion.}
  Equation~\eqref{eq:one_sweep} now holds for every $t \geq 0$.
  Substituting the second bound into the first gives $a_{t+1} \leq \alpha_R + \beta_R\alpha_\Lambda + \beta_R\beta_\Lambda\,a_t$ for $t \geq 1$, and substituting the first into the second gives $b_{t+1} \leq \alpha_\Lambda + \beta_\Lambda\alpha_R + \beta_R\beta_\Lambda\,b_t$ for $t \geq 0$.
  By \eqref{eq:fixed_points}, $a_\infty$ and $b_\infty$ are the fixed points of these two affine maps, and subtracting them yields \eqref{eq:resultsTheorem}.
  The four constants of \eqref{eq:thm_constants} are positive, so Assumption~(\ref{cond:coupling}) gives $0 < \beta_R\beta_\Lambda < 1$, and iterating gives $a_t - a_\infty \leq (\beta_R\beta_\Lambda)^{t-1}\max(a_1 - a_\infty, 0)$ and $b_t - b_\infty \leq (\beta_R\beta_\Lambda)^{t}\max(b_0 - b_\infty, 0)$, whence the two limit superiors.
\end{proof}

\section{Competing estimators}\label{app:sim-estimators}

This appendix gives the definitions, implementations and grids of hyperparameters of the estimators compared with PCSS in \Cref{sec:simulations}.
For the estimators other than PCSS, $\vartheta > 0$ denotes the hyperparameter, and $\rho_n$ is the sample correlation matrix.

\paragraph{Graphical lasso (GLASSO).}
The graphical lasso~\citep{friedman2008} estimates the precision matrix by maximizing an $\ell_1$-penalized Gaussian log-likelihood.
Applied to $\rho_n$, it gives
\begin{equation*}
  \widehat\Omega(\vartheta) = \argmax_{\Omega \succ 0} \Big\{ \log\det\Omega - \tr(\rho_n\Omega) - \vartheta \sum_{i \neq j} |\Omega_{ij}| \Big\},
\end{equation*}
where the penalty acts on the off-diagonal entries only, and larger values of $\vartheta$ set more of them to zero.
Unlike PCSS, which penalizes the partial correlations, GLASSO penalizes the entries of the precision matrix, and its estimates of the partial correlations follow from $\widehat\Omega(\vartheta)$ through \eqref{eq:partial}.
We select
\begin{equation*}
  \hat\vartheta = \argmin_{\vartheta \in \mathcal G} \mathrm{EBIC}(\vartheta),
  \qquad
  \mathcal G := \big\{ 10^{-3u/29}\, \vartheta_{\max} : u = 0, 1, \dots, 29 \big\},
\end{equation*}
where $\mathrm{EBIC}(\vartheta)$ is \eqref{eq:ebic} with $S_n$ replaced by $\rho_n$ and $\widehat\Omega_\gamma$ by $\widehat\Omega(\vartheta)$, and $\vartheta_{\max} := \max_{i \neq j}|(\rho_n)_{ij}|$ is the smallest value of $\vartheta$ for which $\widehat\Omega(\vartheta)$ is diagonal.

\paragraph{Partial correlation graphical lasso (PCGLASSO).}
The partial correlation graphical lasso of~\citet{carter2023pcglasso} moves the $\ell_1$ penalty of GLASSO from the entries of the precision matrix to the partial correlations, as in PCSS.
In the parametrization \eqref{eq:omega-factorization}, it gives
\begin{equation*}
  \begin{aligned}
    \big(\widehat\Lambda(\vartheta), \widehat R(\vartheta)\big) = \argmax_{\Lambda, R} \Big\{ &\log\det(I_p - R) + 2\Big(1 - \frac{4}{n}\Big) \log\det\Lambda \\
    &- \tr\big(\rho_n \Lambda (I_p - R) \Lambda\big) - \vartheta \sum_{i \neq j} |r_{ij}| \Big\},
  \end{aligned}
\end{equation*}
where the maximum is over positive diagonal matrices $\Lambda$ and over $R \in \mathcal R_p$.
Multiplied by $n/2$, the coefficient of $\log\det\Lambda$ is $n - 4$, against $n$ in the Gaussian log-likelihood and $n - 1$ in \eqref{eq:map-objective}.
The coefficient $1 - 4/n$ is the one proposed in~\citet{carter2023pcglasso}.
We compute the estimator with the R package \texttt{pcglassoFast} of~\citet{bogdan2025}, and select
\begin{equation*}
  \hat\vartheta = \argmin_{\vartheta \in \mathcal G} \mathrm{EBIC}(\vartheta),
  \qquad
  \mathcal G := \Big\{ 5 \cdot 10^{-2u/19} \sqrt{\frac{\log p}{n}} : u = 0, 1, \dots, 19 \Big\},
\end{equation*}
where $\mathrm{EBIC}(\vartheta)$ is \eqref{eq:ebic} with $S_n$ replaced by $\rho_n$ and $\widehat\Omega_\gamma$ by $\widehat\Lambda(\vartheta)\big(I_p - \widehat R(\vartheta)\big)\widehat\Lambda(\vartheta)$.
The grid is traversed from its largest value, with each estimate computed starting from the previous one, and the traversal stops after three consecutive values of $\vartheta$ that do not decrease the smallest EBIC found so far.

\paragraph{Smoothly clipped absolute deviation (SCAD) and minimax concave penalty (MCP).}
The SCAD penalty of~\citet{fan2001}, used for graphical models in~\citet{fan2009}, and the MCP of~\citet{zhang2010} replace the $\ell_1$ penalty of GLASSO by nonconvex penalties $p_\vartheta$.
Both are defined by their derivative on $\omega \geq 0$,
\begin{equation*}
  p'_\vartheta(\omega) = \min\Big\{\vartheta, \frac{(a\vartheta - \omega)_+}{a-1}\Big\} \quad \text{(SCAD)},
  \qquad
  p'_\vartheta(\omega) = \Big(\vartheta - \frac{\omega}{b}\Big)_+ \quad \text{(MCP)},
\end{equation*}
with hyperparameters $a > 2$ and $b > 1$.
The derivative equals $\vartheta$ near $\omega = 0$, as for the $\ell_1$ penalty, and vanishes for large $\omega$.
We compute both estimators with the R package \texttt{GGMncv}~\citep{williams2020}, and use its default values $a = 3.7$, recommended by~\citet{fan2001}, and $b = 2$.
Following the local linear approximation of~\citet{zou2008}, as in~\citet{fan2009}, the estimate at each $\vartheta$ is computed by solving a sequence of weighted $\ell_1$-penalized problems, with weights set at the previous iterate,
\begin{equation*}
  \Omega^{(t+1)} = \argmax_{\Omega \succ 0} \Big\{ \log\det\Omega - \tr(\rho_n\Omega) - \sum_{i,j} p'_\vartheta\big(|\Omega^{(t)}_{ij}|\big)\, |\Omega_{ij}| \Big\},
  \qquad t = 0, 1, \dots
\end{equation*}
As in \texttt{GGMncv}, the sum includes the diagonal entries, whose weights vanish unless $\vartheta > \Omega^{(t)}_{ii}/a$ for SCAD or $\vartheta > \Omega^{(t)}_{ii}/b$ for MCP.
The sequence starts from $\Omega^{(0)} := 0$, for which all weights equal $\vartheta$ and $\Omega^{(1)}$ is the GLASSO estimate with the penalty also on the diagonal entries.
The iterations stop when two consecutive iterates are close, and $\widehat\Omega(\vartheta)$ is the last iterate.
We select
\begin{equation*}
  \hat\vartheta = \argmin_{\vartheta \in \mathcal G} \mathrm{EBIC}(\vartheta),
  \qquad
  \mathcal G := \big\{ 10^{-2u/49}\, \vartheta_{\max} : u = 0, 1, \dots, 49 \big\},
\end{equation*}
with $\mathrm{EBIC}(\vartheta)$ and $\vartheta_{\max} := \max_{i \neq j}|(\rho_n)_{ij}|$ as for GLASSO.
As for PCGLASSO, the grid is traversed from its largest value, and the traversal stops after three consecutive values of $\vartheta$ that do not decrease the smallest EBIC found so far.

\paragraph{Joint sparse regression (SPACE).}
Let $\boldsymbol y_1, \dots, \boldsymbol y_p$ be the columns of the data matrix $X$, each centred and scaled to unit variance.
\citet{peng2009} proposed SPACE, which estimates the partial correlations by fitting the $p$ regressions jointly, with an $\ell_1$ penalty on the partial correlations,
\begin{equation*}
  \widehat R(\vartheta) = \argmin_{R} \Big\{ \frac{1}{2} \sum_{i=1}^{p} \Big\| \boldsymbol y_i - \sum_{j \neq i} r_{ij} \Big(\frac{\widehat\Omega_{jj}}{\widehat\Omega_{ii}}\Big)^{1/2} \boldsymbol y_j \Big\|_2^2 + \vartheta \sum_{i < j} |r_{ij}| \Big\},
\end{equation*}
where the minimum is over symmetric $R$ with zero diagonal.
Each $\widehat\Omega_{ii}$ is estimated as the inverse of the mean squared residual of the regression of $\boldsymbol y_i$.
Starting from $\widehat\Omega_{ii} = 1$, the two estimates are updated in turn three times, as proposed in~\citet{peng2009}.
We use the R package \texttt{space}, with equal weights for the $p$ regressions and no ridge penalty.
We select
\begin{equation*}
  \hat\vartheta = \argmin_{\vartheta \in \mathcal G} \mathrm{EBIC}(\vartheta),
  \qquad
  \mathcal G := \big\{ 10^{u/19}\, \vartheta_0/2 : u = 0, 1, \dots, 19 \big\},
\end{equation*}
where $\vartheta_0 := \sqrt n\, \Phi_N^{-1}(1 - 0.05/p^2)$.
Here $\mathrm{EBIC}(\vartheta)$ is \eqref{eq:ebic} with $S_n$ replaced by $\rho_n$ and $\widehat\Omega_\gamma$ by the precision matrix with diagonal entries $\widehat\Omega_{ii}$ and off-diagonal entries $-\widehat r_{ij}(\vartheta)(\widehat\Omega_{ii}\widehat\Omega_{jj})^{1/2}$.

\paragraph{Nodewise regression (NR).}

\citet{meinshausen2006} proposed to estimate the graph by $p$ separate lasso regressions, one for each variable on the others.
With the columns $\boldsymbol y_j$ now scaled to $\|\boldsymbol y_j\|_2^2 = n$, the regression of variable $i$ gives
\begin{equation*}
  \widehat{\boldsymbol\beta}_i(\vartheta_i) = \argmin_{\boldsymbol\beta_i} \Big\{ \frac{1}{2n}\Big\|\boldsymbol y_i - \sum_{j \neq i} \beta_{ij}\, \boldsymbol y_j\Big\|_2^2 + \vartheta_i \sum_{j \neq i} |\beta_{ij}| \Big\},
\end{equation*}
where $\boldsymbol\beta_i$ collects the $p-1$ coefficients $\beta_{ij}$, $j \neq i$, and $\vartheta_i > 0$ controls the number of variables in the estimated neighbourhood $\mathcal N_i(\vartheta_i) := \{j : \hat\beta_{ij}(\vartheta_i) \neq 0\}$ of $i$.
The estimated graph contains the edge $(i,j)$ if and only if $\hat\beta_{ij} \neq 0$ or $\hat\beta_{ji} \neq 0$, which is known as the OR rule.

The first choice of hyperparameter minimizes, separately for each $i$, the EBIC of~\citet{chen2008} for regression,
\begin{equation*}
  \mathrm{EBIC}_i(\mathcal N) = n \log\frac{\mathrm{RSS}_i(\mathcal N)}{n} + |\mathcal N| \log n + 2\gamma_{\mathrm{ebic}} \log\binom{p-1}{|\mathcal N|},
\end{equation*}
where $\mathrm{RSS}_i(\mathcal N)$ is the residual sum of squares of the least-squares regression of $\boldsymbol y_i$ on the columns indexed by $\mathcal N$.
We select
\begin{equation*}
  \hat\vartheta_i = \argmin_{\vartheta \in \mathcal G_i} \mathrm{EBIC}_i\big(\mathcal N_i(\vartheta)\big),
  \qquad
  \mathcal G_i := \big\{ 10^{-cu/39}\, \vartheta_{i,\max} : u = 0, 1, \dots, u_i - 1 \big\},
\end{equation*}
where $\vartheta_{i,\max} := \max_{j \neq i}|(\rho_n)_{ij}|$ is the smallest value of $\vartheta_i$ for which $\mathcal N_i(\vartheta_i)$ is empty, and $c = 4$ if $n \geq p$ and $c = 2$ if $n < p$.
Following~\citet{fan2013}, we only consider neighbourhoods with at most $\lfloor 3\sqrt n \rfloor$ elements, so $u_i$ is the first value of $u$ for which $|\mathcal N_i(10^{-cu/39}\vartheta_{i,\max})| > \lfloor 3\sqrt n \rfloor$, or $u_i = 40$ if there is none.
Without such a limit, the regressions at the smallest values of $\vartheta_i$ nearly interpolate $\boldsymbol y_i$ when $n \le p$, and the first term of $\mathrm{EBIC}_i$ decreases without bound.
We refer to nodewise regression with this choice of hyperparameter as NR (EBIC).

The second choice of hyperparameter is that of~\citet{meinshausen2006}, the same for every $i$,
\begin{equation*}
  \vartheta_i = \frac{1}{\sqrt n}\, \Phi_N^{-1}\Big(1 - \frac{0.025}{p^2}\Big).
\end{equation*}
It bounds by $0.05$ the probability that some estimated neighbourhood connects two connected components of the true graph.
We refer to nodewise regression with this choice of hyperparameter as NR (MB).

\section{Degrees and clustering of the ground truths}\label{app:sim-truths}

This appendix gives the parameters of the preferential attachment construction, defines the two vertex properties shown in \Cref{fig:truths}, and discusses how the three models differ in them.
For $\langle d\rangle = 2$, the preferential attachment construction starts from a complete graph on $3$ vertices and joins each of the $253$ new vertices to $1$ earlier vertex, which gives $m = 256$ edges.
For $\langle d\rangle = 17.5$, it starts from a complete graph on $10$ vertices, and joins $227$ of the $246$ new vertices, chosen uniformly at random, to $9$ earlier vertices and the other $19$ to $8$, which gives $m = 2240$ edges.

The degree of vertex $i$ is its number of neighbours, $d_i := |\{j : (i,j) \in \mathcal S\}|$, so that $\langle d\rangle$ is the average of the $d_i$ and the largest degree $d$ of \Cref{sec:theory} is their maximum.
The local clustering coefficient of $i$~\citep{watts1998} is the fraction of pairs of neighbours of $i$ that are themselves neighbours,
\begin{equation*}
  C_i := \frac{\big|\{(j,k) \in \mathcal S : (i,j) \in \mathcal S,\ (i,k) \in \mathcal S\}\big|}{d_i(d_i - 1)},
  \qquad d_i \geq 2.
\end{equation*}
The Erd\H{o}s--R\'enyi and torus models give almost the same degree distribution, concentrated around $\langle d\rangle$, and differ in their clustering.
In the Erd\H{o}s--R\'enyi graph two neighbours of a vertex are joined with the same probability as any other pair of vertices, so the $C_i$ are small.
In the torus graph two neighbours of a vertex are close to it, and hence often close to each other, so the $C_i$ are much larger.
Preferential attachment gives a wider degree distribution.
No degree is below $8$, most degrees are close to this value, and a few vertices reach degrees of about $100$.
The clustering coefficients of the preferential attachment graph lie between those of the other two models.

\section{Additional simulation figures}\label{app:sim-figures}

\begin{figure}[h]
  \begin{center}
    \includegraphics[width=0.95\textwidth]{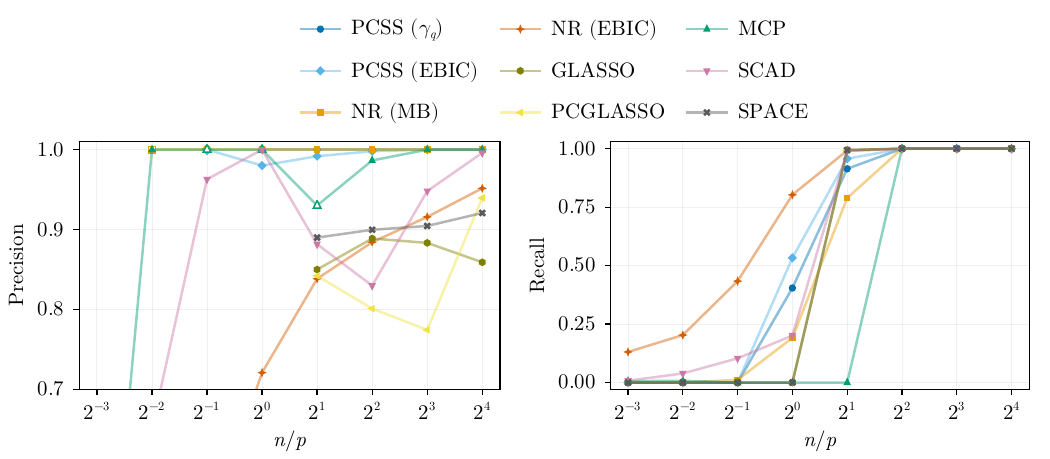}
  \end{center}
  \caption{
    As \Cref{fig:overview}, for the Erd\H{o}s--R\'enyi ground truths with $\langle d\rangle = 2$ and $r^* = 4 r_{\max}/5 = 0.2253$.
  }
  \label{fig:overview_er_d2}
\end{figure}

\begin{figure}[h]
  \begin{center}
    \includegraphics[width=0.95\textwidth]{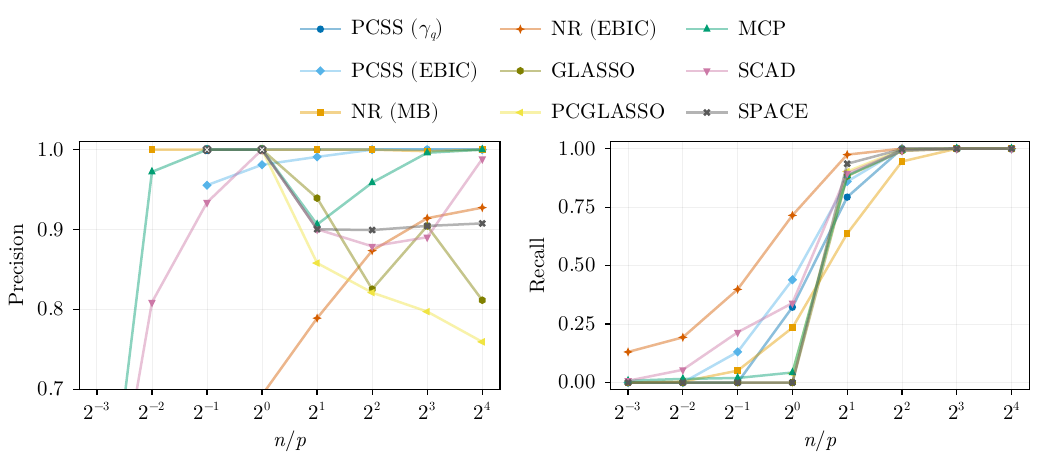}
  \end{center}
  \caption{
    As \Cref{fig:overview}, for the torus ground truths with $\langle d\rangle = 2$ and $r^* = 4 r_{\max}/5 = 0.2253$.
  }
  \label{fig:overview_torus_d2}
\end{figure}

\begin{figure}[h]
  \begin{center}
    \includegraphics[width=0.95\textwidth]{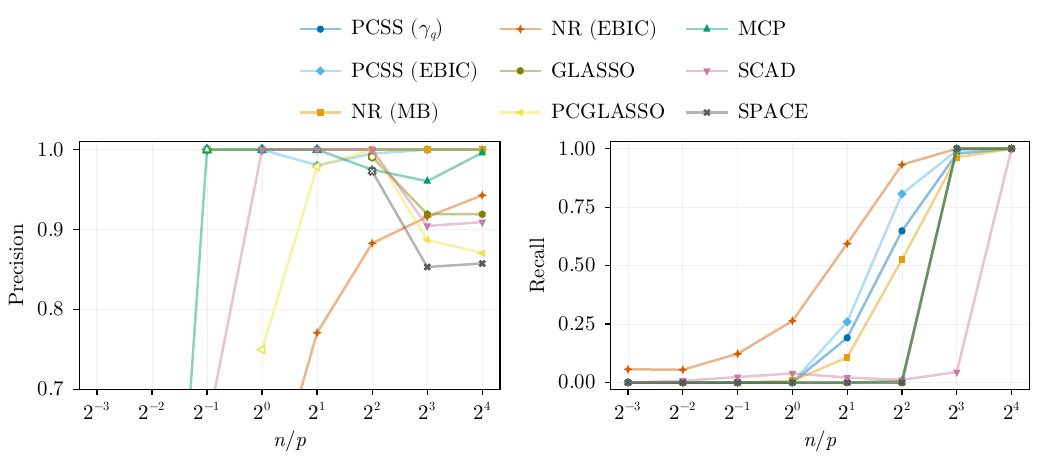}
  \end{center}
  \caption{
    As \Cref{fig:overview}, for the preferential attachment ground truths with $\langle d\rangle = 2$ and $r^* = 4 r_{\max}/5 = 0.1409$.
  }
  \label{fig:overview_pa_d2}
\end{figure}

\begin{figure}[h]
  \begin{center}
    \includegraphics[width=0.95\textwidth]{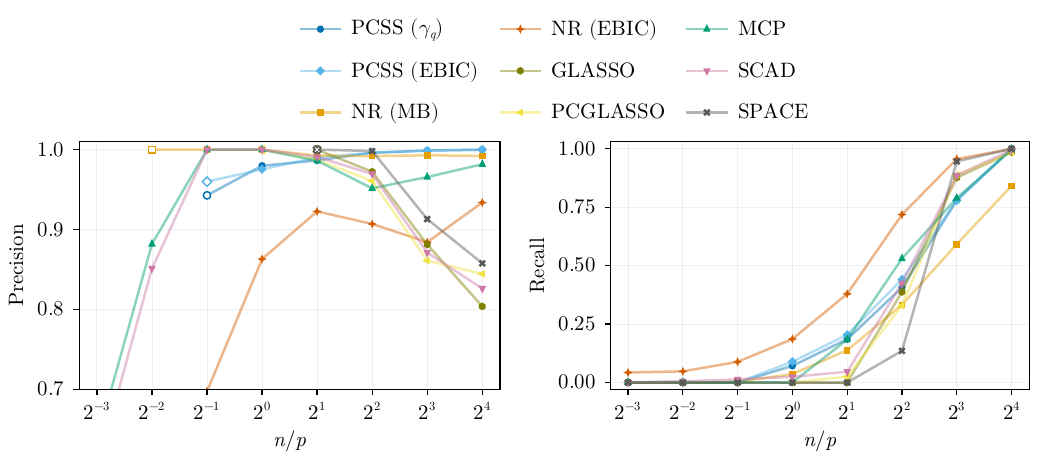}
  \end{center}
  \caption{
    As \Cref{fig:overview}, for the torus ground truths with $\langle d\rangle = 17.5$ and $r^* = r_{\max} = 0.1053$.
  }
  \label{fig:overview_torus_d17p5}
\end{figure}

\begin{figure}[h]
  \begin{center}
    \includegraphics[width=0.95\textwidth]{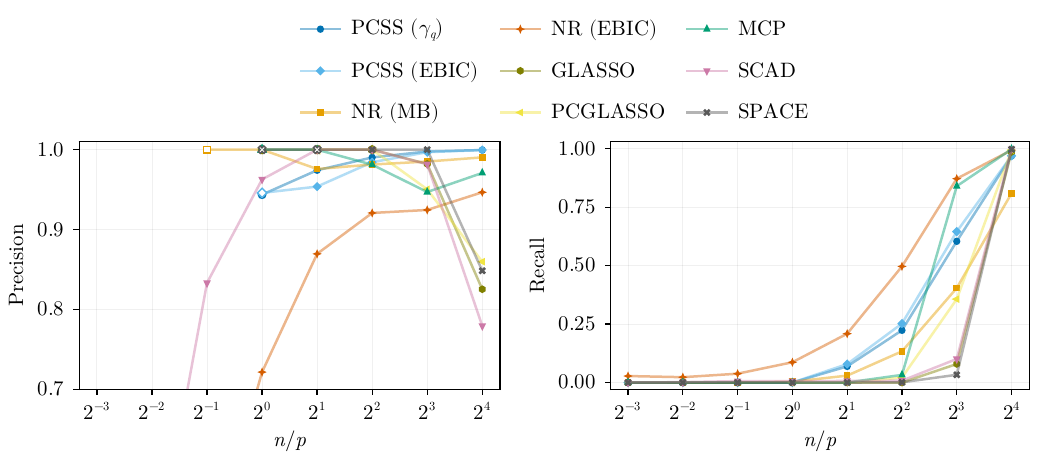}
  \end{center}
  \caption{
    As \Cref{fig:overview}, for the preferential attachment ground truths with $\langle d\rangle = 17.5$ and $r^* = r_{\max} = 0.0822$.
  }
  \label{fig:overview_pa_d17p5}
\end{figure}

\begin{figure}[h]
  \begin{center}
    \includegraphics[width=0.95\textwidth]{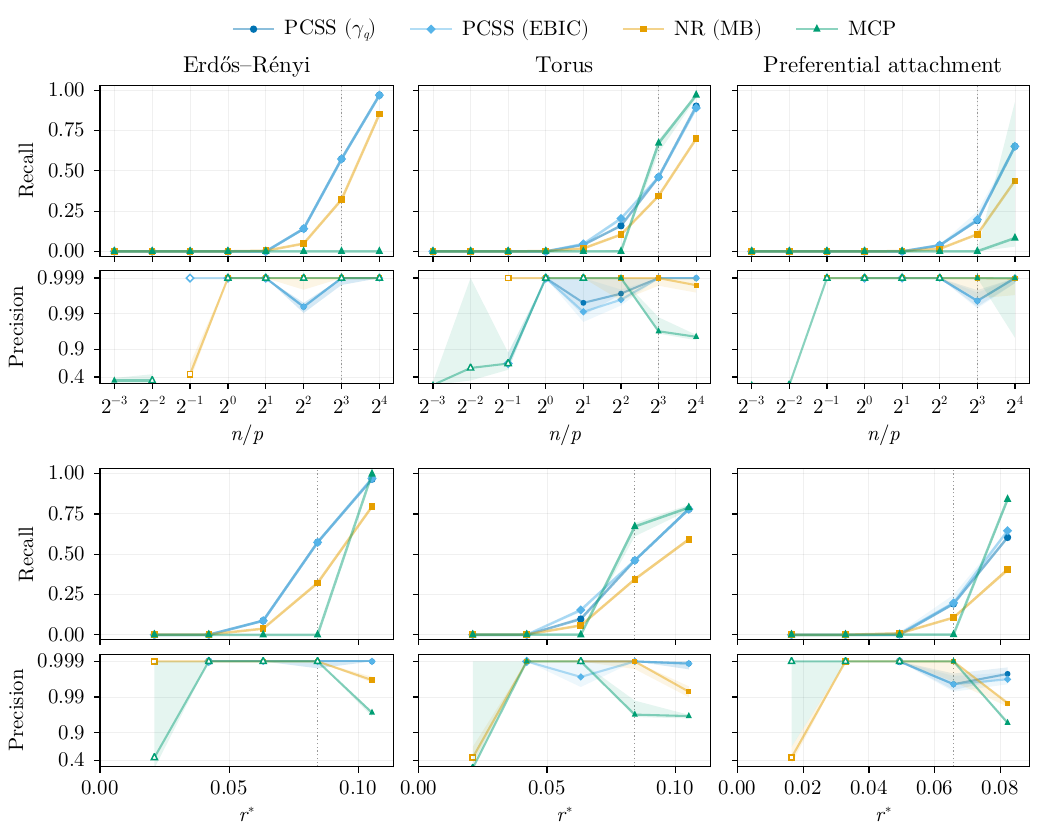}
  \end{center}
  \caption{
    As \Cref{fig:pr_main_d2}, for the ground truths with $\langle d\rangle = 17.5$.
    The top row varies $n/p$ at $r^* = 4 r_{\max}/5$, which is $0.0843$ for the Erd\H{o}s--R\'enyi and torus models and $0.0658$ for the preferential attachment model.
    The bottom row varies $r^*$ at $n = 2^3 p$.
  }
  \label{fig:pr_main_d17p5}
\end{figure}

\Cref{fig:overview_er_d2,fig:overview_torus_d2,fig:overview_pa_d2,fig:overview_torus_d17p5,fig:overview_pa_d17p5} show precision and recall of the estimators of \Cref{sec:simulations} for the combinations of model and $\langle d\rangle$ not shown in \Cref{fig:overview}, each at a single magnitude $r^*$.
They follow the conventions of \Cref{fig:overview}, with $r^* = r_{\max}$ at $\langle d\rangle = 17.5$ and $r^* = 4 r_{\max}/5$ at $\langle d\rangle = 2$.
At $\langle d\rangle = 2$ and $r^* = r_{\max}$, GLASSO, PCGLASSO, SCAD, SPACE and MCP recover most of the true edges of the Erd\H{o}s--R\'enyi and torus ground truths already at $n = p$, with a precision below $0.95$.
The smaller magnitude $4 r_{\max}/5$ is therefore shown, to separate the three regimes of \Cref{subsec:sim-overview}.
\Cref{fig:pr_main_d17p5} is the counterpart of \Cref{fig:pr_main_d2} for the ground truths with $\langle d\rangle = 17.5$.

The three regimes of \Cref{subsec:sim-overview} appear in all these figures, at values of $n/p$ that depend on the model, on $\langle d\rangle$ and on $r^*$.
Three differences from \Cref{fig:overview} stand out.
On the torus at $\langle d\rangle = 17.5$ (\Cref{fig:overview_torus_d17p5}), the recall of GLASSO, PCGLASSO, SCAD and SPACE rises over two doublings of $n$, and that of MCP gradually, so that at $n = 2^2 p$ MCP recovers more edges than PCSS with a lower precision, and SCAD about as many with a slightly lower precision.
At $\langle d\rangle = 2$, SCAD and MCP estimate a few edges at $n = 2^{-2} p$, where PCSS estimates none, in some conditions with a precision above $0.95$.
Also at $\langle d\rangle = 2$, the precision of SCAD falls below $0.9$ once its recall rises, but then recovers and reaches $0.99$ at $n = 2^4 p$ for the Erd\H{o}s--R\'enyi and torus ground truths (\Cref{fig:overview_er_d2,fig:overview_torus_d2}).

\end{document}